\RequirePackage{fix-cm}
\documentclass[twocolumn,epjc3]{svjour3}  

\smartqed
\journalname{Eur. Phys. J. C}

\RequirePackage{graphicx}
\RequirePackage{amsmath}
\RequirePackage{mathtools}
\RequirePackage{booktabs}
\RequirePackage[tableposition=bottom]{caption}
\RequirePackage{hyperref}
\RequirePackage{makecell}
\RequirePackage{nicefrac}
\RequirePackage{physics}
\RequirePackage{cite}
\RequirePackage[a4paper, total={6.5in, 8.5in}]{geometry}
\RequirePackage{xcolor}
\RequirePackage{amssymb}
\RequirePackage{multirow}
\RequirePackage{hepunits}
\RequirePackage{afterpage}
\RequirePackage[utf8]{inputenc}
\RequirePackage{enumitem}
\RequirePackage[normalem]{ulem}

\newcommand{\xBj}{x_{\mathrm{Bj}}}
\newcommand{\eg}{\textit{e.g.}\xspace}
\newcommand{\ie}{\textit{i.e.}\xspace}
\newcommand{\cf}{\textit{cf.}\xspace}
\newcommand{\etc}{\textit{etc.}\xspace}
\def\arrowedvec{\mathaccent"017E}
\newcommand*\stepsymbol{\includegraphics{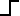}}
\def\figWidth{0.48\textwidth}
\def\figWidthSmall{0.45\textwidth}
\def\figSpace{10px}

\newcommand{\reffig}[1]{\textrm{Fig}.~\ref{#1}}
\newcommand{\refcite}[1]{\textrm{Ref}.~\cite{#1}}
\newcommand{\refsec}[1]{\textrm{Sec}.~\ref{#1}}

\newcounter{comment}

{\refstepcounter{comment}%
\begin{quote}
\small$\blacksquare$ \textbf{\underline{Comment} $\sharp$\thecomment:}}%
{\end{quote}}

{\refstepcounter{comment}%
\begin{quote}
\textcolor{red}{\small$\blacksquare$ \textbf{\underline{Comment} $\sharp$\thecomment:~HM:}}}%
{\end{quote}}

{\refstepcounter{comment}%
\begin{quote}
\textcolor{blue}{\small$\blacksquare$ \textbf{\underline{Comment} $\sharp$\thecomment:~HM:}}}%
{\end{quote}}

{\refstepcounter{comment}%
\begin{quote}
\textcolor{green}{\small$\blacksquare$ \textbf{\underline{Comment} $\sharp$\thecomment:~HM:}}}%
{\end{quote}}

\begin{document}

\title{Unbiased determination of DVCS Compton Form Factors}

\author{
H.~Moutarde\thanksref{e1,addr1}
\and
P.~Sznajder\thanksref{e2,addr2} 
\and
J.~Wagner\thanksref{e3,addr2} 
}

\thankstext{e1}{e-mail: herve.moutarde@cea.fr}
\thankstext{e2}{e-mail: pawel.sznajder@ncbj.gov.pl}
\thankstext{e3}{e-mail: jakub.wagner@ncbj.gov.pl}


\institute{
IRFU, CEA, Universit\'e Paris-Saclay, F-91191 Gif-sur-Yvette, France \label{addr1}
\and
~National Centre for Nuclear Research (NCBJ), Pasteura 7, 02-093 Warsaw, Poland \label{addr2}
}

\date{Received: date / Accepted: date}

\maketitle

\sloppy

\begin{abstract}
The extraction of Compton Form Factors (CFFs) in a global analysis of almost all Deeply Virtual Compton Scattering (DVCS) proton data is presented. The extracted quantities are DVCS sub-amplitudes and the most basic observables which are unambiguously accessible from this process. The parameterizations of CFFs are constructed utilizing the artificial neural network technique allowing for an important reduction of model dependency. The analysis consists of such elements as feasibility studies, training of neural networks with the genetic algorithm and a careful regularization to avoid over-fitting. The propagation of experimental uncertainties to extracted quantities is done with the replica method. The resulting parameterizations of CFFs are used to determine the subtraction constant through dispersion relations. The analysis is done within the PARTONS framework. 

\end{abstract}

\section{Introduction}
\label{sec:introduction}

An intense experimental activity has been dedicated over the last twenty years to the measurements of observables towards a 3D description of the nucleon.  
Most of the knowledge about the 3D quark and gluon structure is embodied in Generalized Parton Distributions (GPDs) and Transverse Momentum Dependent parton distribution functions (TMDs).
The continuing efforts in understanding and determining GPDs and TMDs are partly driven by the quest for nucleon tomography in mixed position-momentum space or pure momentum space (see \eg \refcite{Diehl:2015uka, Bacchetta:2016ccz} and refs. therein), the access to the nucleon energy-momentum tensor \cite{Ji:1996ek, Ji:1996nm} and the description of the mechanical properties (radial and tangential pressures, energy density, surface tension, \etc) of the nucleon \cite{Polyakov:2002wz, Polyakov:2002yz, Polyakov:2018zvc, Lorce:2018egm}.

Relations between measurements and GPDs or TMDs are derived through factorization theorems, established at all order in QCD perturbation theory (pQCD). 
In particular, this connection can be brought under good theoretical control with a careful and systematic check of various pQCD assumptions.
This task requires a sophisticated and modular computing machinery with fitting features. 
In the context of GPDs, the open-source PARTONS framework \cite{Berthou:2015oaw} has been designed and publicly released to fulfill the needs of the experimental and theoretical GPD communities. 

From a theoretical point of view, Deeply Virtual Compton Scattering (DVCS) is the cleanest channel to access GPDs. 
Recent status of the phenomenology of this channel can be found in \refcite{Kumericki:2016ehc} and of the related collected data sets in \refcite{dHose:2016mda}. 
The DVCS cross section is described in terms of Compton Form Factors (CFFs), which are convolutions of GPDs with coefficient functions computed order by order in pQCD.
While GPDs are in principle renormalization and factorization scale-dependent, CFFs do not depend on any arbitrary scales, and are the off-forward analogs of the classical structure functions of Deeply Inelastic Scattering (DIS) -- with GPDs being the off-forward extensions of the Parton Distribution Functions (PDFs) extracted from DIS.

Global GPD fits over the whole DVCS kinematic domain, from the glue to the valence region, have not been achieved yet. 
However, a great deal is already known about CFFs, in particular in the valence region.
This is indeed precious knowledge: due to the scale dependence, GPDs can only be promoted to the status of quasi-observables thanks to universality (the same GPDs should allow the simultaneous analysis of several independent exclusive processes), while CFFs are genuine observables. 
Hence measuring CFFs is an interesting task \textit{per se}. 
Moreover, having a unified description of all existing DVCS measurements in terms of CFFs is important in view of the definition of future experimental programmes: one has to select an observable to be measured, specify the kinematic region, and evaluate the required uncertainty in order to make substantial progress in the understanding of the DVCS process.
This is particularly relevant for future experimental campaigns at Jefferson Lab or for Electron-Ion Collider (EIC) design studies.

As mentioned in \refcite{Kumericki:2016ehc}, the attempts to determine CFFs from experimental data mostly fall into two categories: 
\begin{description}
  \item[\textbf{Local fits}] CFFs are independently determined from measurements between different kinematic bins. 
  This amounts to a sampling of CFFs over the probed kinematic domain. 
  The model-dependence of the result is low, but most of the time, the problem is ill-posed by lack of uniqueness.
  \item[\textbf{Global fits}] The free coefficients of a CFF parameterization are matched to experimental data. 
  Kinematic bins are no more treated independently. 
  Interpolating between measurements of the same observable on neighboring kinematic bins is feasible. 
  Extrapolating out of the domain constrained by measurements becomes possible, paving the way for impact studies.
  However the estimation of the systematic uncertainty associated to the choice of a parameterization is an extremely difficult task, and this is a limit to potential impact studies.
\end{description}
A possible solution to the tension between the advantages and drawbacks of these two approaches may be found in Artificial Neural Networks (ANNs) \cite{Hassoun:1995:FAN:526717}. 
While being of common use for PDFs today, their role in GPD-related physics has been assessed (to the best of our knowledge) only in the pioneering work of Kumeri\v{c}ki, M\"uller and Sch\"afer \cite{Kumericki:2011rz}. 
ANNs offer the joint promise of a great flexibility and of a common description of data sets in different kinematic regions.
This difficulty of this last point should not be underestimated. 
In our previous fit of CFFs from almost all existing DVCS data \cite{Moutarde:2018kwr}, our physically-motivated choice of CFF parameterizations restricted the scope of our study to the quark sector.
There is a need for flexible parameterizations while escaping at the same time the \emph{curse of dimensionality}, and in this respect, ANNs are quite appealing solutions.

We present here the first global fit of most existing DVCS measurements in terms of CFFs described by neural networks. 
The paper is organized as follows. 
In \refsec{sec:theory_framework} we briefly introduce the DVCS channel, define CFFs and stress their role in dispersion relations. 
\refsec{sec:methodology} is a reminder of ANN technique and of the genetic algorithm. 
Our implementation of those techniques and the main elements of our analysis are detailed in \refsec{sec:implemenation}. 
We point out the experimental data being used in this fit in \refsec{sec:data}, and outline our results with a focus on the DVCS subtraction constant in \refsec{sec:results}. This discussion is of major contemporary interest since it is related to the measurability of the distribution of pressure forces in the proton \cite{Polyakov:2018zvc, Lorce:2018egm, Burkert:2018bqq, Kumericki:2019ddg}. At last we summarize in \refsec{sec:summary}.   

\section{Theory framework}
\label{sec:theory_framework}

The golden channel in the GPD extraction programme is the leptoproduction of a single real photon off a nucleon:
\begin{equation}
l(k) + N(p) \rightarrow l(k') + N(p') + \gamma(q') \, .
\end{equation}
Here, the symbols between parentheses denote the four-momentum vectors of the lepton, $l$, the nucleon, $N$, and of the real photon, $\gamma$. The amplitude for this process, $\mathcal{T}$, is given by a sum of amplitudes for two processes having the same initial and final states: $\mathcal{T}^{\mathrm{BH}}$ for the purely electromagnetic Bethe-Heitler process and $\mathcal{T}^{\mathrm{DVCS}}$ for the hadronic DVCS process, such as:
\begin{equation}
\mathcal{T} = \mathcal{T}^{\mathrm{BH}} +\mathcal{T}^{\mathrm{DVCS}} \, .
\end{equation}
The Bethe-Heitler part can be expressed with a great precision in terms of the nucleon electromagnetic form factors. The DVCS part is generally parameterized by twelve helicity amplitudes \cite{Kroll:1995pv}, or equivalently twelve CFFs. However, in this analysis we restrict ourselves only to four of them, which can be related to the leading-twist, chiral-even GPDs. Those CFFs are usually denoted by $\mathcal{H}$, $\mathcal{E}$, $\widetilde{\mathcal{H}}$ and $\widetilde{\mathcal{E}}$, and nowadays they are the most extensively studied ones in the context of GPD phenomenology. The exploration of other CFFs, which can be related to either higher-twist or chiral-odd GPDs, suffers from the sparsity of data collected in the kinematic domain where the factorization theorem applies.

The cross section for single photon production is usually expressed in terms of four variables: (\emph{i}) the usual Bjorken variable, $\xBj = Q^{2}/\left(2 p \cdot \left( k - k' \right)\right)$, (\emph{ii}) the square of the four-momentum transfer to the target nucleon, $t = (p' - p)^{2}$, (\emph{iii}) the virtuality of the exchanged photon, $Q^2 = -q^2 = -(k'-k)^2$, and (\emph{iv}) the azimuthal angle between the leptonic (spanned by incoming and outgoing lepton momenta) and production (spanned by virtual and outgoing photon momenta) planes. In the case of a transversely polarized target one also introduces the angle $\phi_S$ between the leptonic plane and the nucleon polarization vector. It is also convenient to exchange the usual Bjorken variable for the generalized one:
\begin{equation}
\xi = -\frac{(q+q')^2}{2(p+p')(q+q')} \, ,
\end{equation}
which in the case of DVCS is approximately equal to the skewness variable:
\begin{equation}
\eta = \frac{(q'-q)(q+q')}{(p+p')(q+q')}\, .
\end{equation}

In this paper a set of lengthy formulae relating CFFs with cross sections is omitted for brevity, but those can be easily found \eg in Ref. \cite{Belitsky:2012ch}. The unpublished analytic expressions of BH and DVCS amplitudes by Guichon and Vanderhaeghen are used in this analysis, as implemented and publicly available in the PARTONS framework \cite{Berthou:2015oaw}. Let us note at this moment that the leptoproduction of a single real photon is sensitive to the beam charge, but also to the beam and target polarization states. This gives rise to exclusive measurements performed with variety of experimental setups, allowing for a combination of cross sections to probe specific sub-processes (BH, DVCS or the interference between them) or combinations of CFFs. Experimental data used in this analysis will be introduced in \refsec{sec:data}.

Our goal here is a global analysis of DVCS data avoiding any model-dependency whatsoever. We separately extract the real and imaginary parts of the four aforementioned CFFs in the three-dimensional phase-space of ($\xi$, $t$, $Q^2$). During the extraction we do not exploit the fact that the real and imaginary parts of a single CFF can be related together by a fixed-$t$ dispersion relation (see \refcite{Diehl:2007jb} for a study at all orders in pQCD):
\begin{flalign}
&\mathrm{Re} \mathcal{G}(\xi) = C_{G} + \nonumber \\
&\phantom{xx}\frac{1}{\pi}\int_{0}^1 \mathrm{d}\xi'~
\mathrm{Im} \mathcal{G}(\xi')\left[\frac{1}{\xi-\xi'} \mp \frac{1}{\xi+\xi'}\right] \, ,
\label{eq:theory_framework:dr}
\end{flalign}
Here, $\mathcal{G} \in \{ \mathcal{H}, \mathcal{E}, \widetilde{\mathcal{H}}, \widetilde{\mathcal{E}} \}$ denotes a single CFF with $t$ and $Q^2$ dependencies suppressed for brevity, one has minus sign for $\mathcal{G} \in \{ \mathcal{H}, \mathcal{E} \}$ and plus sign for $\mathcal{G} \in \{ \widetilde{\mathcal{H}}, \widetilde{\mathcal{E}} \}$ in the square brackets, and where $C_G$ is a subtraction constant associated to the corresponding CFF (and GPD), where:
\begin{flalign}
&C_{H} = - C_{E} \, ,\\
&C_{\widetilde{H}} = C_{\widetilde{E}} = 0 \, .
\end{flalign}
The dispersion relation is only used afterwards to determine the subtraction constant from the extracted CFF parameterizations. This quantity has on its own important interpretation in terms of strong forces acting on partons inside the nucleon as shown in the recent review \refcite{Polyakov:2018zvc}. We also use the requirement of having $C_{G}$ independent on $\xi$ as a consistency check of our analysis. 

Although hadron tomography \cite{Burkardt:2000za, Burkardt:2002hr, Burkardt:2004bv} is a very important motivation for the phenomenology of the DVCS process, in the present analysis we restrict ourselves to (quasi) model-independent statements. Therefore we will only sketch what we can foresee from the obtained results and leave detailed quantitative tomographic interpretation for future studies.

\section{Methodology}
\label{sec:methodology}

To make this paper self-contained, we remind here some elements of the theory and practice of ANNs and genetic algorithms, and introduce the associated terminology.

\subsection{Artificial neural networks}
\label{sec:methodology:artificial_neural_networks}

ANNs \cite{Hassoun:1995:FAN:526717} are systems processing information. What distinguishes those systems from other information techniques is a unique inspiration from Nature, namely by biological neural networks making human and animal brains. All neural networks have a similar structure made out of simple, but highly connected elements called neurons. In this work we will exclusively focus on feed-forward neural networks, but many other types of ANNs exist. Their usefulness in the study of hard exclusive reactions has not been assessed yet.

\begin{figure}[!ht]
\begin{center}
\includegraphics[width=0.45\textwidth]{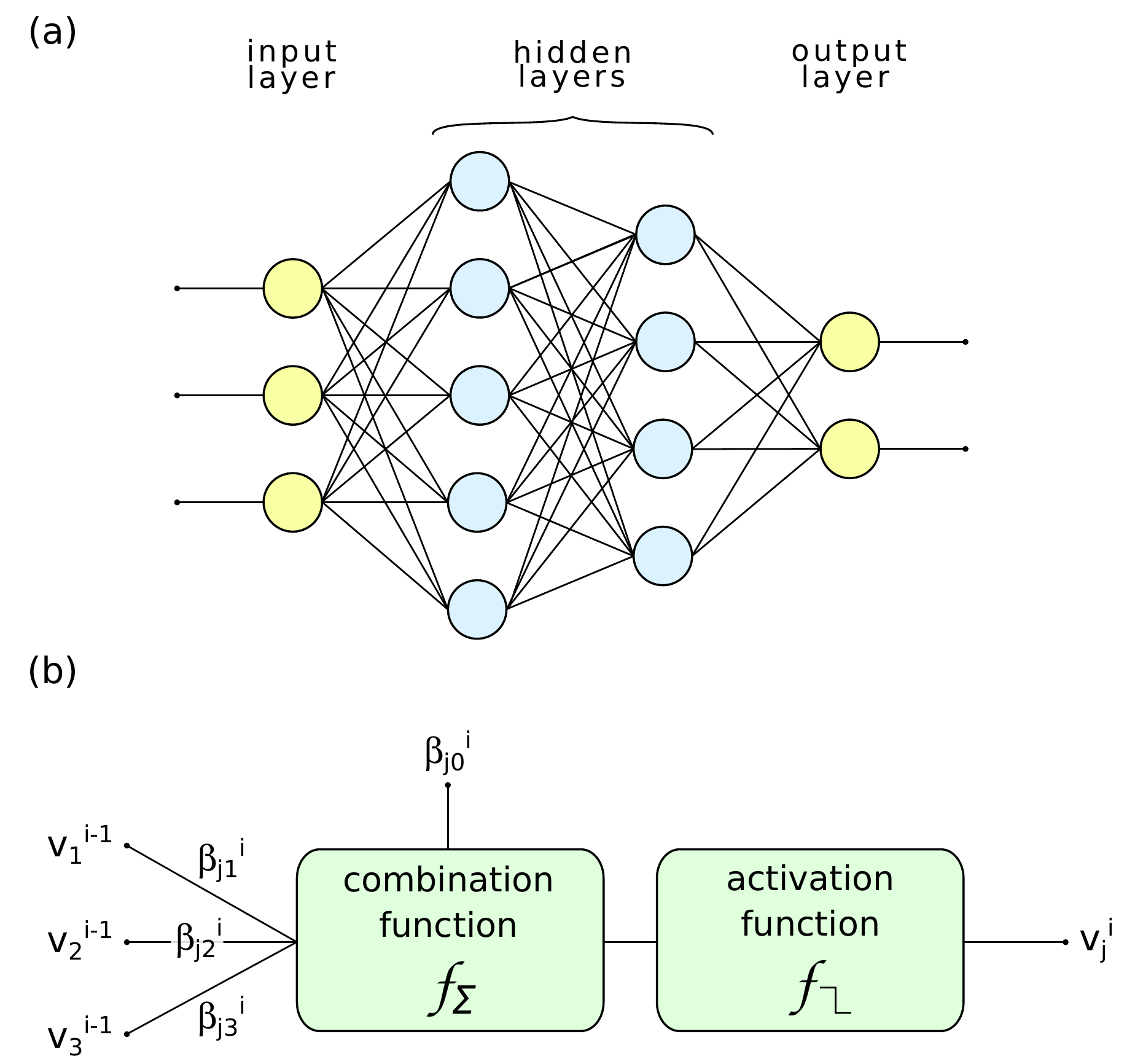}
\caption{(a) Example of a typical feed-forward neural network. The network processes three input variables to give two output values ($3 \to 2$), with the architecture made out of two hidden layers. The first hidden layer contains five neurons, while the second one is made out of four neurons. (b) Schematic illustration of a typical neuron. In this example it is connected to three different neurons belonging to a previous layer of the network. For the explanation of symbols see the text.}
\label{fig:methodology:artificial_neural_networks:nn_1}
\end{center}
\end{figure}

An example of a typical feed-forward neural network structure is shown in \reffig{fig:methodology:artificial_neural_networks:nn_1}a. The data are processed layer-by-layer. One can distinguish three type of layers: (\emph{i}) one input layer, storing and distributing input information, being the first layer of the network, (\emph{ii}) a number of hidden layers processing information and (\emph{iii}) one output layer, aggregating and returning output information, being the last layer of the network. The number of hidden layers is a parameter of the network architecture. 

In machine learning neurons share a similar structure shown in \reffig{fig:methodology:artificial_neural_networks:nn_1}b. A single neuron processes information by evaluating two functions: (\emph{i}) combination function, $f_{\Sigma, j}^{i}(\arrowedvec{\beta}_{j}^{i}, \arrowedvec{v}^{i-1})$, and (\emph{ii}) activation function, $f_{\stepsymbol, j}^{i}(x)$, so that the overall neuron's output, $v_{j}^{i}$, is given by:
\begin{equation}
v_{j}^{i} = f_{\stepsymbol, j}^{i}\left( f_{\Sigma, j}^{i}(\arrowedvec{\beta}_{j}^{i}, \arrowedvec{v}^{i-1}) \right) \, .
\end{equation}
Here, the index $i$ identifies a specific layer of the neural network, while $j$ denotes a single neuron in that layer. The vector $\arrowedvec{\beta}_{j}^{i}$ contains a so-called \emph{bias}, $\beta_{j0}^{i}$, and all \emph{weights} associated to the input nodes, $\beta_{jk}^{i}$, with $k=1, 2, \dots$ The vector $\arrowedvec{v}^{i-1}$ contains values returned by neurons in the previous layer, $v_{k}^{i-1}$, where again $k=1, 2, \dots$ The role of the \emph{combination function} is the aggregation of input information to a single value to be processed by the \emph{activation function}. A typical combination function is based on the dot product of two vectors:
\begin{equation}
f_{\Sigma, j}^{i} (\arrowedvec{\beta}_{j}^{i}, \arrowedvec{v}^{i-1}) = \beta_{j0}^{i} + \beta_{j1}^{i} v_{1}^{i-1} + \beta_{j2}^{i} v_{2}^{i-1} + \dots \,
\end{equation}

\begin{figure}[!ht]
\begin{center}
\includegraphics[width=0.45\textwidth]{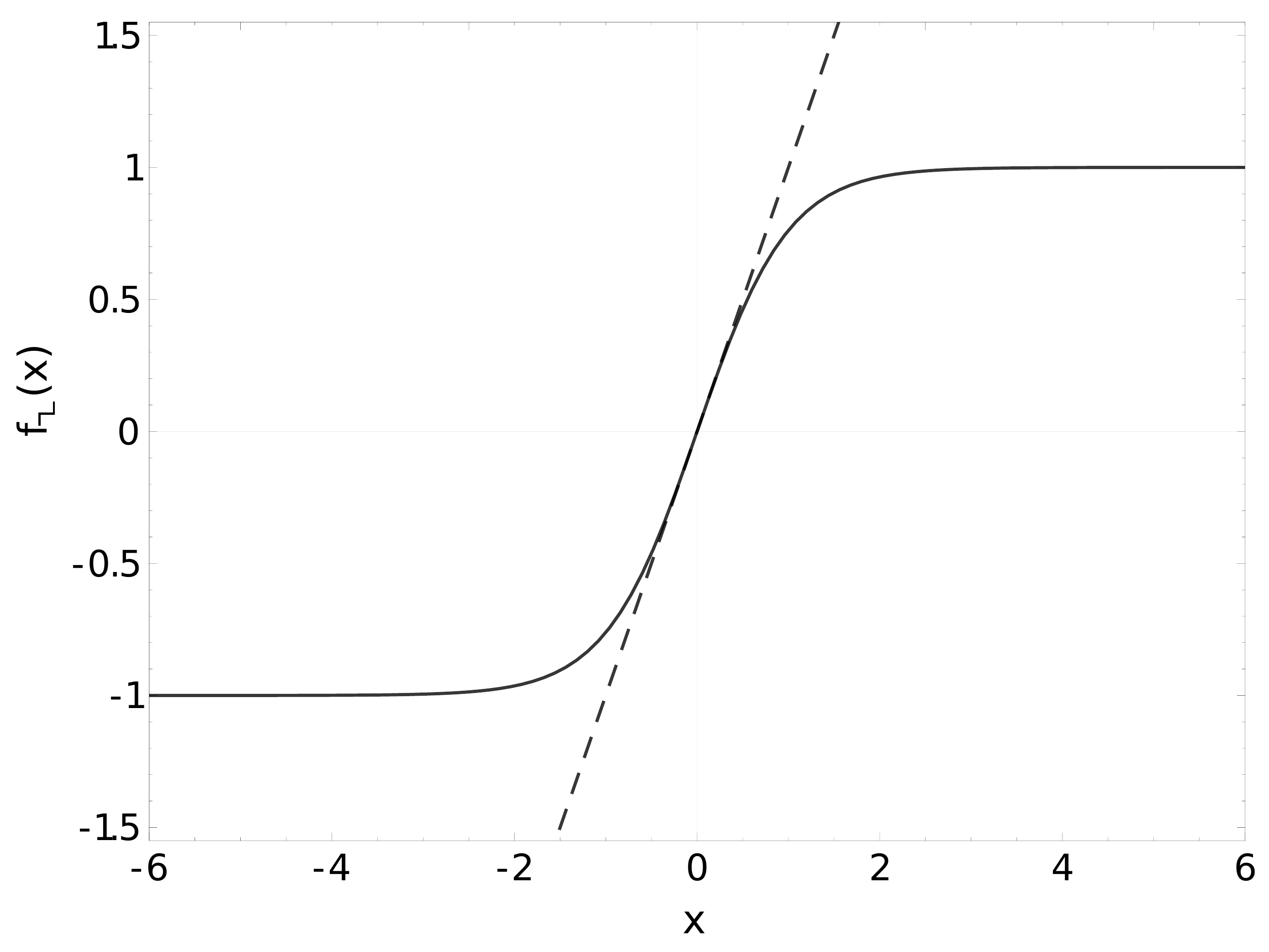}
\caption{Hyperbolic tangent (solid curve) and linear (dashed curve) activation functions.}
\label{fig:methodology:artificial_neural_networks:activation_function}
\end{center}
\end{figure}

The activation function gives an answer returned by the neuron. It can be a numeric value, but it can be also a binary answer (``yes'' or ``no''), which is typically used in classification applications. Many types of activation functions exist. In this work we use the hyperbolic tangent function for neurons in the hidden layers,
\begin{equation}
f_{\stepsymbol, j}^{i}(x) = \tanh(x) \, ,
\end{equation} 
and the linear function (identity) for neurons in the output layer, 
\begin{equation}
f_{\stepsymbol, j}^{i}(x) = x \, .
\end{equation} 
We illustrate both functions with \reffig{fig:methodology:artificial_neural_networks:activation_function}. For the hyperbolic tangent function one can spot an active part near $-1 < x < 1$ where the signal is processed almost linearly. Outside this range a saturation appears, so that the input signal is blocked, \ie it is not processed by a neuron in an effective way. Those features put extra constraints on input data to be processed by the network. In particular, those data must be normalized to the range of $(-1, 1)$, so they can be processed by the first layer of the network.

Biases and weights are free parameters of neural networks. Instead of predefining those parameters a training is done, and only during that training biases and weights are fixed to accommodate the network's answer to the stated problem. Such training is done with a \emph{training sample}, which should be as representative as possible for the problem under consideration. Both input and output information are known for the training sample. 

In a wide variety of applications neural networks are trained with the back-propagation algorithm \cite{goodfellow2016deep}. This algorithm minimizes the so-called \emph{cost function} (\eg root mean square error, RMSE) against the training sample. Such a minimization usually is straightforward, as typically the cost function is analytically differentiable with respect to each parameter of the network, so that its gradient is known in the space of those parameters. However, there are cases where the back-propagation algorithm (or similar) cannot be used. Those are in particular problems with the differentiation of the cost function being practically difficult. In those cases one can evaluate the gradient numerically. Eventually, one can use the genetic algorithm \cite{Mitchell:1998:IGA:522098}, which we will describe in the next paragraph. It is a popular replacement for the back-propagation algorithm in the context of machine learning. In particular, the genetic algorithm is used in this analysis, as a differentiation of cross sections and other observables with respect to the free parameters of CFF parameterizations is not straightforward. 

\subsection{Genetic algorithm}
\label{sec:methodology:learning_strategies}

The genetic algorithm \cite{Mitchell:1998:IGA:522098} is a heuristic technique used for search and optimization purposes. It is applicable in particular whenever standard minimization techniques fail because of the complexity of the problems under consideration. Because of that, but also because of a unique ability to avoid local minima, the genetic algorithm is commonly used in physics, mathematics, engineering and manufacturing. It is frequently used for the learning of neural networks, where it supervises the search of biases and weights against the training sample.

The genetic algorithm mimics mechanisms of reproduction, natural selection and mutation, all being the cornerstones of the evolution process known from Nature. It utilizes nomenclature inspired by the natural evolution process. A \emph{population} is a set of possible solutions to the stated problem. A single solution is referred to as a \emph{candidate} and it is characterized by a set of \emph{genes} making up a \emph{genotype}. In the following, we will thus speak equivalently of candidates and genotypes. The usability of single candidates in solving problems is examined by a \emph{fitness function}, whose argument is the genotype. To help the reader understand this nomenclature we explain it with an example of a multidimensional fit to experimental data. In such a fit a single fitted parameter may be identified as one gene, while a complete set of those parameters as the genotype. The ability to describe experimental data by a given set of parameters is typically examined with a $\chi^{2}$ function, which in this case we will refer to as the fitness function. A set of possible solutions makes the population. We point out that in general the implementation of introduced terms strongly depends on the problem under consideration.

\begin{figure}[!ht]
\begin{center}
\includegraphics[width=0.45\textwidth]{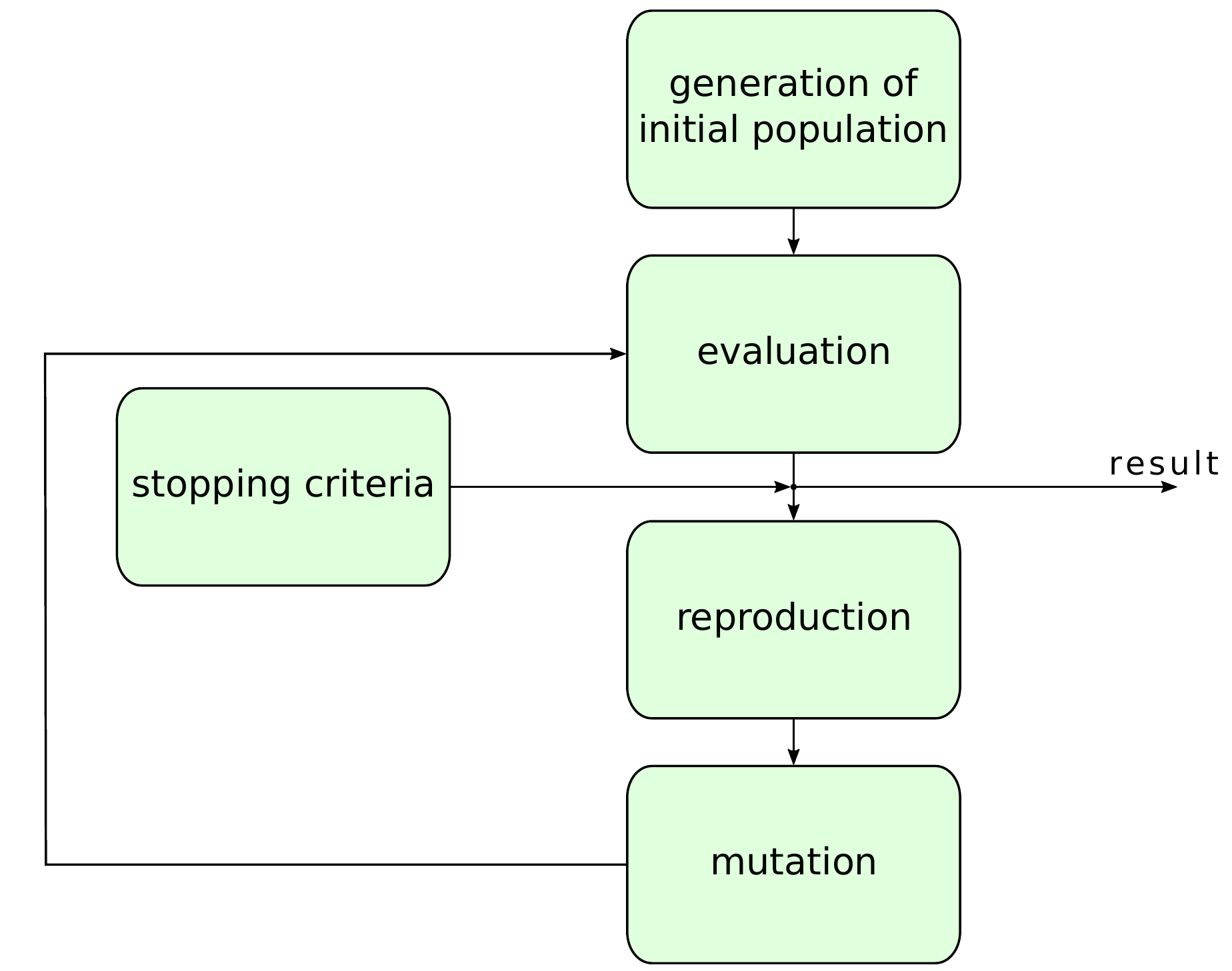}
\caption{Scheme of the genetic algorithm.}
\label{fig:methodology:artificial_neural_networks:nn_2}
\end{center}
\end{figure}

The genetic algorithm is iterative, that is, it consists of several steps being consecutively executed in a loop, until one of well defined conditions, referred to as \emph{stopping criteria}, is fulfilled. The algorithm is described with the help of \reffig{fig:methodology:artificial_neural_networks:nn_2}. (\emph{i}) In the first step an initial population is generated. A number of candidates making up this initial population is a parameter of this algorithm. The creation of a new candidate consists for instance of the random generation of its genotype with gene values laying in defined ranges. (\emph{ii}) Each candidate is evaluated by acting on its genotype with the fitness function. The resulting fitness score indicates how good the candidate is accommodated to the stated problem. Then, all the candidates are sorted according to those scores. The one having the most preferable score is considered as the best one. (\emph{iii}) Stopping criteria are checked and the algorithm is stopped if one of them is fulfilled. In such a case the genotype of the best candidate is considered as the solution. A possible stopping criterion is for instance a threshold on the fitness score of the best candidate or a threshold on the number of iteration epochs that has passed already. (\emph{iv}) A fraction of candidates being identified by the worst fitness scores is removed from the population and forgotten. Their place in the population is taken by new candidates carrying genes of the remaining candidates. That is, two or more remaining candidates pass a fraction of their genotypes to a new candidate in a hope of creating even a better adapted individual. This step is known as the reproduction or cross-over. (\emph{v}) A random fraction of genes in the population is modified. This modification may consist of the random generation of a new gene value within a specified range (as during the initialization) or it can be a small shift with respect to the original gene value. The purpose of this step, which is known as the mutation, is to keep the population diverse. The mutation probability cannot be too high, as in such a case the genetic algorithm becomes a random search. The mutation ends a single epoch, so that the algorithm proceeds with the evaluation as the next step. 
 
\subsection{Regularization}
\label{sec:methodology:regularization}

\begin{figure}[!ht]
\begin{center}
\includegraphics[width=0.45\textwidth]{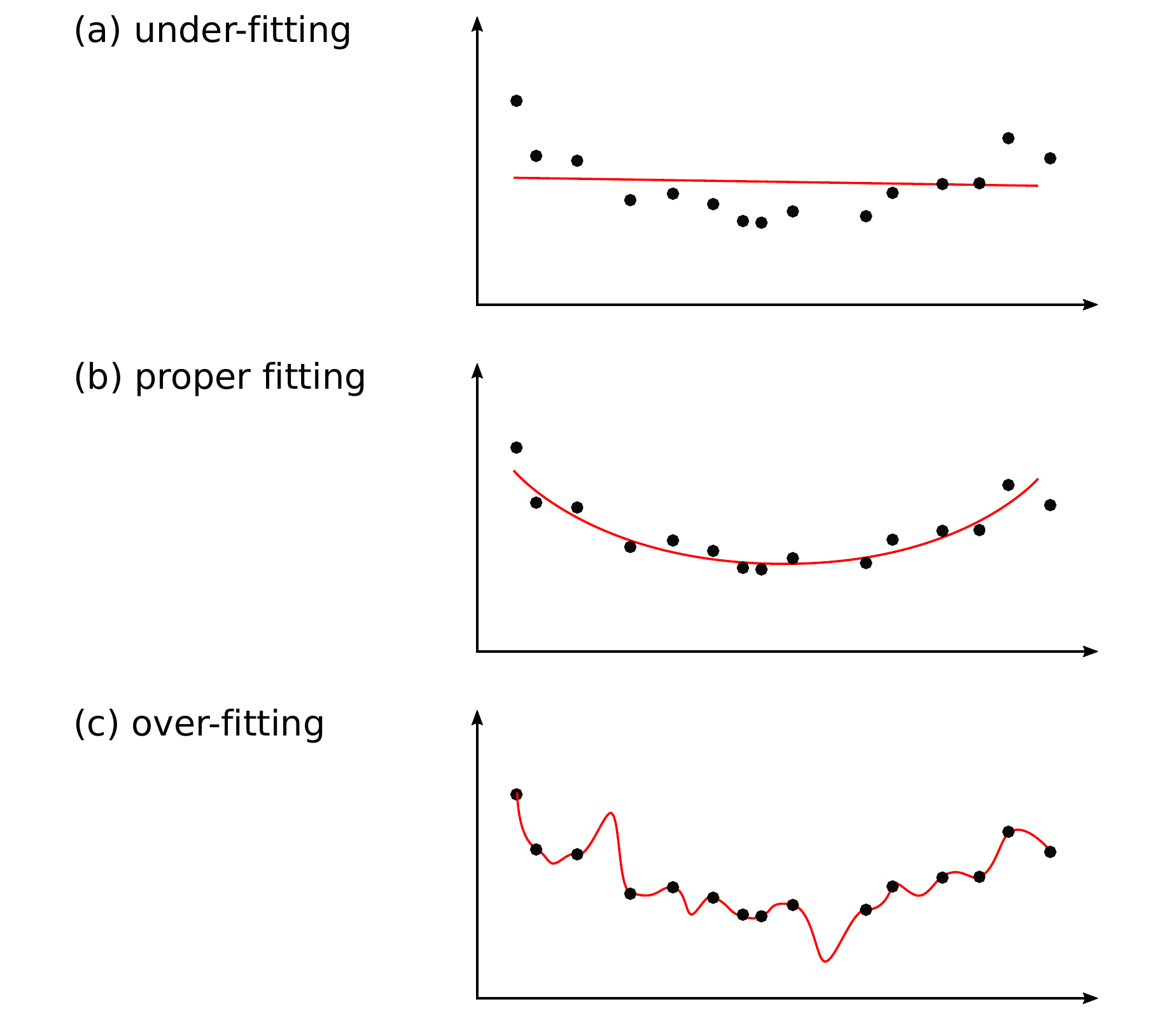}
\caption{Illustration of under-fitting, proper fitting and over-fitting. Training data are represented by points, while the effect of a training procedure is represented by red solid curves. The same training data are used in all three cases.}
\label{fig:methodology:artificial_neural_networks:nn_3}
\end{center}
\end{figure}

Two extreme cases of failure in the learning procedure are known as \emph{under-fitting} and \emph{over-fitting}. A compact explanation of those effects goes as follows. Under-fitting results in a lack of success in describing the training sample, as well as any other sample, while over-fitting is having a model being too flexible, which is not able to describe data other than the training sample. Over-fitting usually happens either because the model memorizes a noise present in the training sample, or because it becomes too flexible in a range between training points. The extrapolation does not count here, as neural networks typically become pliant outside ranges covered by the training sample. Both effects, together with the case of proper training, are graphically illustrated in \reffig{fig:methodology:artificial_neural_networks:nn_3}. At this point we should stress that both under-fitting and over-fitting are not specific features of neural network training, but are known and common problems met in data fitting whenever the number of free parameters becomes large.

Regularization is the introduction of additional information to improve the resolution of badly behaved problems. In the case of neural networks, the regularization helps to avoid over-fitting. In the following we describe the \emph{early stopping} regularization that is used in this analysis, however many other types of regularization exist. We note that in this analysis we have found regularization techniques using a penalty term in the cost function impractical. Such techniques, like Lasso or Ridge regressions \cite{goodfellow2016deep}, make the values of neural network weights small and as a consequence replicas similar to each other (the replication method will be introduced in \refsec{sec:methodology:replica_method}). While this may be seen as a feature in some analyses, in this one it does not lead to a reliable estimation of uncertainties, as replicas are not spread significantly in a domain that is not sufficiently constrained by data.    

Our procedure of dealing with over-fitting makes the training \emph{supervised}. It means that we randomly divide our initial sample into two independent sub-samples, which in the following we will refer to as \emph{training sample} and \emph{validation sample}. While the purpose of the training sample remains the same, \ie it is used for the training, the validation sample is exclusively used for the detection of over-fitting.

\begin{figure}[!ht]
\begin{center}
\includegraphics[width=0.45\textwidth]{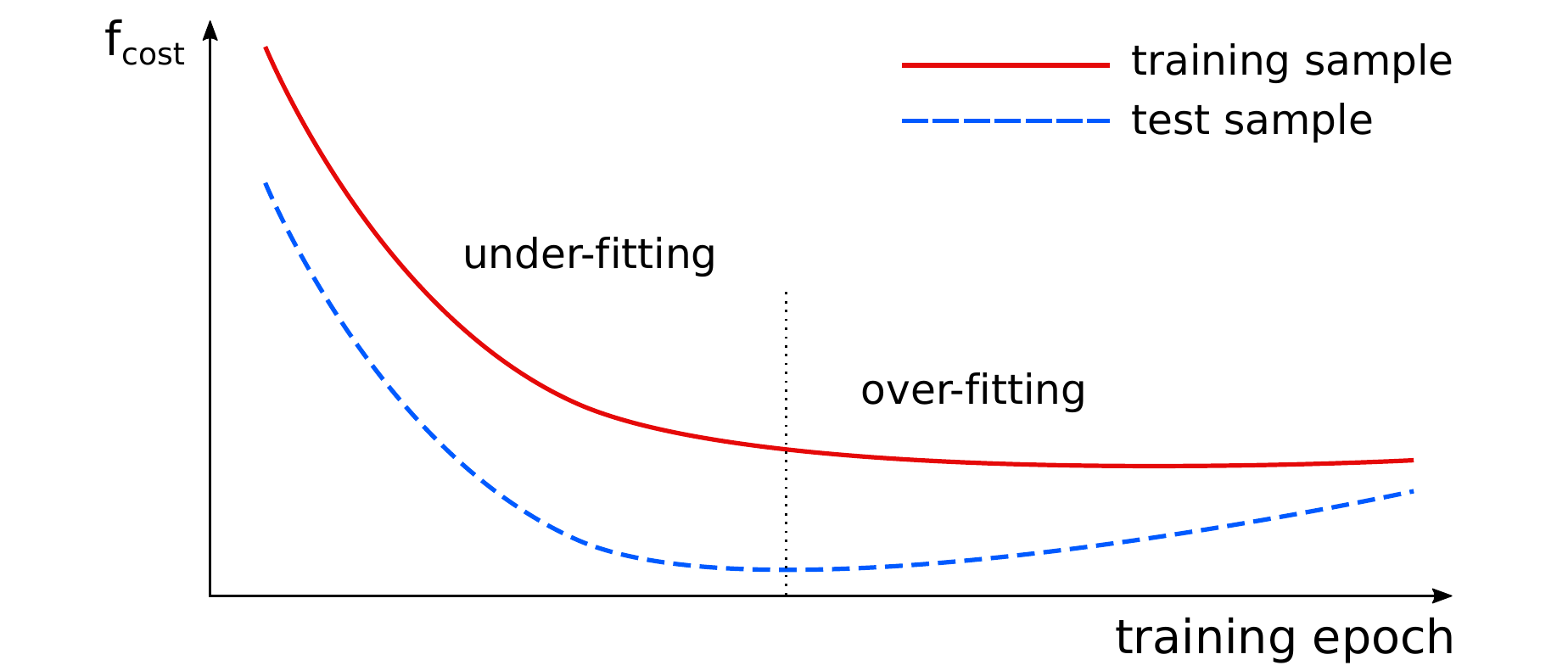}
\caption{Demonstration of early stopping regularization technique: the cost function for training (red solid curve) and test (blue dashed curve) samples as a function of training epoch.}
\label{fig:methodology:artificial_neural_networks:nn_5}
\end{center}
\end{figure}

As a principle, too short a learning leads to under-fitting, while a too long one is responsible for over-fitting, unless the network's architecture prevents it. Therefore, between those two extremes there must be an optimal epoch in the training corresponding to a most favorable solution. The search of such an optimal epoch is the goal of the early stopping technique. The principle of this method is to monitor the validation sample with the cost function, $f_{\mathrm{cost}}$, and to stop the training whenever the value of this function starts to grow. This idea is illustrated in \reffig{fig:methodology:artificial_neural_networks:nn_5}. In practice, such a growth is difficult to be detected due to fluctuations of the cost function values. Therefore, various quantities are used for its detection. In this analysis we monitor the following quantity evaluated for the test sample, which is inspired by \refcite{Prechelt2012}:
\begin{equation}
f_{\mathrm{stop}}(t_{i}) = \frac{f_{\mathrm{cost}}^{\mathrm{av}}(t_{i})}{\displaystyle\min_{j < i} f_{\mathrm{cost}}^{\mathrm{av}}(t_{j})} - 1 \; .
\label{eq:methodology:regularization:early_stopping:eq1}
\end{equation}
Here, $t_{i}$ and $t_{j}$ denote different epochs in the training. In the numerator of Eq. \eqref{eq:methodology:regularization:early_stopping:eq1} one has the value of the cost function for the epoch under consideration, and in the denominator the minimal value of this function obtained in all previous epochs. To minimize the impact of random fluctuations on $f_{\mathrm{stop}}(t_{i})$ the cost function is averaged using the running average technique: 
\begin{equation}
f_{\mathrm{cost}}^{\mathrm{av}}(t_{i}) = \frac{1}{2n + 1}\displaystyle\sum_{j=-n}^{n}  f_{\mathrm{cost}}(t_{i-j}) \; ,
\label{eq:methodology:regularization:early_stopping:eq2}
\end{equation}
where $n$ is the length of a strip of numbers that is used for the averaging. In this analysis $n = 25$. With those definitions $f_{\mathrm{stop}}(t_{i}) > 0$ indicates a need of stopping the training. However, as fluctuations of the cost function cannot be completely avoided we stop the training only when $f_{\mathrm{stop}}(t_{i}) > 0$ for more than one hundred consecutive iterations. We take the set of parameters obtained in the first iteration of such a sequence as the result of our training procedure. In rare cases the stopping criterion is not fulfilled during the whole training procedure. In those cases the result is obtained from the last allowed epoch of this procedure, $t_{\max} = 1000$.

\subsection{Replica method}
\label{sec:methodology:replica_method}

As in our last analysis \cite{Moutarde:2018kwr}, the \emph{replica method} is used to propagate uncertainties coming from experimental data to CFF parameterizations. In addition to our nominal extraction of CFFs, the fit is repeated one hundred times, each time independently modifying the central values of experimental points using the following prescription:
\begin{flalign}
&v_{ij} \pm {\Delta}_{ij}^{\mathrm{tot}} \xrightarrow[]{\mathrm{replica}~k} \nonumber \\
&\phantom{xx}\left(\mathrm{rnd}_{k}(v_{ij}, {\Delta}_{ij}^{\mathrm{tot}}) \pm {\Delta}_{ij}^{\mathrm{tot}} \right) \times \mathrm{rnd}_{k}(1, {\Delta}_{i}^{\mathrm{norm}}) \;.
\label{eq:methodology:replica_method:eq1}
\end{flalign}
Here, $v_{ij}$ is the measured value associated to the experimental point $j$ coming from the data set $i$. It is linked to statistical, ${\Delta}_{ij}^{\mathrm{stat}}$, systematic, ${\Delta}_{ij}^{\mathrm{sys}}$, and normalization, ${\Delta}_{i}^{\mathrm{norm}}$, uncertainties. The latter one is correlated across bins of data, like for instance uncertainties related to beam and target polarization measurements. The total uncertainty, which is used to evaluate $\chi^2$ function utilized in this analysis to perform the fit, is given by:
\begin{equation}
\Delta_{ij}^{\mathrm{tot}} = \sqrt{\left({\Delta}_{ij}^{\mathrm{stat}}\right)^{2} + \left({\Delta}_{ij}^{\mathrm{sys}}\right)^{2}} \;.
\label{eq:methodology:replica_method:eq2}
\end{equation}
The generator of random numbers following a specified normal distribution, $f(x | \mu, \sigma)$, is denoted by $\mathrm{rnd}_{k}(\mu, \sigma)$, where $k$ both identifies the replica and is a unique random seed.

\section{Implementation}
\label{sec:implemenation}

\subsection{Neural network architecture}
\label{sec:implemenation:architecture}

In this subsection we describe the architecture of the used neural networks. In particular a determination of the number of hidden neurons in those networks is described. We start with a general statement: any neural network is a container to store information and as such it cannot be too small to serve that purpose. It can be larger than needed though, however such neural networks require a careful regularization to avoid over-fitting and their training is more time-consuming. We note that usually the optimal network's architecture is selected prior the training, however there exist regularization techniques like \emph{dropout} \cite{Srivastava:2014:DSW:2627435.2670313}, which alter an initial architecture during the training to obtain best results. In those techniques a number of 	neurons is either added or dropped as the training proceeds.

We extract four CFFs, each one being a complex quantity having both real and imaginary parts. Although the real and imaginary parts of the a given CFF can be connected together \emph{via} a dispersion relation (with the subtraction constant being involved), in the extraction presented in this paper we keep them fully independent quantities. The dispersion relation is only used for an interpretation of the obtained results, which is presented in \refsec{sec:results}. Therefore eight independent neural networks are used.

\begin{figure}[!ht]
\begin{center}
\includegraphics[width=\columnwidth]{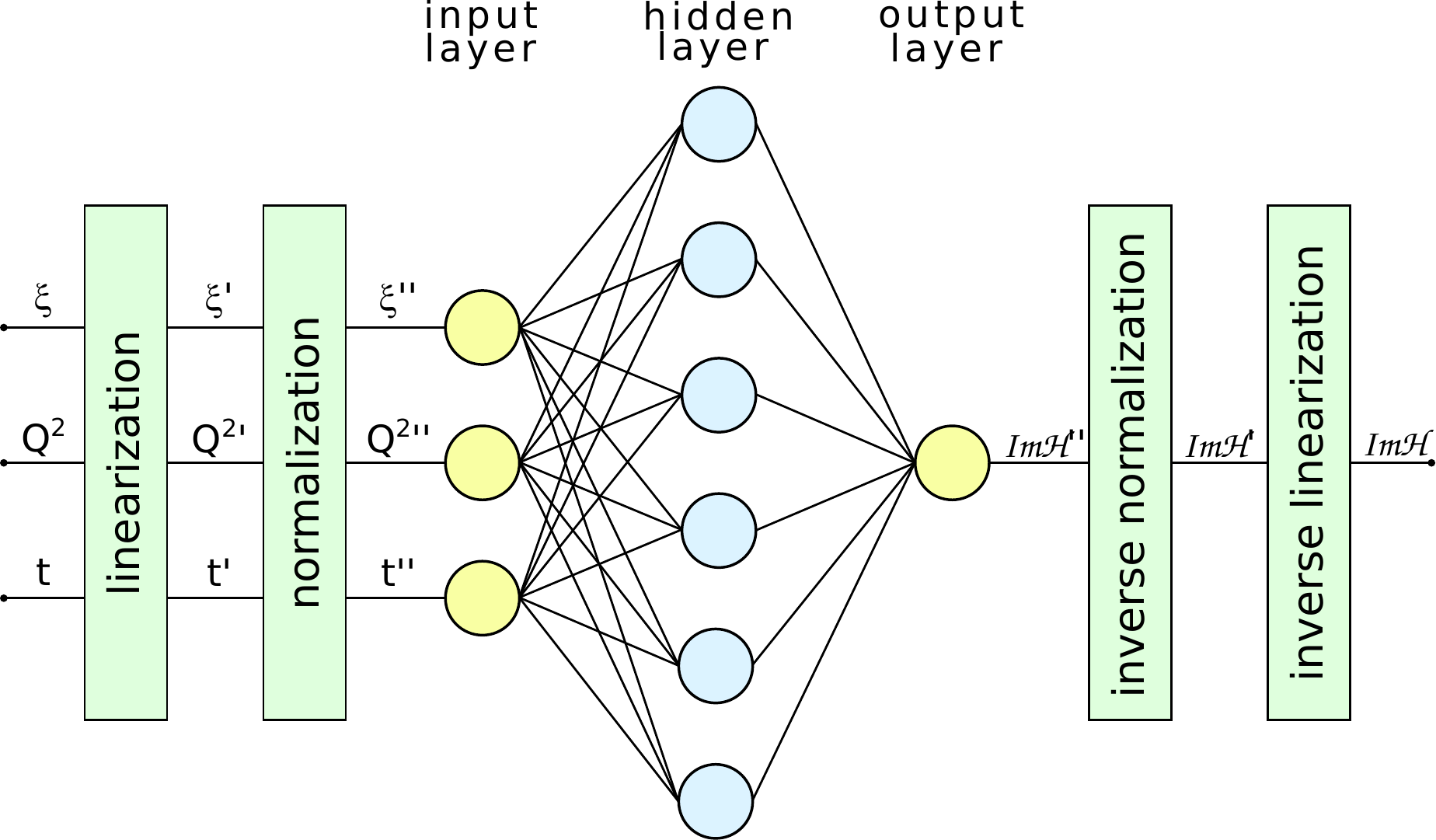}
\caption{Scheme of a single neural network that is used in this analysis to represent either the real or the imaginary part of a single CFF.}
\label{fig:implementation:architecture}
\end{center}
\end{figure}

The architecture of a single network used in this analysis is shown in \reffig{fig:implementation:architecture}. It consists of three input neurons, one hidden layer containing six neurons, and one output neuron. Both input and output variables of the network are linearized and normalized (as much as possible). The linearization is achieved with a logarithmic projection and $\xi$ pre-factors:
\begin{gather}
\xi' = \log_{10}\xi \, , \\
t' = t \, , \\
{Q^{2}}' = \log_{10}Q^{2} \, , \\
\mathrm{Re}\mathcal{G}' = \xi\,\mathrm{Re}\mathcal{G} \, ,\\
\mathrm{Im}\mathcal{G}' = \xi\,\mathrm{Im}\mathcal{G} \, ,
\end{gather}
where the prime symbol is used to distinguish between original and linearized variables. For the normalization we utilize the min-max method:
\begin{gather}
v''= -1 + 2\frac{v'-v_{\mathrm{min}}'}{v_{\mathrm{max}}'-v_{\mathrm{min}}'} \,,
\end{gather}
where $v'$ and $v''$ is a given linearized variable before and after the normalization, respectively, and where $v_{\mathrm{min}}'$ and $v_{\mathrm{max}}'$ are normalization parameters specified in Table \ref{tab:implementation:scaling_ranges}. The min-max values used for the normalization of $\xi'$, $t'$ and ${Q^{2}}'$ have been selected to cover the target phase-space, while those for $\mathrm{Re}\mathcal{G}'$ and $\mathrm{Im}\mathcal{G}'$ roughly correspond to $v_{\mathrm{min, GK}}' - \Delta v_{\mathrm{GK}}'$ and $v_{\mathrm{max, GK}}' + \Delta v_{\mathrm{GK}}'$ values, where $\Delta v_{\mathrm{GK}}' = v_{\mathrm{max, GK}}' - v_{\mathrm{min, GK}}'$. Here, $v_{\mathrm{min, GK}}'$ and $v_{\mathrm{max, GK}}'$ are min-max values found in a poll of CFFs evaluated from the Goloskokov-Kroll (GK) GPD model \cite{Goloskokov:2005sd, Goloskokov:2007nt, Goloskokov:2009ia} for the experimental data kinematics that are used in this analysis, see  \refsec{sec:data}. Both linearization and normalization significantly improve the performance of ANNs. We point out that the ranges specified in Table \ref{tab:implementation:scaling_ranges} are not absolute, \ie the networks may still reasonably describe data covering the exterior of domains defined by $v_{\mathrm{min}}'$ and $v_{\mathrm{max}}'$ values. In particular in this analysis the activation function of output neurons is set to the identity and thus does not show any saturation effects.  

\begin{table}[!ht]
\caption{Values used in the scaling of input and output variables.}
\label{tab:implementation:scaling_ranges}
\begin{center}
\begin{tabular}{@{}lccclcc@{}}
\toprule
\makebox[0.11\columnwidth][l]{$v'$} & 
\makebox[0.11\columnwidth][c]{$v_{\mathrm{min}}'$} &
\makebox[0.11\columnwidth][c]{$v_{\mathrm{max}}'$} &  
& 
\makebox[0.11\columnwidth][l]{$v'$} & 
\makebox[0.11\columnwidth][c]{$v_{\mathrm{min}}'$} &
\makebox[0.11\columnwidth][c]{$v_{\mathrm{max}}'$}
 \\ \midrule
$\log_{10}\xi$ 							& $-6$ 		& $1$  	& & 		$t$ 										& $-1$ 		& $0.5$ \\
$\log_{10}{Q^{2}}$						& $-1$	 	& $2$ 	& & & & \\ \addlinespace[0.5em]
$\xi\mathrm{Re}\mathcal{H}$				& $-1$ 		& $1.5$ & & 		$\xi\mathrm{Im}\mathcal{H}$ 				& $-4$ 		& $6$ \\
$\xi\mathrm{Re}\widetilde{\mathcal{H}}$	& $-0.6$ 	& $0.9$ & & 		$\xi\mathrm{Im}\widetilde{\mathcal{H}}$	& $-1$ 		& $1.5$ \\
$\xi\mathrm{Re}\mathcal{E}$				& $-1.5$ 	& $1$ 	& & 		$\xi\mathrm{Im}\mathcal{E}$ 				& $-6$ 		& $4$ \\ 
$\xi\mathrm{Re}\widetilde{\mathcal{E}}$	& $-120$ 	& $180$ & &		$\xi\mathrm{Im}\widetilde{\mathcal{E}}$ 	& $-8$	 	& $12$ \\ \bottomrule
\end{tabular}
\end{center}
\end{table}

\begin{figure}[!ht]
\begin{center}
\includegraphics[width=\columnwidth]{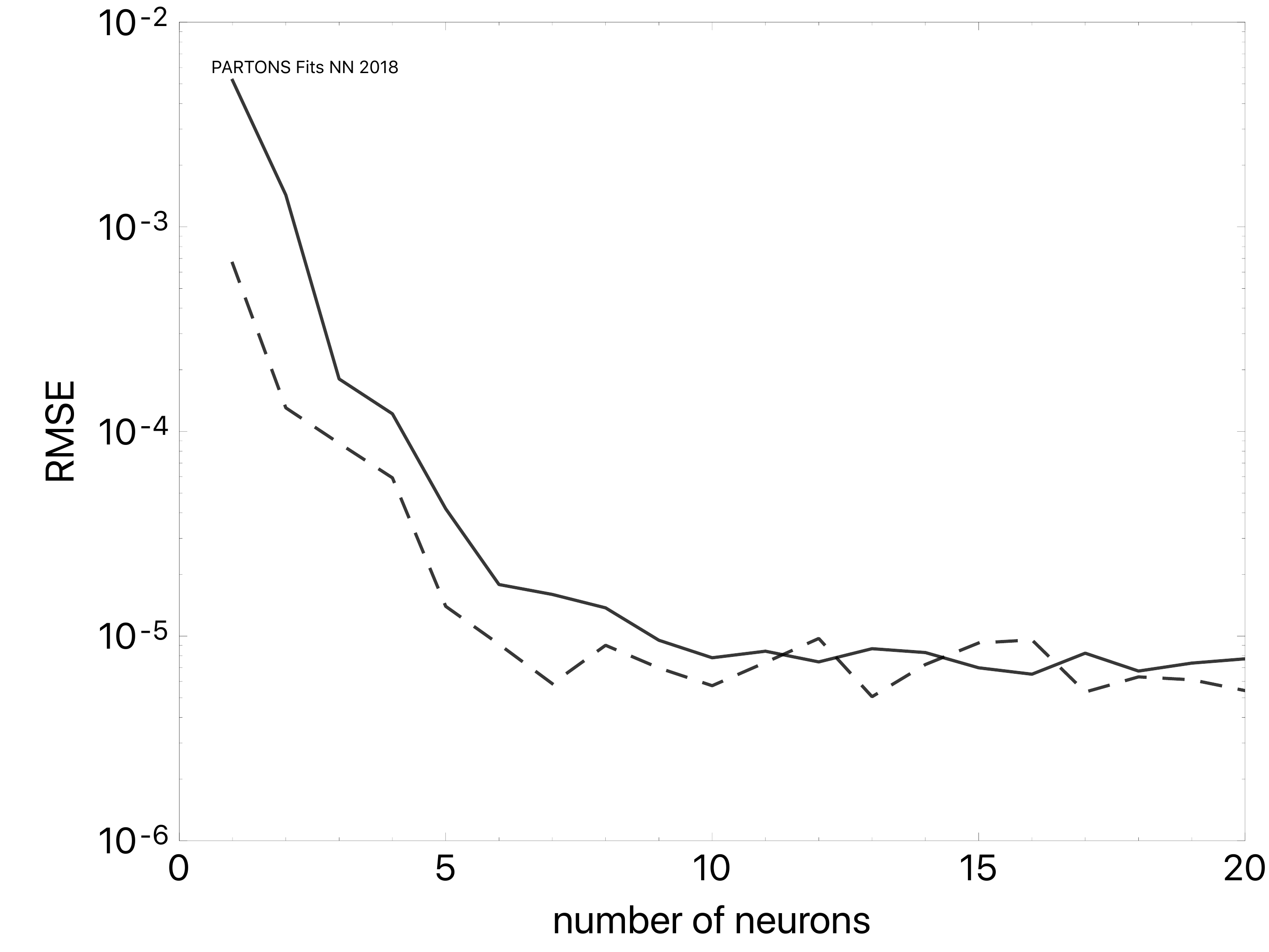}
\caption{Average root mean square error (RMSE) for the neural network describing the benchmark sample (see the text) as a function of the number of hidden neurons in that network for the real (solid line) and imaginary (dashed line) parts of CFF $\mathcal{H}$. The RMSE values correspond to the normalized variables, see Table \ref{tab:implementation:scaling_ranges}.}
\label{fig:implementation:number_of_neurons}
\end{center}
\end{figure}

The number of neurons in the hidden layer is determined with a benchmark sample made out of one thousand CFF points evaluated with the GK GPD model. Those CFF points are randomly generated in a broad range of $10^{-6} < \xi < 1$, $0~\mathrm{GeV}^{2} < |t| < 1~\mathrm{GeV}^{2}$ and $1~\mathrm{GeV}^{2} < Q^{2} < 100~\mathrm{GeV}^{2}$. We have checked how many neurons in the hidden layer are needed to provide a good description of the benchmark sample. This test is performed with the FANN library \cite{nissen03} for the neural network implementation and with the \emph{quickprop} training algorithm \cite{fahlman:faster} available in that library. The post-training RMSE for CFF $\mathcal{H}$ is shown in \reffig{fig:implementation:number_of_neurons} as a function of the number of hidden neurons. From this figure one may conclude that the addition of a new neuron to the network made already out of six neurons does not significantly improve the performance. We note that a high number of neurons slows down the training and makes the regularization more difficult. That is why in this analysis the number of hidden neurons in each network is set to six, which we consider to be a sufficient number.

\subsection{Initial fit}
\label{sec:implementation:initial_fit}

Without any improvement the fit either does not converge or converges very slowly. This is due to a large number of parameters to be constrained in the training and a widely opened phase-space for CFF values, \cf Table \ref{tab:implementation:scaling_ranges}. To overcome this difficulty the network is initially trained on values obtained in a local extraction of CFFs. That is, CFF values multiplied by $\xi$ are independently extracted in each kinematic bin of data, analogously as in \refcite{Dupre:2016mai, Dupre:2017hfs, Burkert:2018bqq}, and then used in a training based on the back-propagation algorithm. 

The local extraction procedure includes all four CFFs, so effectively eight values per each kinematic bin of $(\xBj, t, Q^{2})$ are fitted. This may make a single fit underconstrained, in particular if a given kinematic bin is populated only by a single experimental point. To overcome this difficulty it is imperative to set sensible ranges in which the fitted values are allowed to vary. In this analysis those ranges are set to ($5 \times v_{\mathrm{min}}', 5 \times v_{\mathrm{max}}'$), where the values of $v_{\mathrm{min}}'$ and $v_{\mathrm{max}}'$ are specified in Table \ref{tab:implementation:scaling_ranges}. The ranges are wider than those used in the construction of our neural networks, which prevents from introducing a bias on the final extraction of CFF parameterizations.

The initial fit reduces the $\chi^{2}$ value per single data point from $\mathcal{O}(10^{6})$ to $\mathcal{O}(10)$. The further minimization, which reduces the same quantity to $\mathcal{O}(1)$, is done by our genetic algorithm minimizer, where values of weights and biases obtained in the initial fit are used as starting parameters.

\subsection{Genetic algorithm}
\label{sec:implemenation:genetic_algorithm}

The parameters of our genetic algorithm minimizer are as follows. The gene values (weights and biases) may vary between $-10$ and $10$. The population consists of $1000$ candidates and the fraction of candidates left after the selection process (survivors) is $30\%$. We consider two types of mutation and each of them occurs with the same probability of $0.1\%$ per gene. The mutation type-$A$ is a random generation of gene values from the range of $-10$ and $10$. This type of mutation keeps the population diverse. The mutation type-$B$ is a small shift of gene values. This shift is randomly generated from a distribution of gene values in the population. This distribution is made for the modified gene. The mutation type-$B$ provides a fine-tuning of fitted parameters. The fraction of $20\%$ and $10\%$ best candidates in the population is resistant to the mutation type-$A$ and type-$B$, respectively, so the best results are not destroyed by the mutation process. 

\begin{figure*}[!ht]
\begin{center}
\includegraphics[width=\textwidth]{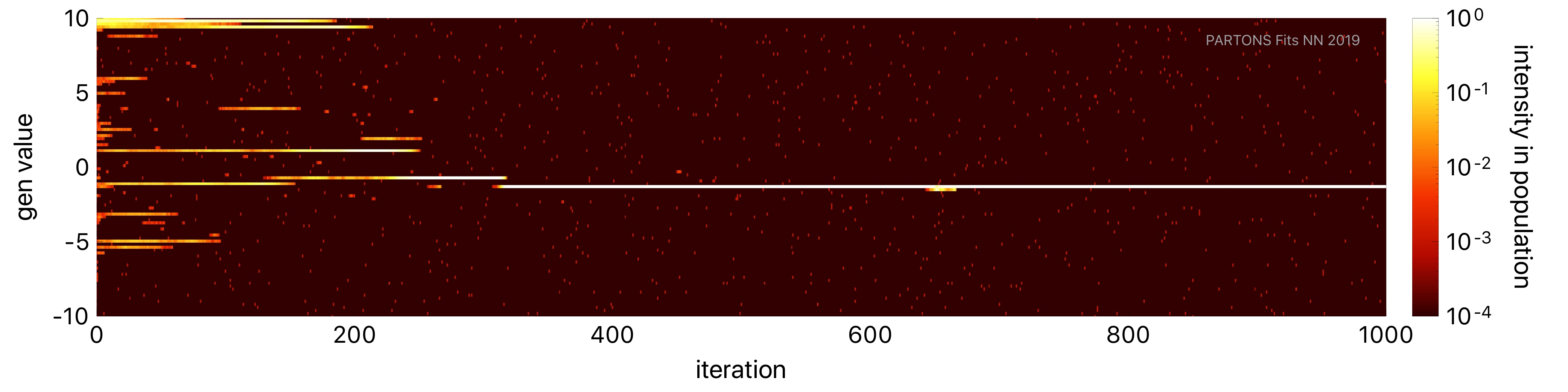}
\caption{Diversity of gene values in the whole population against the training epoch for a single gene. For more details see the text.}
\label{fig:implementation:genetic_algoritm}
\end{center}
\end{figure*}

We demonstrate the performance of our genetic algorithm minimizer with \reffig{fig:implementation:genetic_algoritm}, where values of a single gene are plotted against the training epoch. \emph{Intensity in population} equal $1$ means that all candidates in the population share the same gene value. Low values of this quantity indicate that only few or none candidates share the same gene value. One can see few features in this plot being typical for a performance of genetic algorithms: (\emph{i}) the diversity of values is large at the beginning of minimization, which means that the whole phase-space available is equally scanned to find a cost function minimum, (\emph{ii}) usually, few such local minima are found at the beginning of minimization and simultaneously explored by the algorithm for a number of iterations, (\emph{iii}) close to the actual minimum the algorithm concentrates on its neighborhood, trying to perform a fine-tuning of fitted parameters, (\emph{iv}) even at this stage of minimization, the whole phase-space is homogeneously scanned due to the mutation process to eventually find a new minimum and to avoid the convergence to a local one. 

\subsection{Regularization}
\label{sec:implemenation:regularization}

\begin{figure*}[!ht]
\begin{center}
\includegraphics[width=\textwidth]{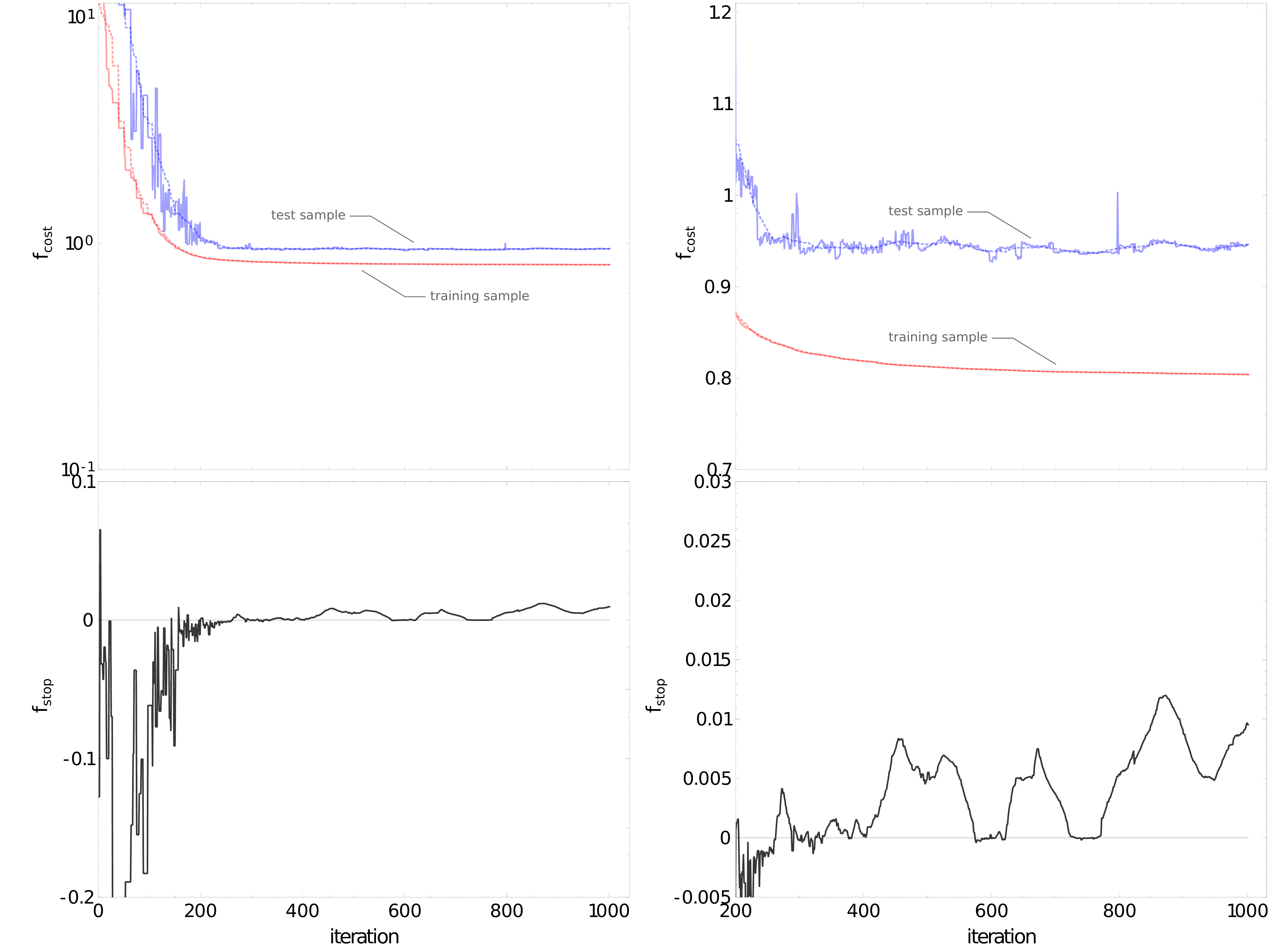}
\caption{Cost function (top plots) and $f_{\mathrm{stop}}$ quantity (bottom plots) used to detect over-fitting in the regularization method based on the early stopping technique, see \refsec{sec:methodology:regularization}, as a function of the training epoch. The left plots show dependencies in the full range of training epoch, while the right ones are for a zoom in the range of $200 < t < 1000$. The red (blue) solid and dashed curves illustrate the values of the cost function for the training (test) sample, before and after averaging, respectively.}
\label{fig:implementation:early_stopping}
\end{center}
\end{figure*}

A typical distribution of the cost function as a function of the training epoch is shown in \reffig{fig:implementation:early_stopping}. In addition, we show the distribution of the $f_{\mathrm{stop}}$ quantity that we use to detect over-fitting, see Eq. \eqref{eq:methodology:regularization:early_stopping:eq1}. With our stopping criterion: $f_{\mathrm{stop}} > 0$ for more than one hundred iterations, for this example we consider the solution obtained in the epoch $382$ as the valid one and we use this solution in the further analysis.

\subsection{Feasibility test}
\label{sec:implementation:feasibility_test}

In order to check our procedure a feasibility test was performed. In this test the extraction of CFF parameterizations is done on a sample of pseudo-data generated in the leading-order formalism with the GK GPD model \cite{Goloskokov:2005sd, Goloskokov:2007nt, Goloskokov:2009ia}. The generation of pseudo-data is straightforward: for each experimental data point, a single point of pseudo-data is generated. This single point of pseudo-data is generated with the same type (differential cross section, or beam spin asymmetry, or beam charge asymmetry, \etc), and at the same kinematics as the original point. The obtained observable value is then smeared according to the uncertainties of the original point. As a result of this procedure, pseudo-data are  faithfully generated, corresponding to the experimental data; however ``input" CFFs are known for this sample. These pseudo-data are used in the same procedure of CFF extraction as the genuine experimental data. 

\begin{figure}[!ht]
\begin{center}
\includegraphics[width=0.49\textwidth]{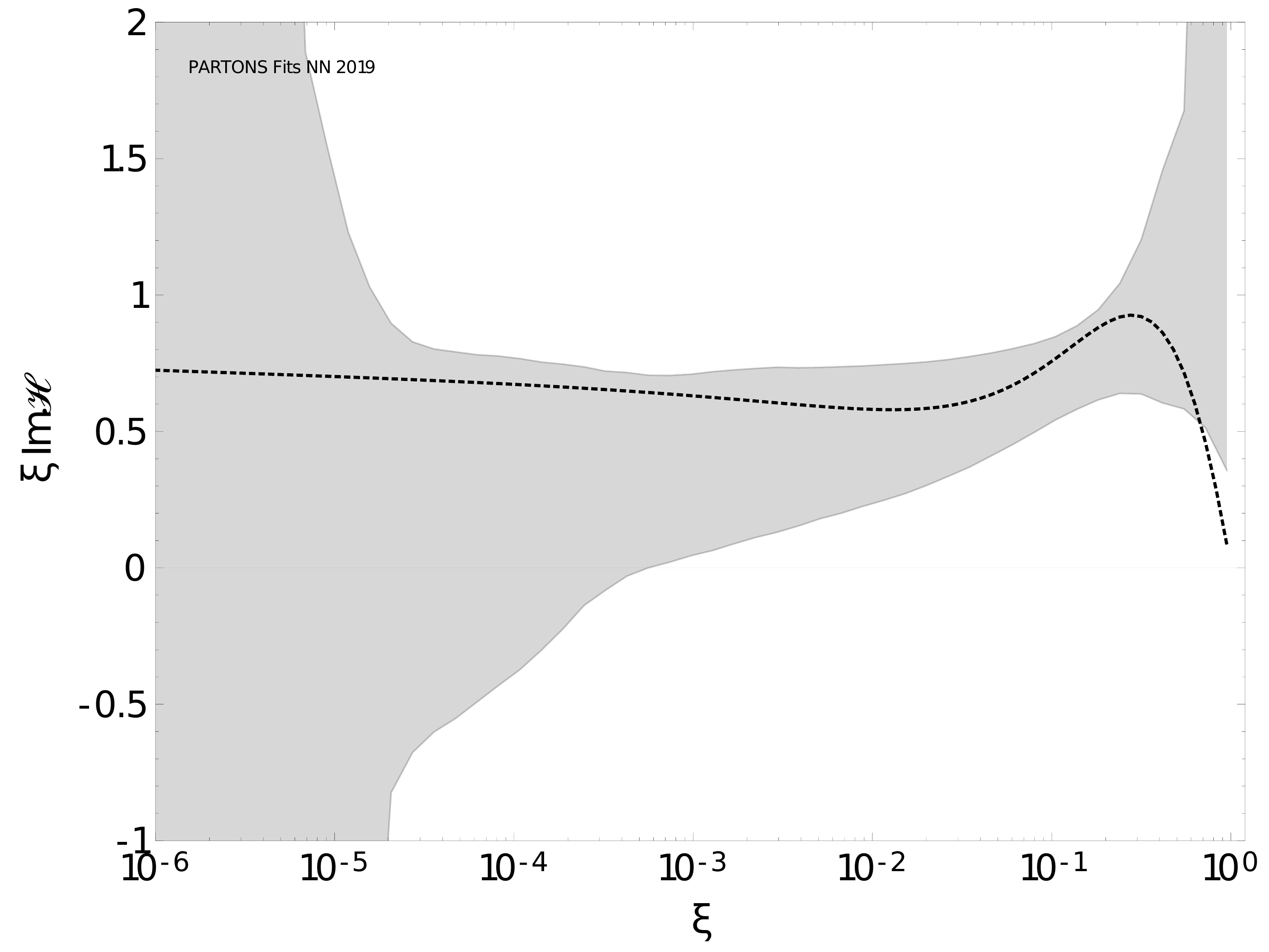}
\caption{Example outcome of the feasibility test. The imaginary part of the CFF $\mathcal{H}$ evaluated from GK GPD model \cite{Goloskokov:2005sd, Goloskokov:2007nt, Goloskokov:2009ia} (dotted line) and the corresponding parameterization of that quantity extracted from the pseudo-data generated with this model (gray band). The plot is shown for $Q^{2} = 2~\mathrm{GeV}^{2}$ and $t = -0.3~\mathrm{GeV}^{2}$.}
\label{fig:implementation:feasibility_test}
\end{center}
\end{figure}

The outcome of this feasibility test is in general positive. The value of the $\chi^{2}$ function per single data point is $1.13$, or $1.06$ for data other than HERA and COMPASS points (see \refsec{sec:results} for an explanation of this behavior). The agreement between the CFF parameterizations obtained from the pseudo-data and the GK GPD model used to generate those data is demonstrated with \reffig{fig:implementation:feasibility_test}, where the imaginary part of CFF $\mathcal{H}$ and the model are compared as a function of $\xi$ for example kinematics of $t = -0.3~\mathrm{GeV}^2$ and $Q^{2} = 2~\mathrm{GeV}^{2}$. The curve representing the model stays within the uncertainty band of the extracted parameterization. The only exception is the region of $\xi \approx 1$, which is not covered by pseudo-data and where the model rapidly goes to zero. However, in this region the difference between the model and the central value of the extracted parameterization is still reasonable (\eg $1.23 \sigma$ for $\xi = 0.95$). 

\section{Experimental data}
\label{sec:data}

Table \ref{tab:data:dvcs_data} summarizes DVCS data used in this analysis. For the explanation of symbols used to distinguish between observable types see Ref. \cite{Moutarde:2018kwr}. Only proton data are used. Available neutron data are sparse ones, but \emph{in principle} they may be used to attempt a flavor separation. This is however beyond the scope of the presented analysis. 

As in our previous analysis \cite{Moutarde:2018kwr}, recent Hall A data \cite{Defurne:2015kxq, Defurne:2017paw} for cross sections measured with unpolarized beam and target, $d^{4}\sigma_{UU}^{-}$, are not used in the final extraction of CFF information. Again, only the corresponding differences of cross sections probing longitudinally polarized beam, $\Delta d^{4}\sigma_{LU}^{-}$, are used. We will elaborate on the inclusion of Hall A data in \refsec{sec:results}. 

The difference with respect to our previous analysis \cite{Moutarde:2018kwr} in terms of used data comes from the inclusion of low-$\xBj$ sets. For COMPASS, instead of a single point for the measurement of the $t$-slope $b$, four points for the cross section measurement, $d^{3}\sigma_{UU}^{\pm}$, are used. The measurement of both slope and cross sections is reported in Ref. \cite{Akhunzyanov:2018nut}. The cross sections provide richer information than the slope itself. In addition, the slope measurement relies on the assumption of the $t$-dependence being exponential, which can be avoided analyzing the cross section points. However, the COMPASS cross sections are measured in broad kinematic bins and because of that they cannot be compared to models evaluated at the average kinematics of events corresponding to those bins. That is why in this analysis, the COMPASS points are compared to respective integrals of cross sections. Those multidimensional integrals are time consuming and significantly extend the computing time of our fits.

HERA data for DVCS \cite{Chekanov:2008vy, Aktas:2005ty, Aaron:2009ac} are included in this analysis. As we will demonstrate in \refsec{sec:results}, those data provide important constraints on CFF parameterizations in the low-$\xBj$ region. However, the sparsity of data covering this region makes our extraction of CFFs difficult, as will be explained in \refsec{sec:results}. 

We apply two kinematic cuts on experimental data\footnote{There is a typo in Eq. (86) of our last publication \cite{Moutarde:2018kwr}. Instead of $-t/Q^{2} < 0.25$ there should be $-t/Q^{2} < 0.2$ , as in Eq. \eqref{eq:data:cut_2} here.}:
\begin{flalign}
&Q^{2} > 1.5~\mathrm{GeV}^{2} \label{eq:data:cut_1} \;, \\
&-t/Q^{2} < 0.2 \label{eq:data:cut_2} \;.
\end{flalign} 
The purpose of those cuts is the restriction of the phase-space covered by experimental data to the deeply virtual region, where one can rely on the factorization between GPDs and the hard scattering kernel. \emph{In principle}, those cuts can be avoided in the extractions of amplitudes. However, we keep them here to allow an interpretation of the extracted CFFs in terms of GPDs. In addition, the cuts are applied to keep a correspondence with our previous analysis \cite{Moutarde:2018kwr}, allowing a straightforward comparison. Let us note at this point, that the neural network approach developed in this work provides an easy way for the addition of CFFs identified with higher-twist contributions. With those contributions included one could relax the cuts \eqref{eq:data:cut_1} and \eqref{eq:data:cut_2}, allowing more data to be included.    

\begin{figure*}[!ht]
\begin{center}
\includegraphics[width=0.45\textwidth]{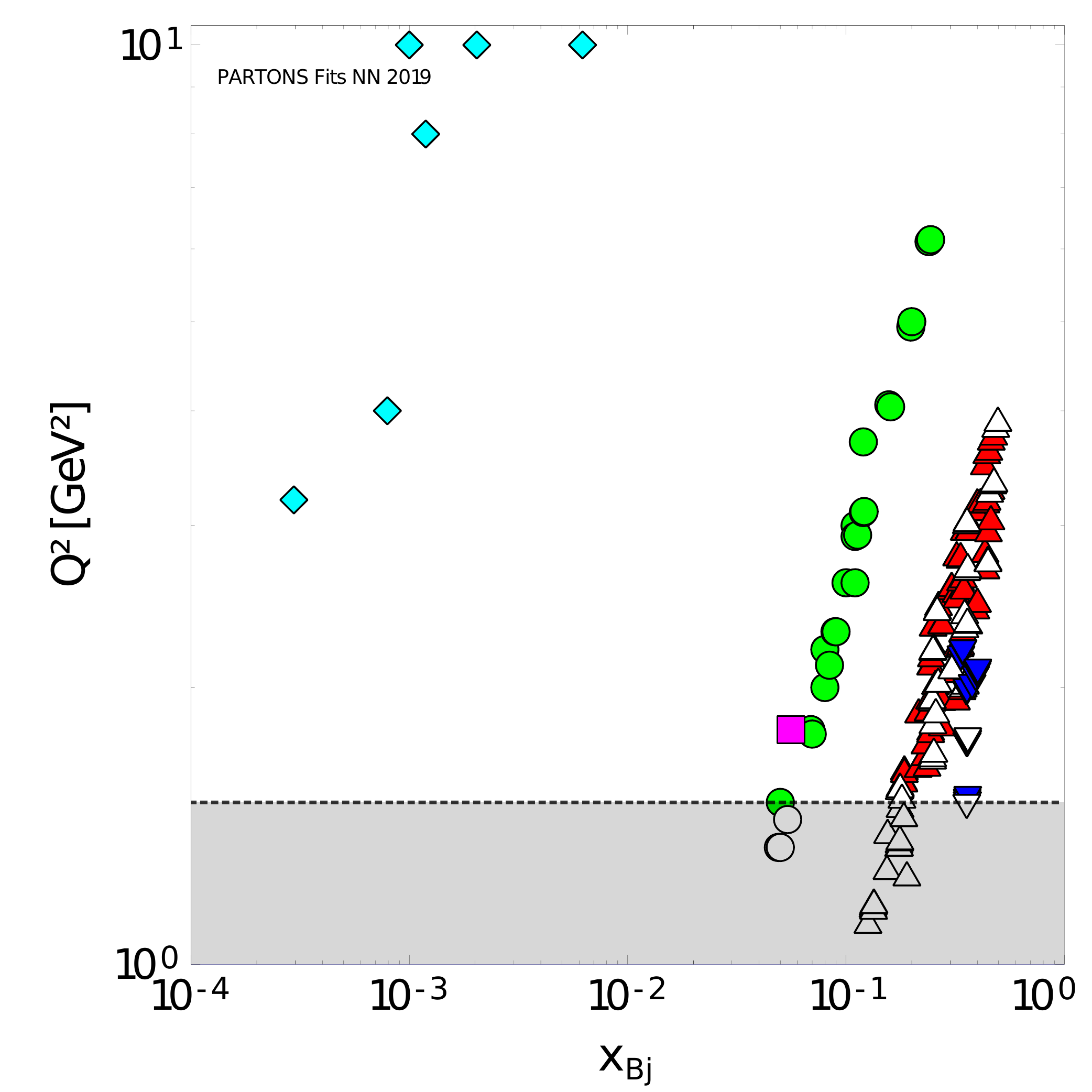}
\includegraphics[width=0.45\textwidth]{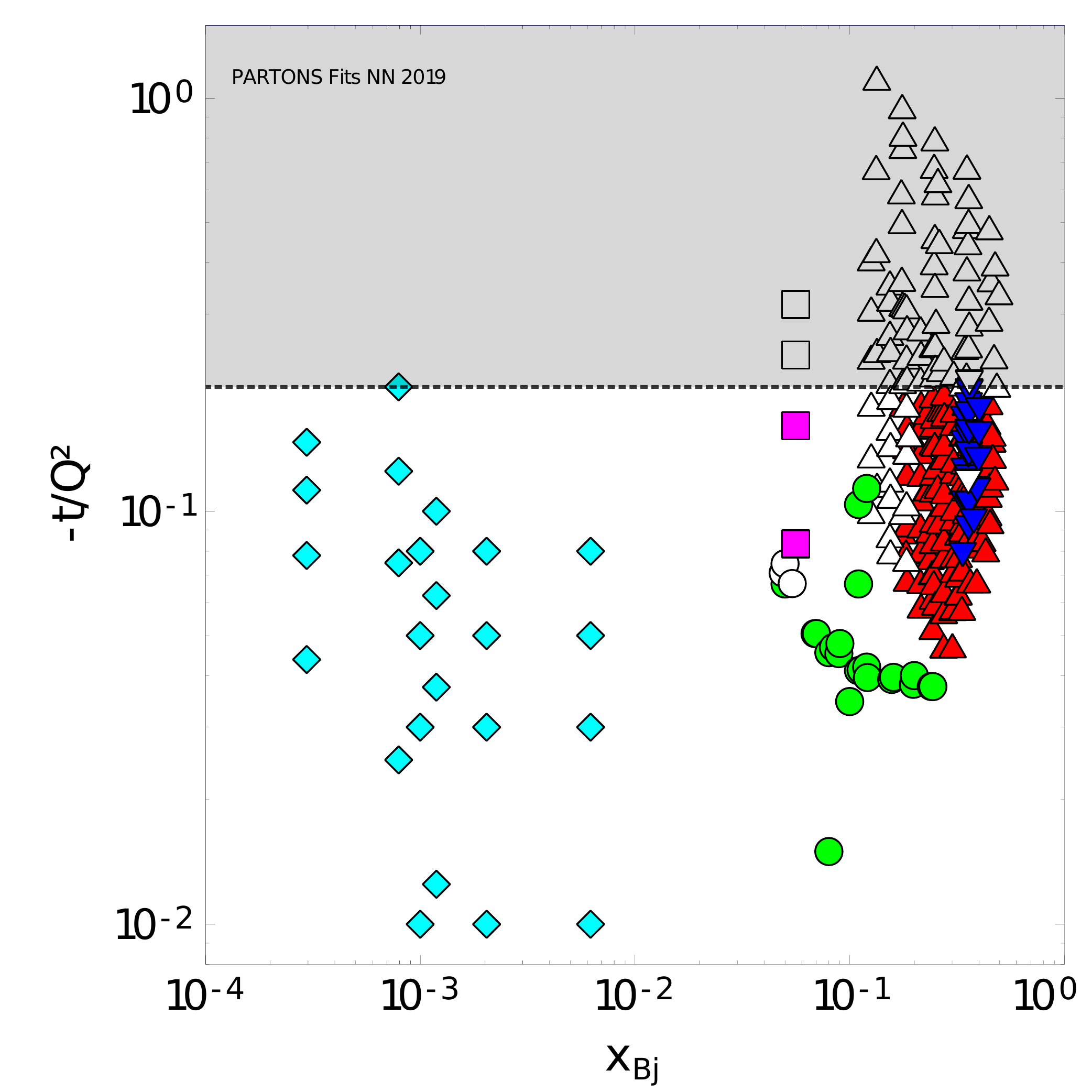}
\caption{Coverage of the $(\xBj, Q^{2})$ (left) and $(\xBj, -t/Q^{2})$ (right) phase-spaces by the experimental data listed in Table \ref{tab:data:dvcs_data}. The data come from the Hall A (\textcolor{blue}{$\blacktriangledown$}, \textcolor{black}{$\triangledown$}), CLAS (\textcolor{red}{$\blacktriangle$}, \textcolor{black}{$\vartriangle$}), HERMES (\textcolor{green}{$\bullet$}, \textcolor{black}{$\circ$}), COMPASS (\textcolor{magenta}{$\blacksquare$}, \textcolor{black}{$\square$}) and HERA H1 and ZEUS (\textcolor{cyan}{$\blacklozenge$}, \textcolor{black}{$\lozenge$}) experiments. The gray bands (open markers) indicate phase-space areas (experimental points) being excluded from this analysis due to the cuts introduced in Eqs. \eqref{eq:data:cut_1} and \eqref{eq:data:cut_2}.}
\label{fig:data:coverage}
\end{center}
\end{figure*}

In total, in this analysis $2624$ points are used out of $3996$ available in all considered data sets. The coverage of phase-space by those data in the sections of $(\xBj, Q^{2})$ and $(\xBj, -t/Q^{2})$  is shown in Fig. \ref{fig:data:coverage}. With those plots one can easily conclude about the actual coverage by various experiments, but also identify parts of phase-space not probed at all by existing data. 

\begin{table*}[!ht]
\centering
\caption{DVCS data used in this analysis.}
\label{tab:data:dvcs_data}
\begin{tabular}{ccccclcc}
\toprule
No. & Collab. 	& Year & Ref. & \multicolumn{2}{c}{Observable} & \makecell{Kinematic \\ dependence} & \makecell{No. of points \\ used /\ all} \\ \midrule
1   & HERMES 	& 2001 & \cite{Airapetian:2001yk} & $A_{LU}^{+}$ & & $\phi$ & 10 /\ 10 \\
2   &  		& 2006 & \cite{Airapetian:2006zr} & $A_{C}^{\cos i \phi}$ & $i = 1$ & $t$ & 4 /\ 4 \\ 
3   &  		& 2008 & \cite{Airapetian:2008aa} & $A_{C}^{\cos i \phi}$ & $i = 0, 1$ & $\xBj$ & 18 /\ 24 \\ 
    &  		&      & 			  & $A_{UT, \mathrm{DVCS}}^{\sin(\phi-\phi_{S})\cos i \phi}$ & $i = 0$ & & \\
    &  		&      & 			  & $A_{UT, \mathrm{I}}^{\sin(\phi-\phi_{S})\cos i \phi}$ & $i = 0, 1$ & & \\
    &  		&      & 			  & $A_{UT, \mathrm{I}}^{\cos(\phi-\phi_{S})\sin i \phi}$ & $i = 1$ & & \\                    
4   &  		& 2009 & \cite{Airapetian:2009aa} & $A_{LU, \mathrm{I}}^{\sin i \phi}$ & $i = 1, 2$ & $\xBj$ & 35 /\ 42 \\
    &  		&      & 			  & $A_{LU, \mathrm{DVCS}}^{\sin i \phi}$ & $i = 1$ & & \\     
    &  		&      & 			  & $A_{C}^{\cos i \phi}$ & $i = 0, 1, 2, 3$ & & \\   
5   &  		& 2010 & \cite{Airapetian:2010ab} & $A_{UL}^{+, \sin i \phi}$ & $i = 1, 2, 3$ & $\xBj$ & 18 /\ 24 \\
    &  		&      & 			  & $A_{LL}^{+, \cos i\phi}$ & $i = 0, 1, 2$ & & \\ 
6   &  		& 2011 & \cite{Airapetian:2011uq} & $A_{LT, \mathrm{DVCS}}^{\cos(\phi-\phi_{S})\cos i \phi}$ & $i = 0, 1$ & $\xBj$ & 24 /\ 32 \\   
    &  		&      & 			  & $A_{LT, \mathrm{DVCS}}^{\sin(\phi-\phi_{S})\sin i \phi}$ & $i = 1$ & & \\
    &  		&      & 			  & $A_{LT, \mathrm{I}}^{\cos(\phi-\phi_{S})\cos i \phi}$ & $i = 0, 1, 2$ & & \\
    &  		&      & 			  & $A_{LT, \mathrm{I}}^{\sin(\phi-\phi_{S})\sin i \phi}$ & $i = 1, 2$ & & \\
7   &  		& 2012 & \cite{Airapetian:2012mq} & $A_{LU, \mathrm{I}}^{\sin i \phi}$ & $i = 1, 2$ & $\xBj$ & 35 /\ 42 \\
    &  		&      & 			  & $A_{LU, \mathrm{DVCS}}^{\sin i \phi}$ & $i = 1$ & & \\ 
    &  		&      & 			  & $A_{C}^{\cos i \phi}$ & $i = 0, 1, 2, 3$ & & \\
8   & CLAS 	& 2001 & \cite{Stepanyan:2001sm}  & $A_{LU}^{-, \sin i \phi}$ & $i = 1, 2$ & --- & 0 /\ 2 \\
9   & 	 	& 2006 & \cite{Chen:2006na} 	  & $A_{UL}^{-, \sin i \phi}$ & $i = 1, 2$ & --- & 2 /\ 2 \\
10  & 	 	& 2008 & \cite{Girod:2007aa} 	  & $A_{LU}^{-}$ & & $\phi$ & 283 /\ 737\\
11  & 	 	& 2009 & \cite{Gavalian:2008aa}   & $A_{LU}^{-}$ & & $\phi$ & 22 /\ 33 \\
12  & 	 	& 2015 & \cite{Pisano:2015iqa}    & $A_{LU}^{-}$, $A_{UL}^{-}$, $A_{LL}^{-}$ & & $\phi$ & 311 /\ 497 \\
13  & 		& 2015 & \cite{Jo:2015ema}	  & $d^{4}\sigma_{UU}^{-}$ & & $\phi$ & 1333 /\ 1933 \\
14  & Hall A 	& 2015 & \cite{Defurne:2015kxq}   & $\Delta d^{4}\sigma_{LU}^{-}$ & & $\phi$ & 228 /\ 228 \\
15  & 		& 2017 & \cite{Defurne:2017paw}   & $\Delta d^{4}\sigma_{LU}^{-}$ & & $\phi$ & 276 /\ 358 \\ 
16  & COMPASS 	& 2018 & \cite{Akhunzyanov:2018nut} & $d^{3}\sigma_{UU}^{\pm}$ & & t & 2 /\ 4\\
17  & ZEUS 	& 2009 & \cite{Chekanov:2008vy} 			& $d^{3}\sigma_{UU}^{+}$ & & t & 4 /\ 4\\
18  & H1 	& 2005 & \cite{Aktas:2005ty} 			& $d^{3}\sigma_{UU}^{+}$ & & t & 7 /\ 8\\
19  &		& 2009 & \cite{Aaron:2009ac} 			& $d^{3}\sigma_{UU}^{\pm}$ & & t & 12 /\ 12\\
    & 	 	&      & & & & & \\
    & 	 	&      & & & & SUM: & 2624 /\ 3996\\ \bottomrule
\end{tabular}
\end{table*}

\section{Results}
\label{sec:results}

\subsection{Performance}
\label{sec:results:performance}

The value of the $\chi^{2}$ function obtained after the training of our neural network system with $2624$ experimental points is $2243.5$ for the central replica. It gives the average deviation of that system's answer from experimental data $2243.5/2624 \approx 0.85$ per single data point. This value per experimental data set is given in Table \ref{tab:results:chi2_per_dataset}. The number of free parameters, that is the number of weights and biases in all networks, is $248$. There is not much sense of studying the goodness of fit with this information taken into account, as we know \emph{a priori} that the size of a single network may be too large, and that is why we are using a regularization. In other words, we are able to increase the number of neurons in our network, and therefore the number of weight and biases, keeping at the same time the same precision of data description. 
\begin{table}[!ht]
\centering
\caption{Values of the $\chi^{2}$ function per data set. For a given data set, \emph{cf}. Table \ref{tab:data:dvcs_data}, given are: $\chi^{2}$ value, the number of experimental points $n$, and the ratio between those two numbers.}
\label{tab:results:chi2_per_dataset}
\begin{tabular}{ccccccc}
\toprule
No. & Collab. 	& Year & Ref. & $\chi^{2}$ & $n$ &  $\chi^{2}/n$\\ \midrule
1   & HERMES 	& 2001 & \cite{Airapetian:2001yk}&  $10.7$  	&   $10$ &  $1.07$  \\
2   &  		& 2006 & \cite{Airapetian:2006zr}   &  $5.5$  	&    $4$ &  $1.38$  \\
3   &  		& 2008 & \cite{Airapetian:2008aa}   &  $18.5$ 	&   $18$ &  $1.03$  \\       
4   &  		& 2009 & \cite{Airapetian:2009aa}   &  $34.7$ 	&   $35$ &  $0.99$  \\
5   &  		& 2010 & \cite{Airapetian:2010ab}   &  $40.7$  	&   $18$ &  $2.26$  \\
6   &  		& 2011 & \cite{Airapetian:2011uq}   &  $16.7$ 	&   $24$ &  $0.70$  \\
7   &  		& 2012 & \cite{Airapetian:2012mq}   &  $22.4$ 	&   $35$ &  $0.64$  \\
8   & CLAS 	& 2001 & \cite{Stepanyan:2001sm}    & ---   	   &    $0$ & ---      \\
9   & 	 	& 2006 & \cite{Chen:2006na} 	    	&  $1.0$  	&    $2$ &  $0.52$  \\
10  & 	 	& 2008 & \cite{Girod:2007aa} 	      &  $376.4$ 	&  $283$ &  $1.33$  \\
11  & 	 	& 2009 & \cite{Gavalian:2008aa}     &  $28.3$ 	&   $22$ &  $1.29$  \\
12  & 	 	& 2015 & \cite{Pisano:2015iqa} 	   &  $306.6$ 	&  $311$ &  $0.99$  \\
13  & 		& 2015 & \cite{Jo:2015ema} 	    	&  $884.7$ 	& $1333$ &  $0.66$  \\
14  & Hall A 	& 2015 & \cite{Defurne:2015kxq}  &  $231.8$ 	&  $228$ &  $1.02$  \\
15  & 		& 2017 & \cite{Defurne:2017paw}     &  $211.4$	&  $276$ &  $0.77$  \\
16  & COMPASS 	& 2018 & \cite{Akhunzyanov:2018nut} &  $3.0$ &    $2$ &  $1.50$  \\  
17  & ZEUS 	& 2009 & \cite{Chekanov:2008vy} 		& $5.49$    &    $4$ &  $1.38$  \\
18  & H1 	& 2005 & \cite{Aktas:2005ty} 			& $22.2$    &    $7$ &  $3.17$  \\
19  &		& 2009 & \cite{Aaron:2009ac} 			   & $23.4$    &    $12$ &  $1.95$ \\
\bottomrule
\end{tabular}
\end{table}

\begin{figure*}[!ht]
\begin{center}
\includegraphics[width=\figWidth]{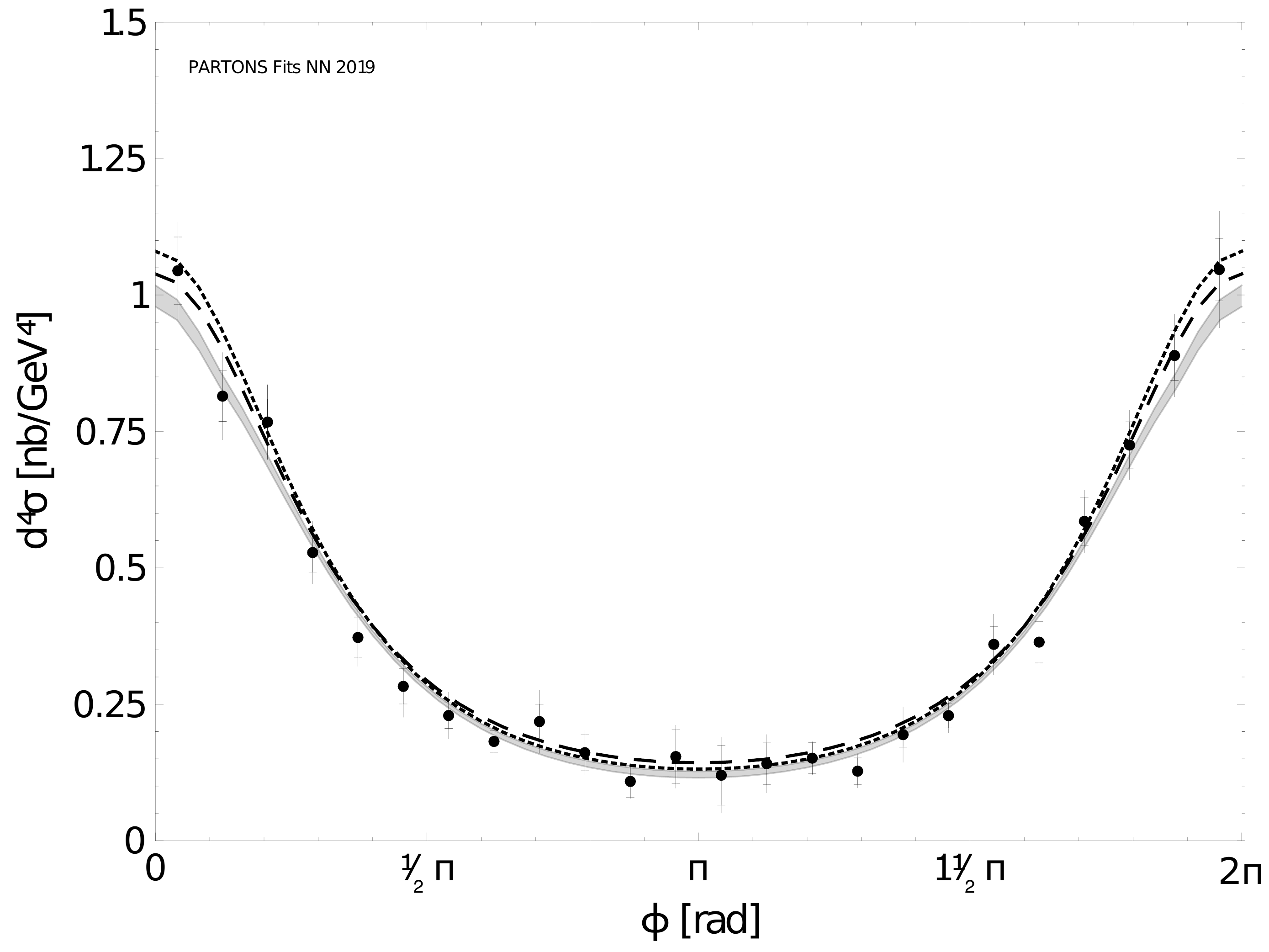}
\hspace{\figSpace}
\includegraphics[width=\figWidth]{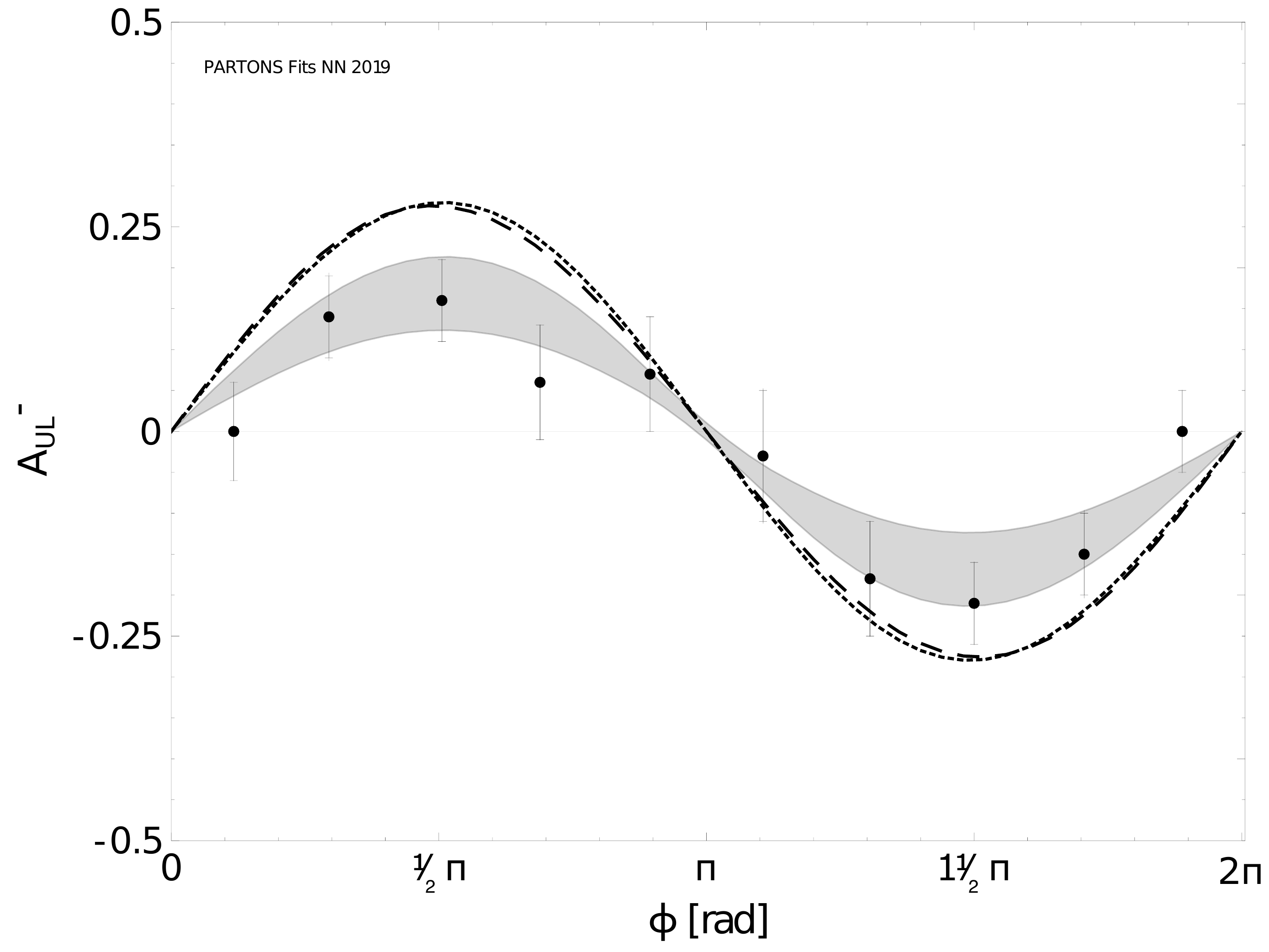}
\caption{Comparison between the results of this analysis, the selected GPD models and experimental data published by CLAS in Refs. \cite{Jo:2015ema, Pisano:2015iqa} for $d^{4}\sigma_{UU}^{-}$ at $\xBj = 0.244$, $t = -0.15~\mathrm{GeV}^2$ and $Q^{2} = 1.79~\mathrm{GeV}^2$ (left) and for $A_{UL}^{-}$ at $\xBj = 0.2569$, $t = -0.23~\mathrm{GeV}^2$, $Q^{2} = 2.019~\mathrm{GeV}^2$ (right). The gray bands correspond to the results of this analysis with 68\% confidence level for the uncertainties coming from DVCS data, respectively. The dotted curve is for the GK GPD model \cite{Goloskokov:2005sd, Goloskokov:2007nt, Goloskokov:2009ia}, while the dashed one is for VGG \cite{Vanderhaeghen:1998uc, Vanderhaeghen:1999xj, Goeke:2001tz, Guidal:2004nd}. The curves are evaluated at the kinematics of experimental data.}
\label{fig:results:clas}
\end{center}
\end{figure*}

\begin{figure*}[!ht]
\begin{center}
\includegraphics[width=\figWidth]{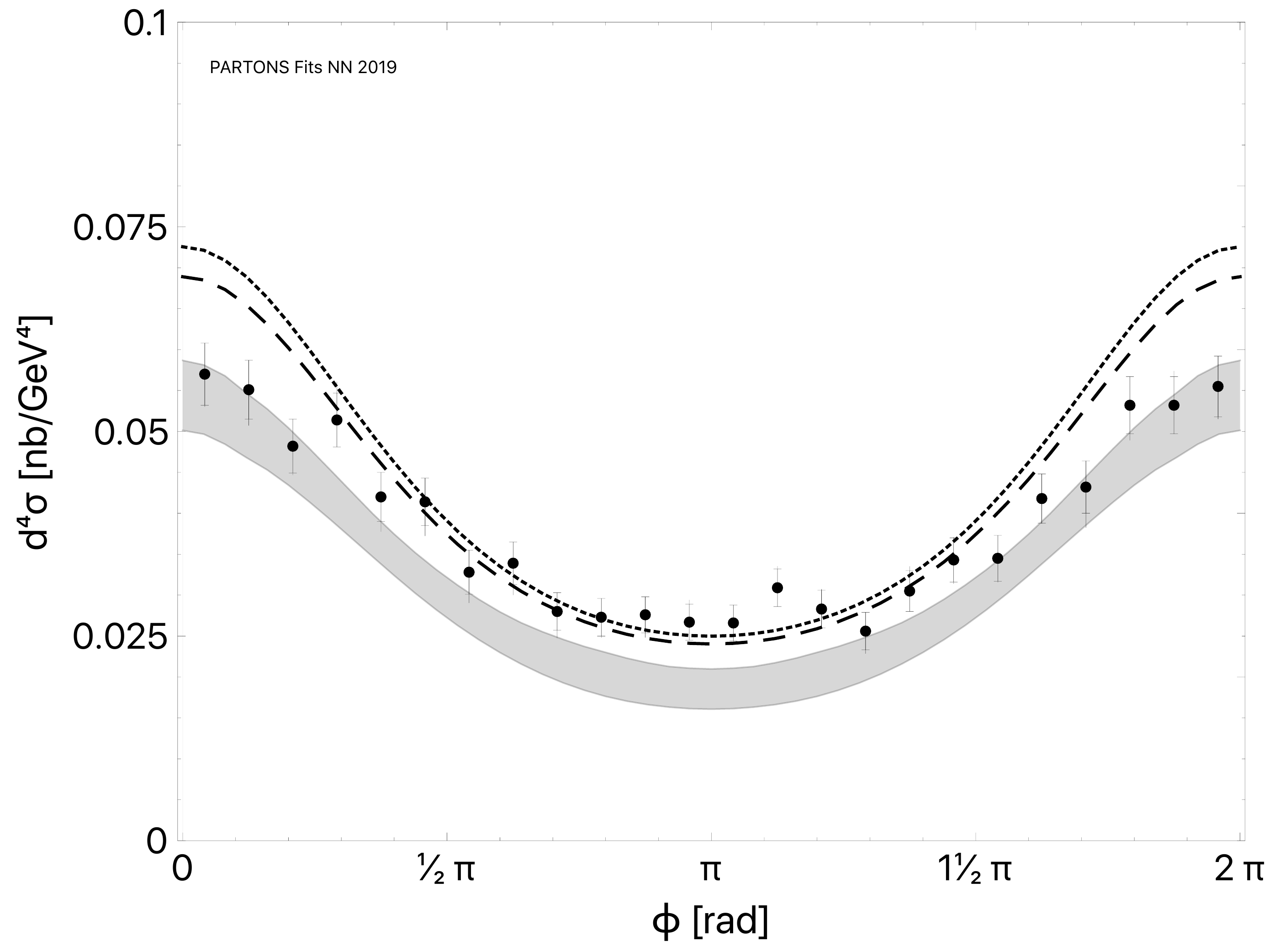}
\hspace{\figSpace}
\includegraphics[width=\figWidth]{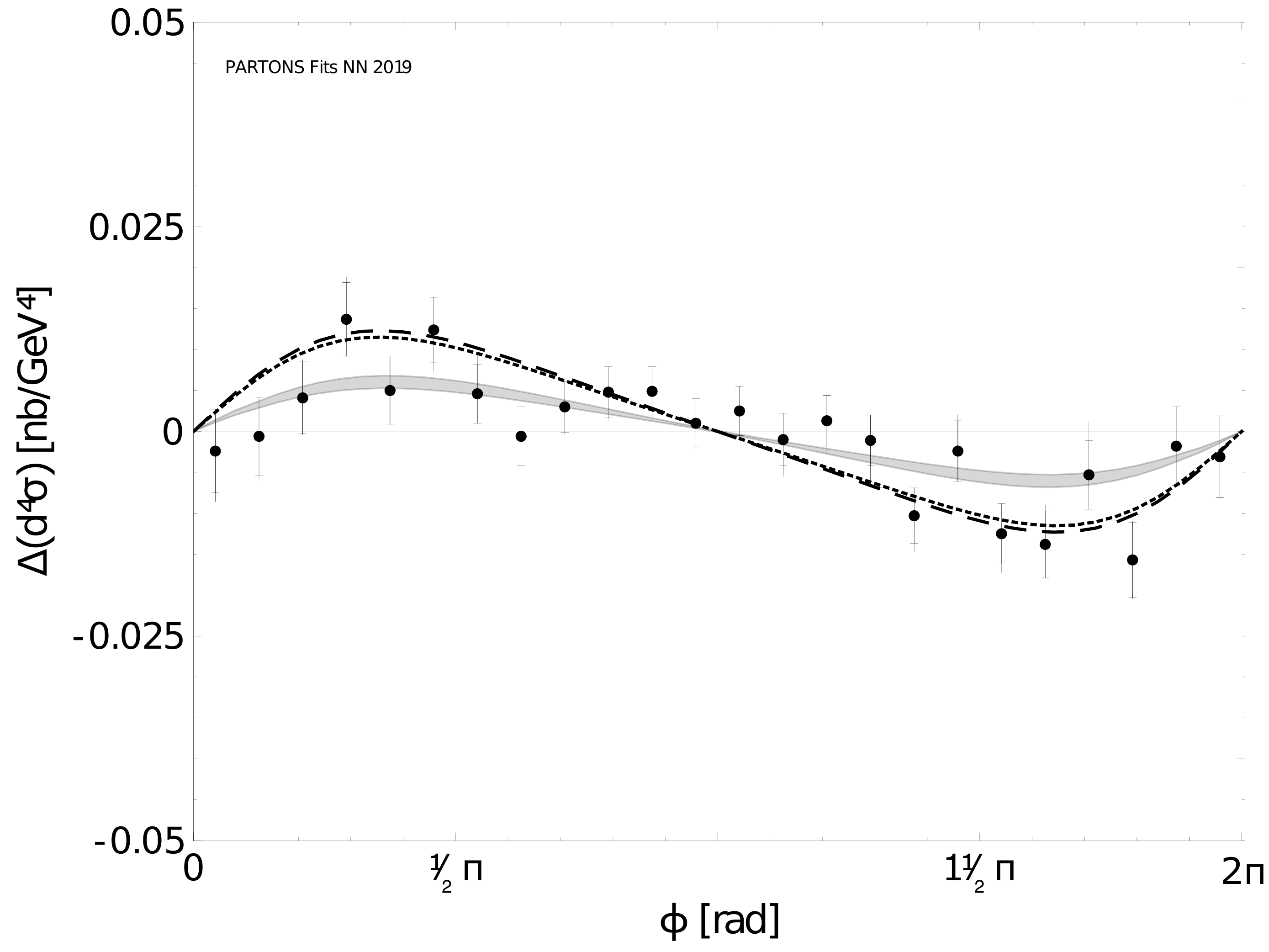}
\caption{Comparison between the results of this analysis, the selected GPD models and experimental data published by Hall A in Ref. \cite{Defurne:2015kxq} for $d^{4}\sigma_{UU}^{-}$ (left) and $\Delta d^{4}\sigma_{LU}^{-}$ (right) at $\xBj = 0.392$, $t = -0.233~\mathrm{GeV}^2$ and $Q^{2} = 2.054~\mathrm{GeV}^2$. For further description see the caption of \reffig{fig:results:clas}.}
\label{fig:results:halla}
\end{center}
\end{figure*}

\begin{figure*}[!ht]
\begin{center}
\includegraphics[width=\figWidth]{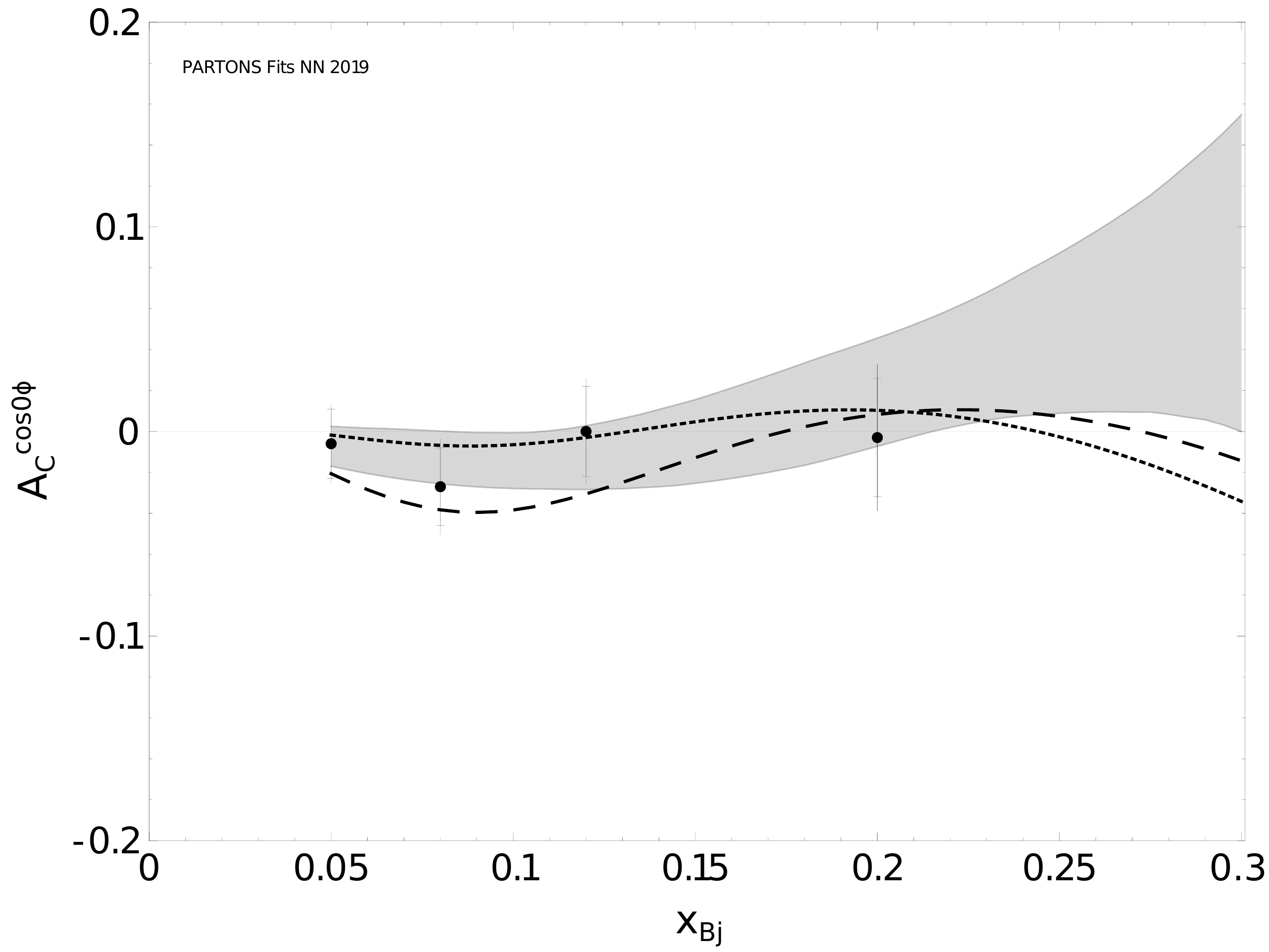}
\hspace{\figSpace}
\includegraphics[width=\figWidth]{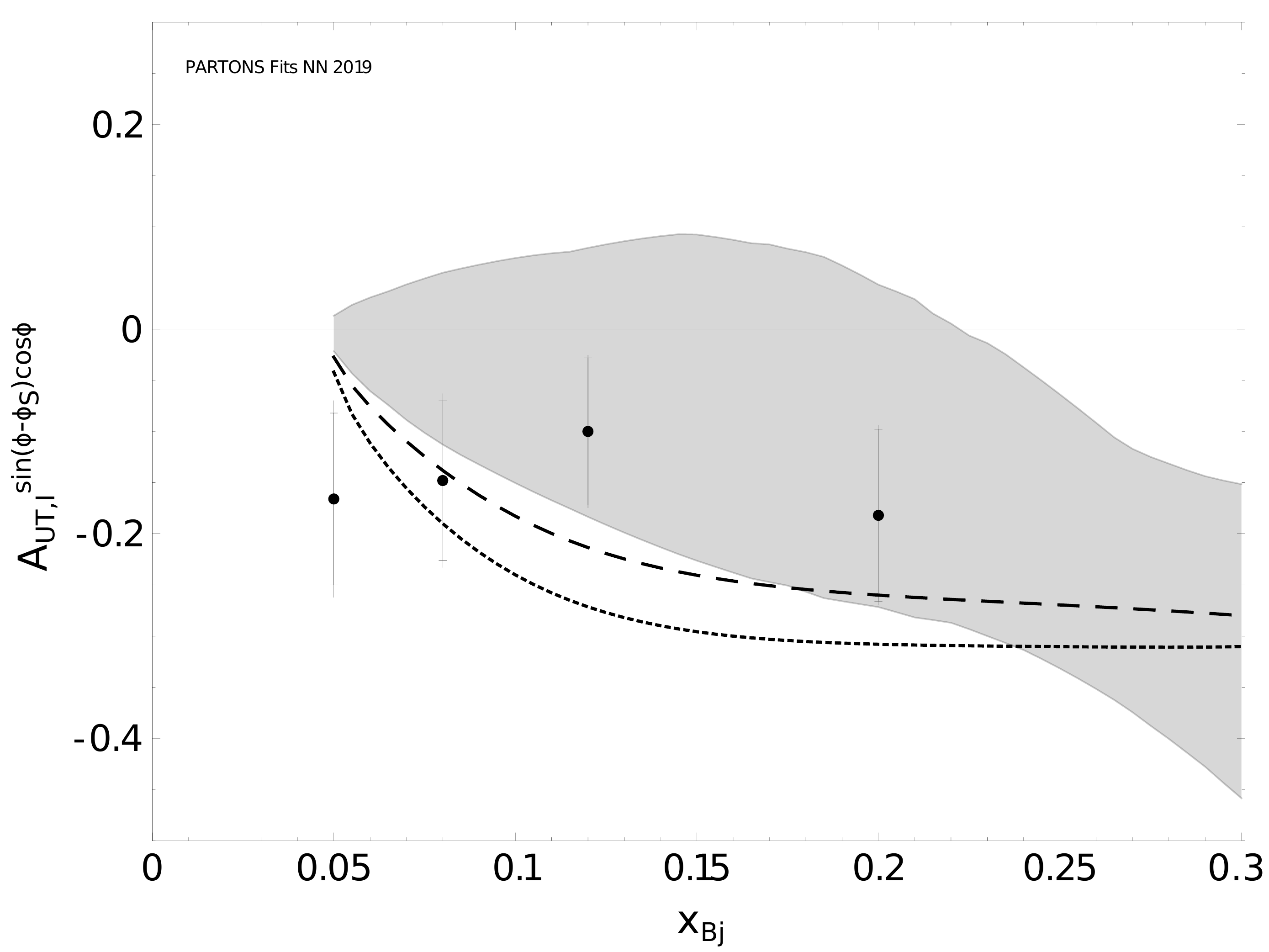}
\caption{Comparison between the results of this analysis, the selected GPD models and the experimental data published by HERMES in Ref. \cite{Airapetian:2008aa} for $A_{C}^{\cos 0 \phi}$ (left) and $A_{UT, \mathrm{I}}^{\sin(\phi-\phi_{S})\cos \phi}$ (right) at $t = -0.12~\mathrm{GeV}^2$ and $Q^{2} = 2.5~\mathrm{GeV}^2$. For further description see the caption of \reffig{fig:results:clas}.}
\label{fig:results:hermess}
\end{center}
\end{figure*}

\begin{figure*}[!ht]
\begin{center}
\includegraphics[width=\figWidth]{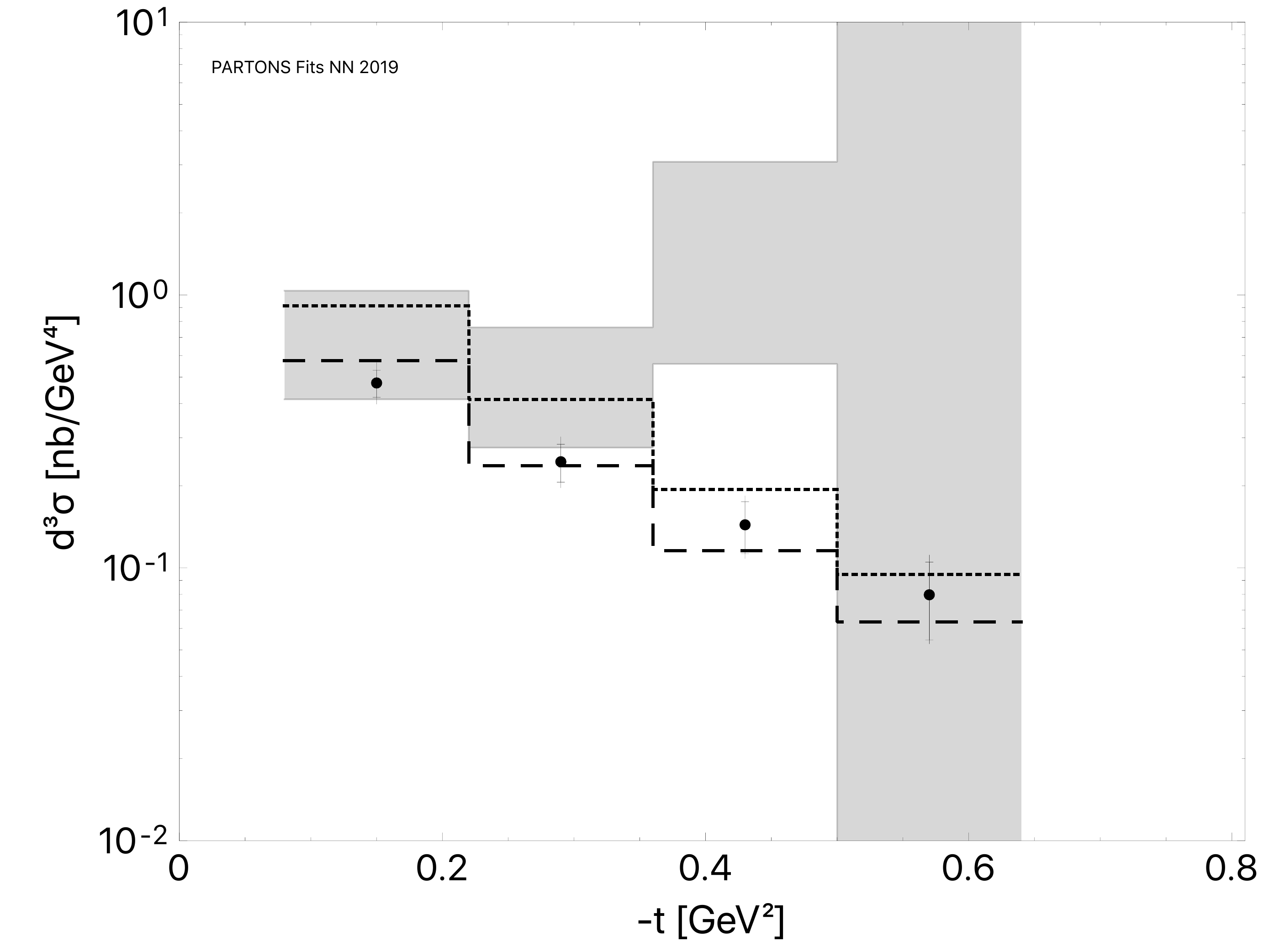}
\includegraphics[width=\figWidth]{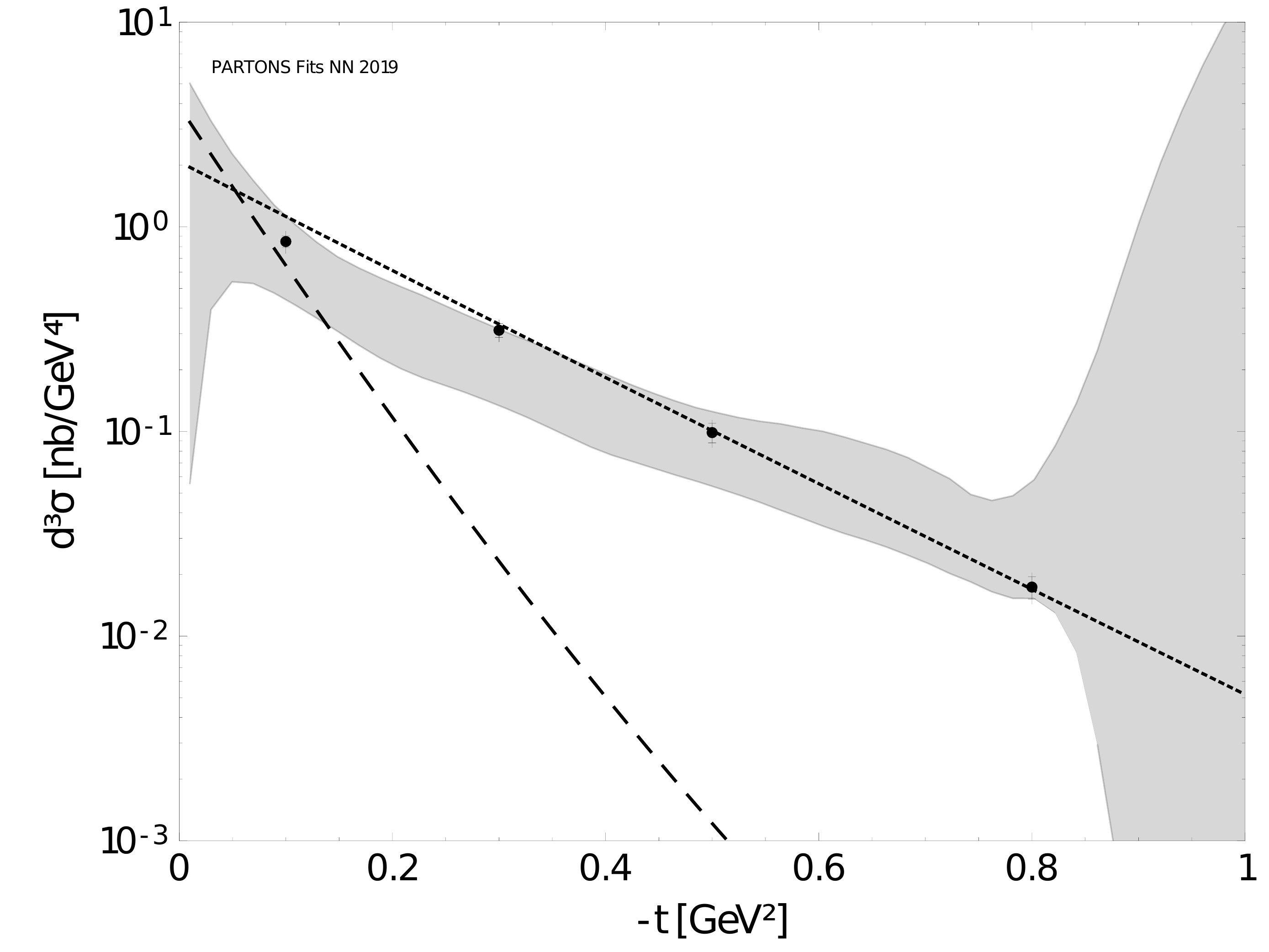}
\caption{Comparison between the results of this analysis, the selected GPD models and the experimental data published by COMPASS in Ref. \cite{Akhunzyanov:2018nut} for $d^{3}\sigma_{UU}^{\pm}$ (left) and by H1 in Ref. \cite{Aaron:2009ac} for $d^{3}\sigma_{UU}^{+}$ at $\xBj = 0.002$ and $Q^{2} = 10~\mathrm{GeV}^{2}$ (right). In the left plot, the experimental data, the results of this analysis and the GPD model evaluations are 3D integrals in four bins of the $(\nu, Q^{2}, t)$ phase-space, where $\nu$ is the virtual-photon energy. The ranges of those bins are specified in Ref. \cite{Akhunzyanov:2018nut}. For further description see the caption of  \reffig{fig:results:clas}.}
\label{fig:results:compass_and_hera}
\end{center}
\end{figure*}

Figures \ref{fig:results:clas} - \ref{fig:results:compass_and_hera} provide a straightforward comparison between our fit and the selected data sets. As indicated by the $\chi^{2}$ values summarized in Table \ref{tab:results:chi2_per_dataset}, we are able to describe the data well within a single phenomenological framework based on the neural network approach. This includes data ranging from HERA to JLab kinematics. Beyond the results of this analysis, predictions coming from the GK \cite{Goloskokov:2005sd, Goloskokov:2007nt, Goloskokov:2009ia} and VGG \cite{Vanderhaeghen:1998uc, Vanderhaeghen:1999xj, Goeke:2001tz, Guidal:2004nd} GPD models are also shown. Those two models originate from the exploratory phase of GPD studies and are able to describe only a general behavior followed by experimental data. This confirms the need for new GPD models constrained in global analyses, preferably multi-channel ones.

As has been already mentioned in \refsec{sec:data}, Hall A cross sections are not included in the nominal data set that is used in this analysis. We point out, that in our last analysis \cite{Moutarde:2018kwr} we were not able to describe those cross sections with the proposed Ansatz and those data were excluded there as well. The question is: can those data be described in this analysis, where a flexible CFF parameterization based on the neural network approach is used? The answer consists of two parts: (\emph{i}) the parameterizations of CFFs obtained in this analysis from the nominal data set do not have the predictive power to describe Hall A cross sections. The value of the $\chi^2$ function for those cross sections is $5916.6$, which for $594$ points gives the reduced value of $9.96$. The bad description is also seen by eye in Fig. \ref{fig:results:halla}. (\emph{ii}) the inclusion of Hall A cross sections in the extraction of CFFs significantly improves the $\chi^{2}$ evaluated for this subset, from aforementioned $9.96$ to $1.12$ per single data point. However, such an inclusion makes the $\chi^{2}$ for the CLAS cross sections worse, from $0.66$ reported in Table \ref{tab:results:chi2_per_dataset} to $0.83$ per single data point. Taking both observations into account we found the situation unclear at this moment. Hall A cross sections will be investigated further in a future analysis with higher-twist contributions taken into account.   

The inclusion of low-$\xBj$ experimental data has significantly extended the coverage of phase-space. However, the sparsity of those data creates problems in the supervised training that we utilize in this analysis, see \refsec{sec:methodology:regularization}. Namely, the random division of the available experimental data into training and test subsets causes an insufficient coverage of some parts of the phase-space by either the training or the test points. This may lead to over-fitting, if a given part of the phase-space is only covered by training points, or under-fitting, if kinematics is only covered by test points. The effect increases the spread of replicas and because of that gives an additional contribution to the estimated uncertainties. The goodness of fit for those data is also worse as shown in Table \ref{tab:results:chi2_per_dataset}. Foreseen data coming from electron-ion facilities will improve this situation. 

\subsection{Compton form factors}
\label{sec:results:cffs}

\begin{figure}[!ht]
\begin{center}
\includegraphics[width=\figWidth]{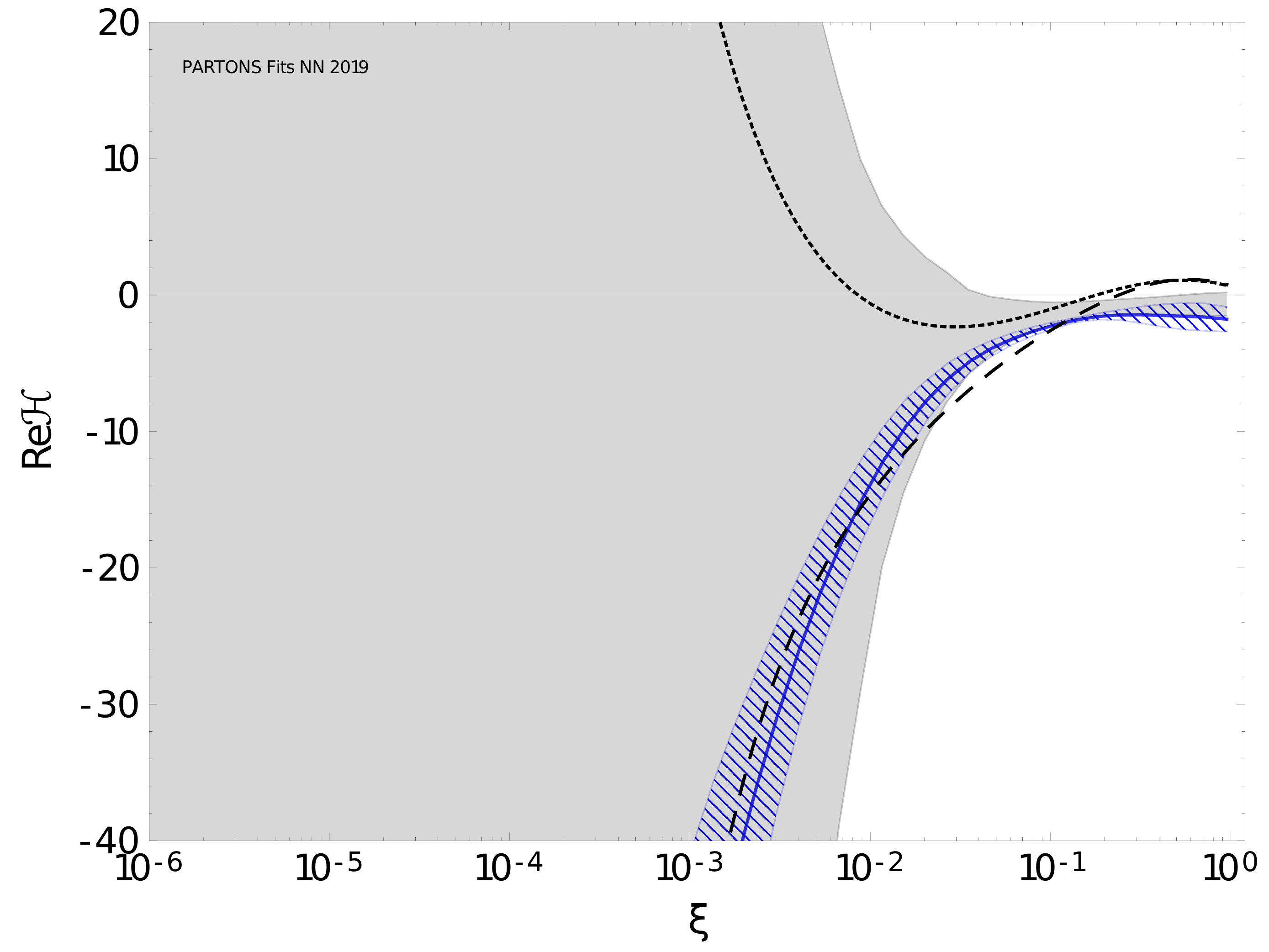}
\hspace{\figSpace}
\includegraphics[width=\figWidth]{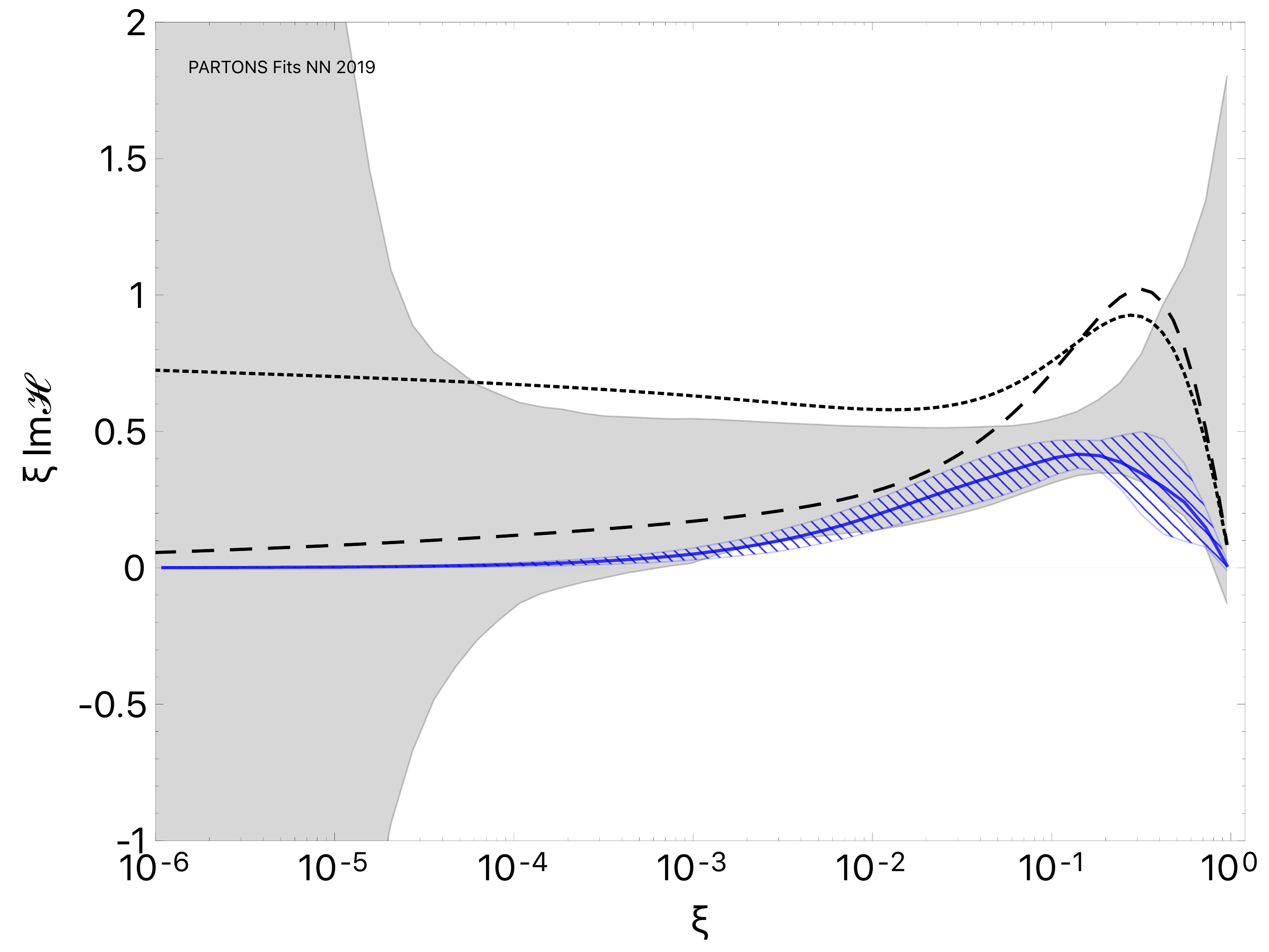}
\caption{Real (left) and imaginary (right) parts of the CFF $\mathcal{H}$ as a function of $\xi$ for $t = -0.3~\mathrm{GeV}^{2}$ and $Q^{2} = 2~\mathrm{GeV}^{2}$. The blue solid line surrounded by the blue hatched band denotes the result of our previous analysis \cite{Moutarde:2018kwr}. The depicted uncertainty accounts for uncertainties estimated in that analysis for experimental data, unpolarized, polarized PDFs and elastic form factors. For further description see the caption of \reffig{fig:results:clas}.}
\label{fig:results:cff_h}
\end{center}
\end{figure}

\begin{figure*}[!ht]
\begin{center}
\includegraphics[width=\figWidth]{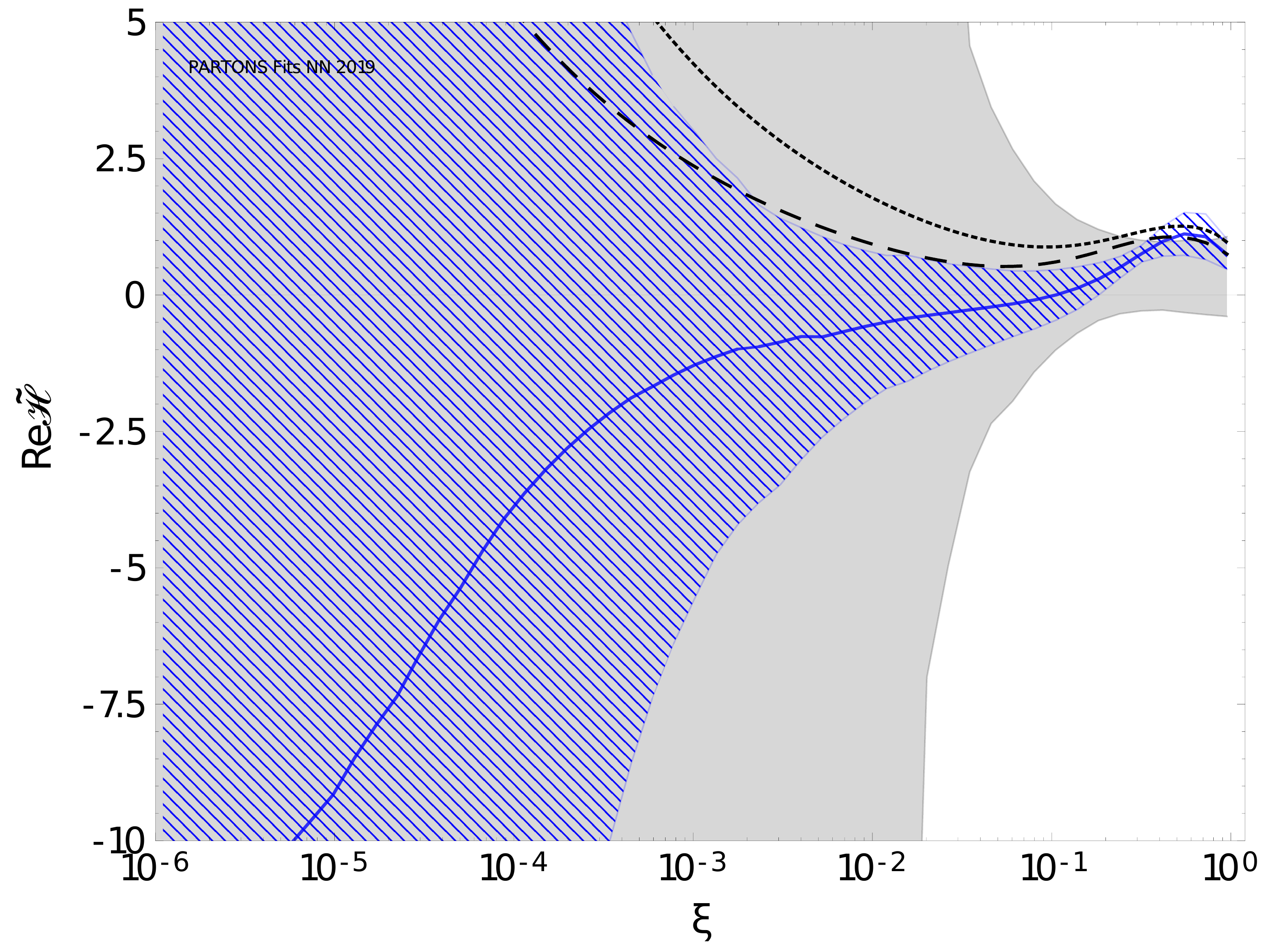}
\hspace{\figSpace}
\includegraphics[width=\figWidth]{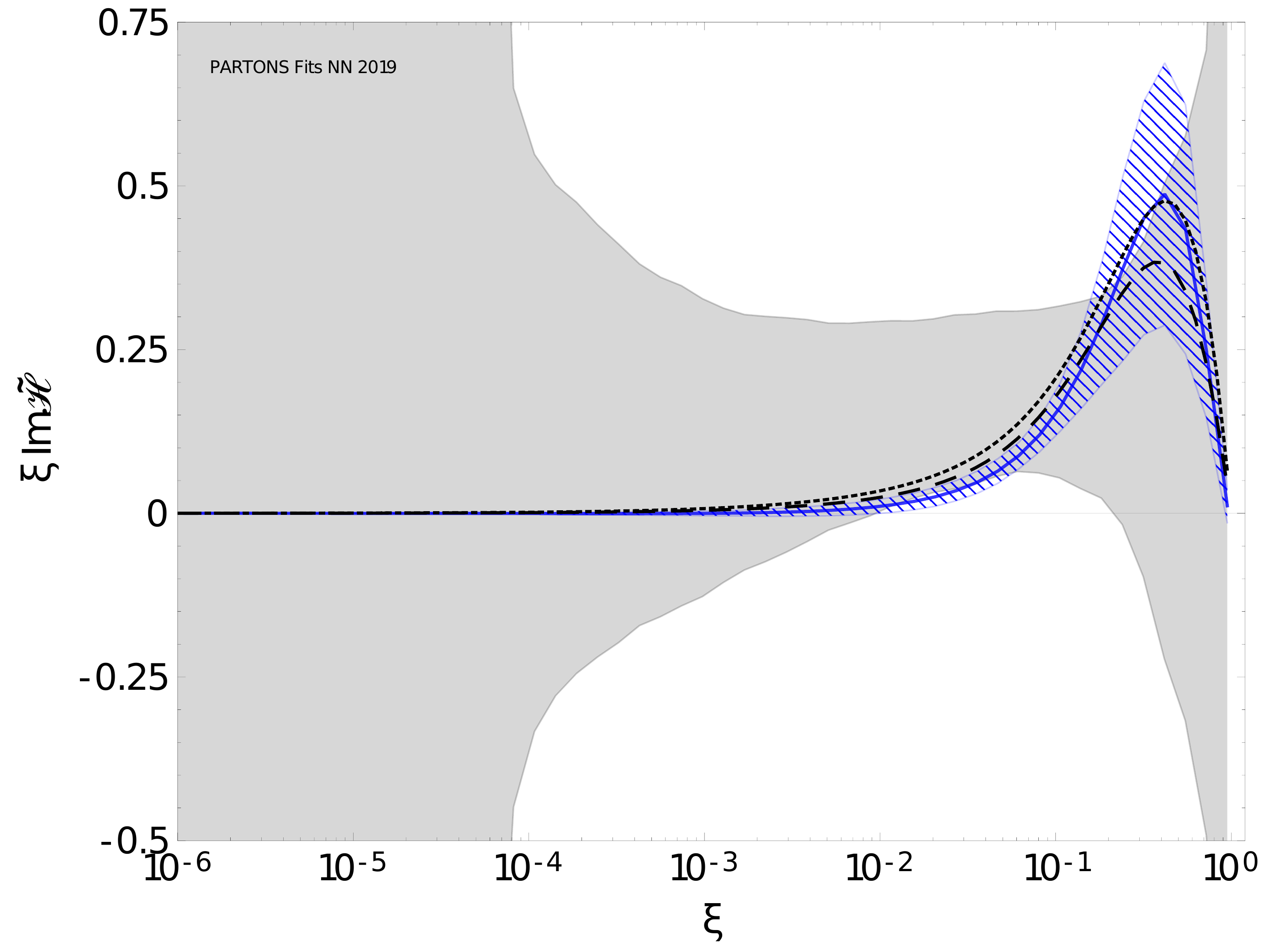}
\caption{Real (left) and imaginary (right) parts of the CFF $\widetilde{\mathcal{H}}$ as a function of $\xi$ for $t = -0.3~\mathrm{GeV}^{2}$ and $Q^{2} = 2~\mathrm{GeV}^{2}$. For further description see the caption of \reffig{fig:results:cff_h}.}
\label{fig:results:cff_ht}
\end{center}
\end{figure*}

\begin{figure*}[!ht]
\begin{center}
\includegraphics[width=\figWidth]{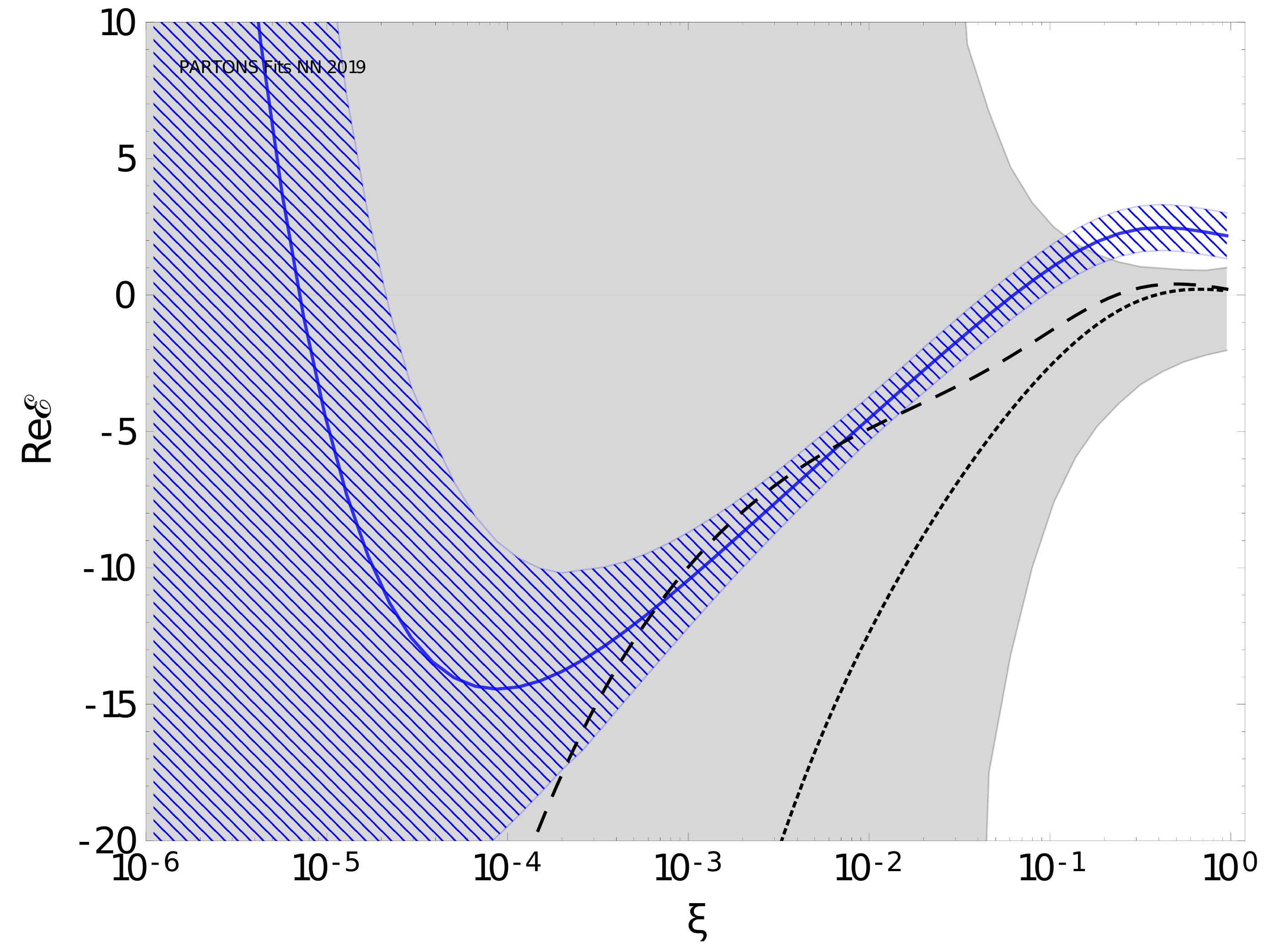}
\hspace{\figSpace}
\includegraphics[width=\figWidth]{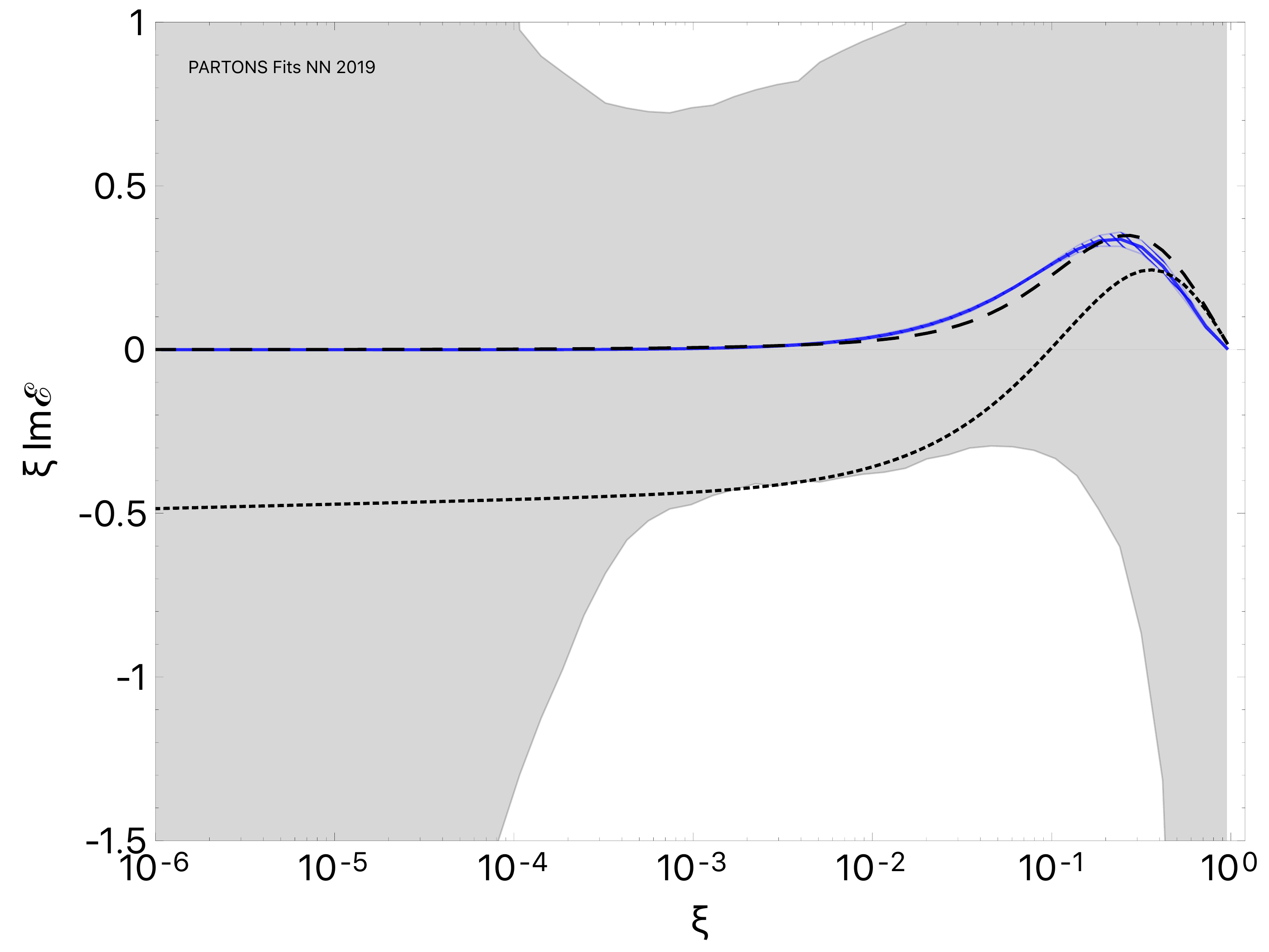}
\caption{Real (left) and imaginary (right) parts of the CFF $\mathcal{E}$ as a function of $\xi$ for $t = -0.3~\mathrm{GeV}^{2}$ and $Q^{2} = 2~\mathrm{GeV}^{2}$. For further description see the caption of \reffig{fig:results:cff_h}.}
\label{fig:results:cff_e}
\end{center}
\end{figure*}

\begin{figure*}[!ht]
\begin{center}
\includegraphics[width=\figWidth]{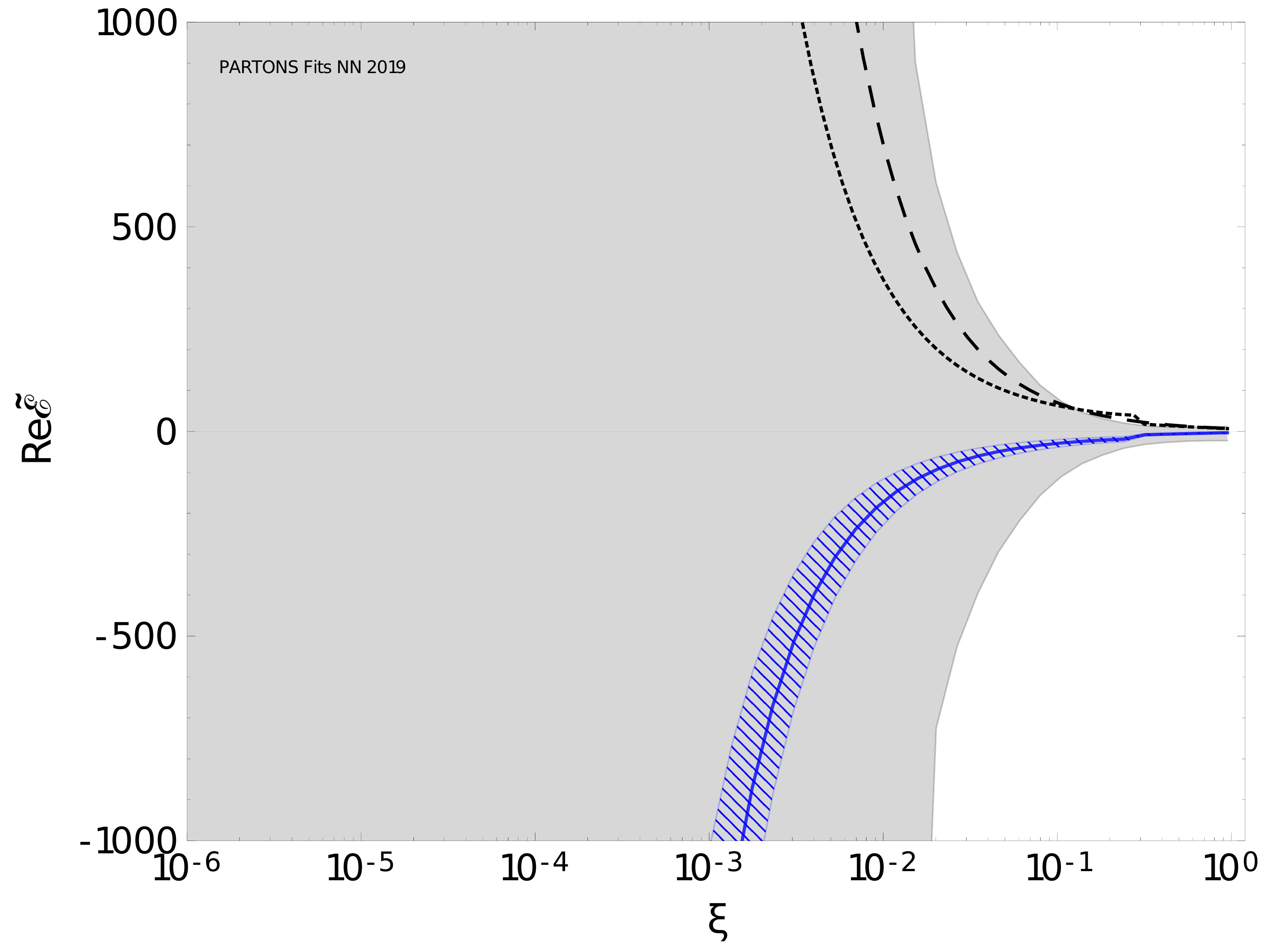}
\hspace{\figSpace}
\includegraphics[width=\figWidth]{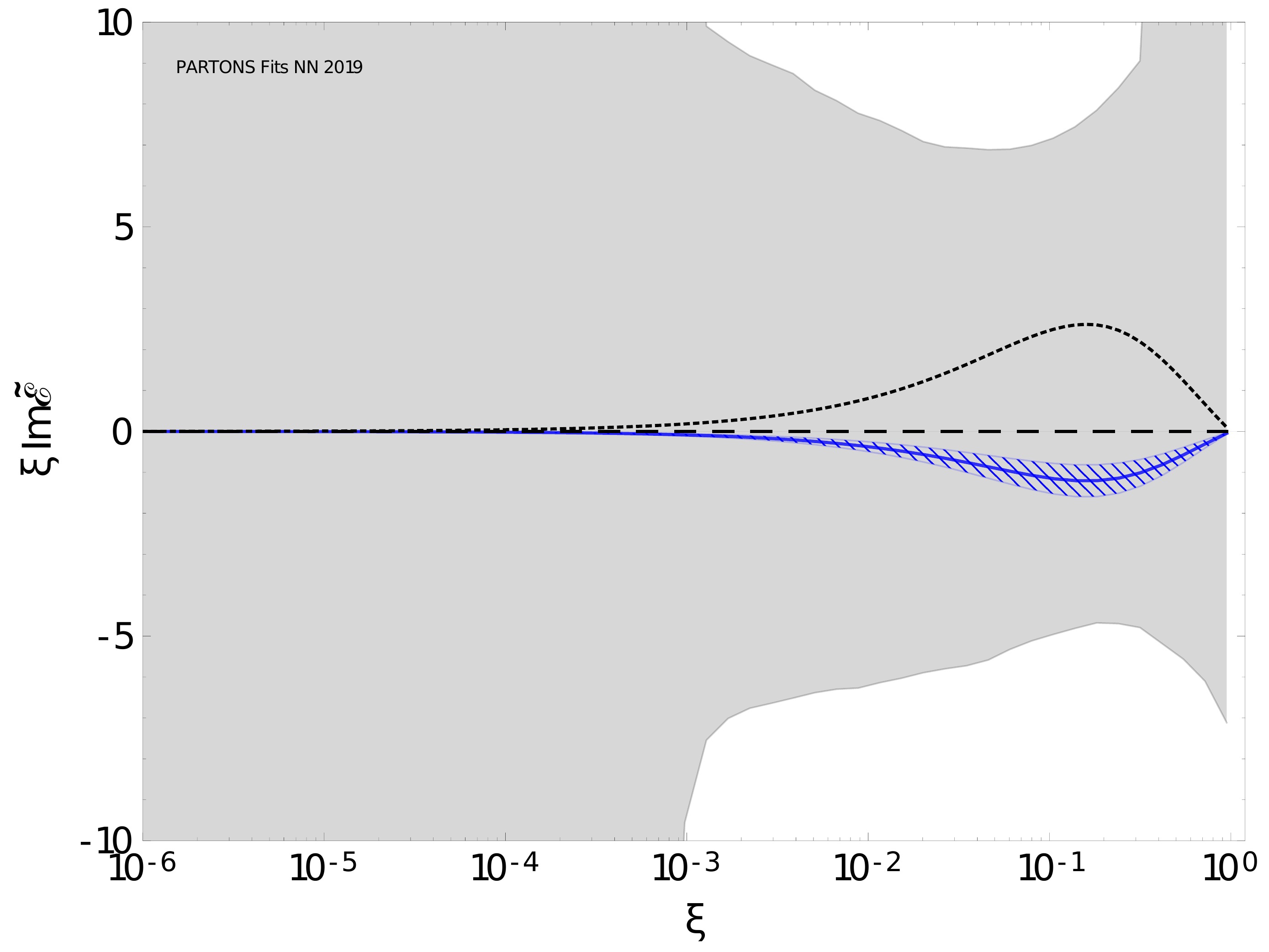}
\caption{Real (left) and imaginary (right) parts of the CFF $\widetilde{\mathcal{E}}$ as a function of $\xi$ for $t = -0.3~\mathrm{GeV}^{2}$ and $Q^{2} = 2~\mathrm{GeV}^{2}$. For further description see the caption of \reffig{fig:results:cff_h}.}
\label{fig:results:cff_et}
\end{center}
\end{figure*}

The parameterizations of CFFs are shown in Figs. \ref{fig:results:cff_h} - \ref{fig:results:cff_et} for example kinematics of $t = -0.3~\mathrm{GeV}^{2}$ and $Q^{2} = 2~\mathrm{GeV}^{2}$ as a function of $\xi$. As expected the data provide the best constraints on $\mathrm{Im}\mathcal{H}$, and some on $\mathrm{Re}\mathcal{H}$, $\mathrm{Im}\widetilde{\mathcal{H}}$ and $\mathrm{Re}\widetilde{\mathcal{E}}$. Other CFFs are poorly constrained by the available data, in particular $\mathcal{E}$ related to GPD $E$, being of a great importance for the study of parton densities in a transversely polarized proton and the determination of the orbital angular momentum through Ji's sum rule.

\begin{figure*}[!ht]
\begin{center}
\includegraphics[width=\figWidth]{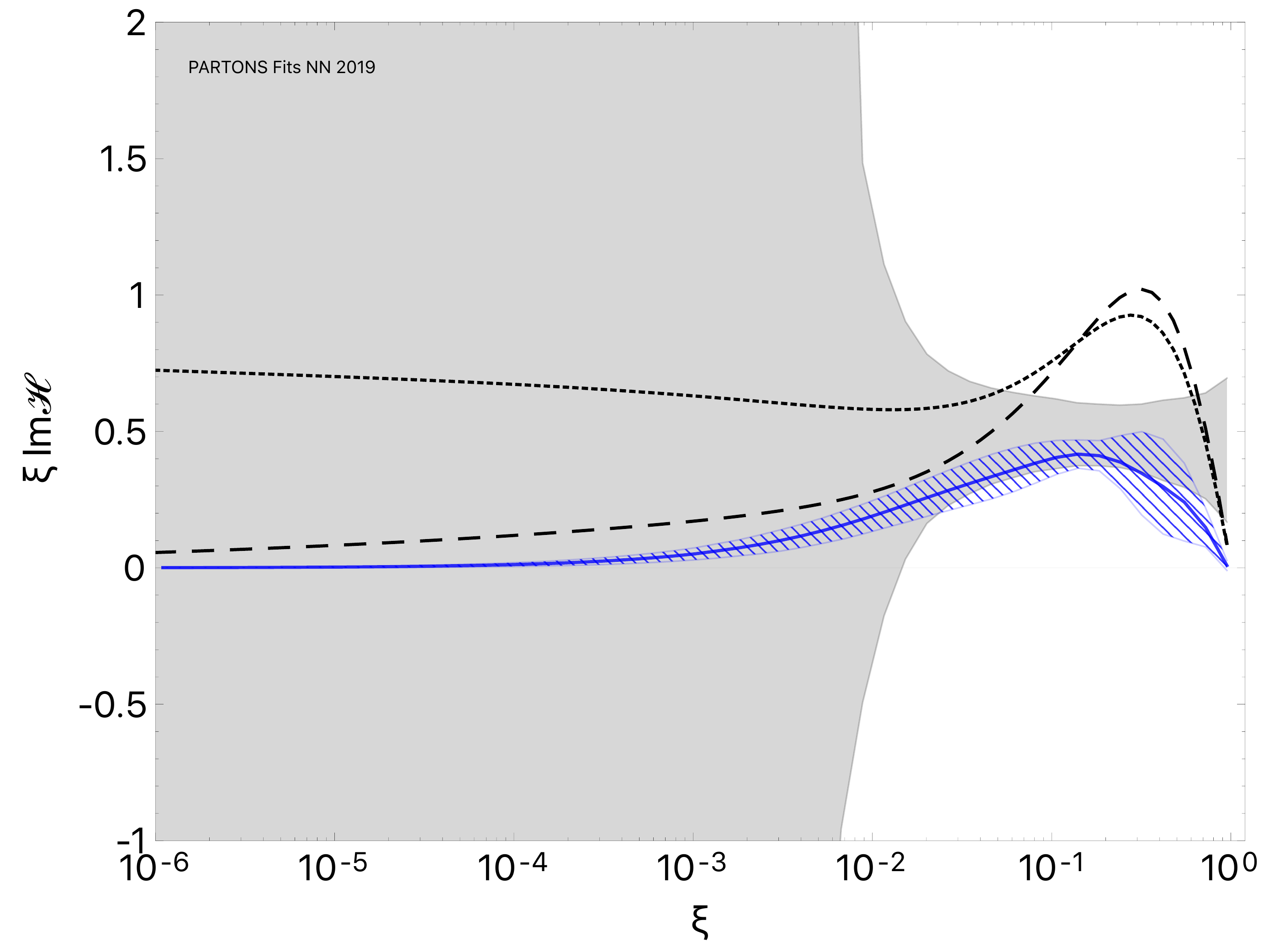}
\caption{The same as \reffig{fig:results:cff_h}, but without the HERA and COMPASS data taken into account in the extraction of CFFs.}
\label{fig:results:cff_h_no_lowX}
\end{center}
\end{figure*}

The inclusion of HERA and COMPASS data in a global extraction of CFFs is not trivial, but those data provide important constraints in the low and intermediate range of $\xBj$. We demonstrate it with the example of  \reffig{fig:results:cff_h_no_lowX}, where the imaginary part of the CFF $\mathcal{H}$ is shown without the inclusion of low-$\xBj$ data. We point out that the gap between the collider and fixed target experiments seen in the coverage of the $(\xBj, Q^{2})$ phase-space in \reffig{fig:data:coverage} is expected to be filled by future experiments at expected electron-ion collider facilities. The precision of foreseen data should allow for a precise phenomenology in that domain.

\begin{figure*}[!ht]
\begin{center}
\includegraphics[width=\figWidth]{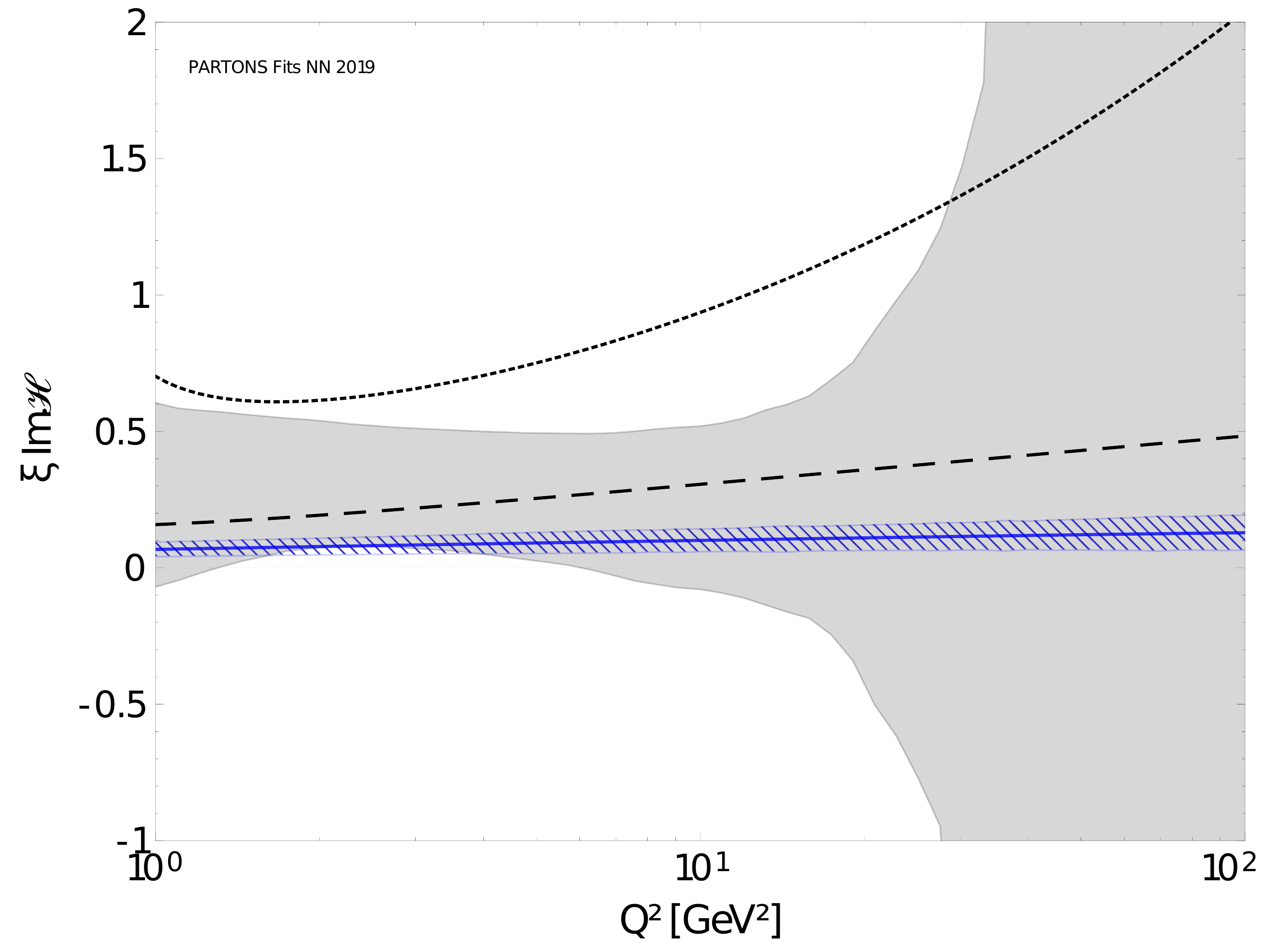}
\hspace{\figSpace}
\includegraphics[width=\figWidth]{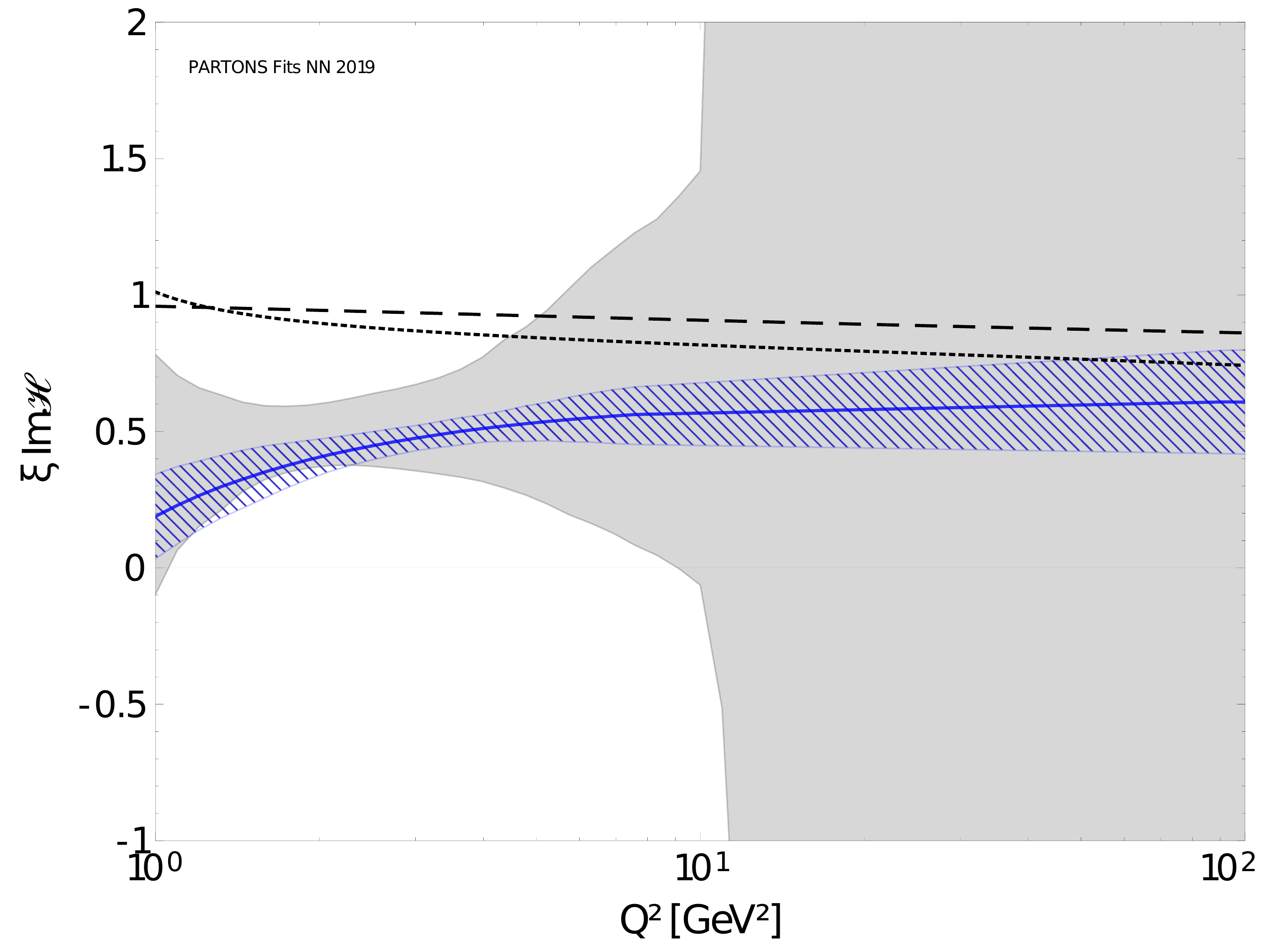}
\caption{Imaginary part of the CFF $\mathcal{H}$ as a function of $Q^{2}$ for $t = -0.3~\mathrm{GeV}^{2}$ and $\xi = 0.002$ (left) and $\xi = 0.2$ (right). For further description see the caption of \reffig{fig:results:cff_h}.}
\label{fig:results:cff_h_Q2_evolution}
\end{center}
\end{figure*}

The $Q^{2}$ evolution of the  CFF $\mathcal{H}$ for $\xi = 0.002$ and $\xi = 0.2$ is shown in  \reffig{fig:results:cff_h_Q2_evolution}. One can note a rather mild $Q^{2}$-dependence followed by the extracted CFF parameterizations, consistent with the expected logarithmic behavior. 

\subsection{Nucleon tomography}
\label{sec:results:nt}

\begin{figure*}[!ht]
\begin{center}
\includegraphics[width=\figWidth]{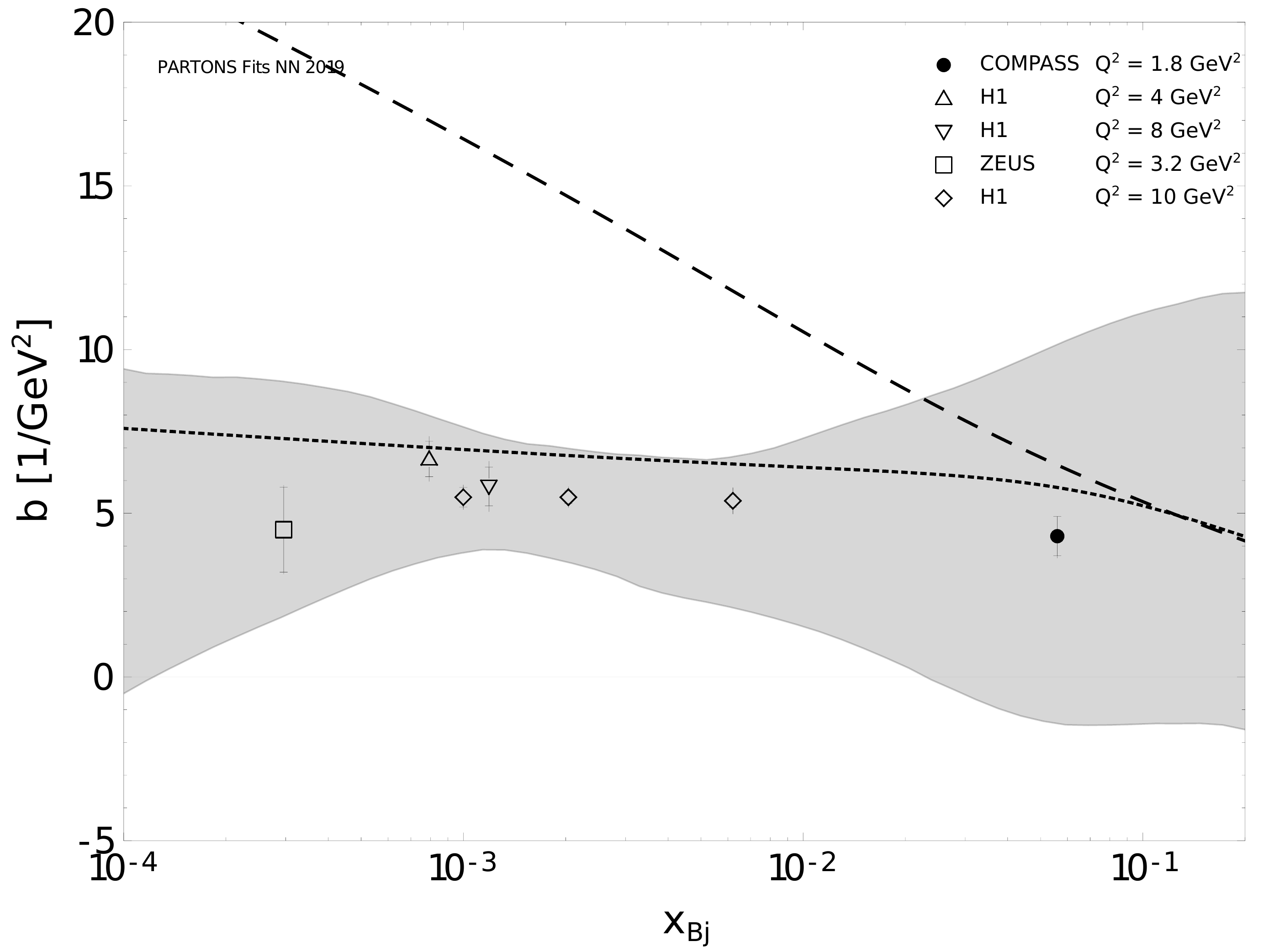}
\caption{Comparison between the results of this analysis, the selected GPD models and the experimental data by the COMPASS \cite{Akhunzyanov:2018nut}, ZEUS \cite{Chekanov:2008vy} and H1 \cite{Aktas:2005ty, Aaron:2009ac} Collaborations for the slope $b$ at $Q^{2} = 10~\mathrm{GeV}^2$. Note that the shown data differ by $Q^{2}$ values indicated in the legend. For further description see the caption of \reffig{fig:results:clas}.}
\label{fig:results:slope}
\end{center}
\end{figure*}

In \reffig{fig:results:slope} we show our results for the slope $b$ of the DVCS cross section described by a single exponential function, $d^{3}\sigma_{UU} \propto \exp(bt)$. This slope is evaluated as indicated in our previous analysis \cite{Moutarde:2018kwr}. It can be converted into a transverse extension of partons under assumptions that are also specified in Ref. \cite{Moutarde:2018kwr}. The uncertainties on the extracted values of $b$ are larger than expected, mainly because of the aforementioned problems with the supervised training caused by the sparsity of the low-$\xBj$ data. In addition, without an explicit assumption about the exponential $t$-behavior of $\mathrm{Im}\mathcal{H}$, the estimation of $b$ from pliant replicas gets additional uncertainties. 

The extraction of tomography information from CFFs is also possible in the high-$\xBj$ range, see for instance Refs. \cite{Dupre:2016mai, Dupre:2017hfs}. However, this requires a ``de-skewing'' of $\mathrm{Im}\mathcal{H}$, \ie one needs to evaluate the GPD $H^{q}(x, 0, t)$ from:
\begin{equation}
\mathrm{Im}\mathcal{H}(\xi, t) \overset{\mathrm{LO}}{=} \pi \sum_q e_{q}^{2} H^{q(+)}(x = \xi, \xi, t) \, ,
\end{equation}
where the sum runs over quark flavors $q$, $e_{q}$ is the electric charge of a specific quark flavor in units of the positron charge $e$ and where $H^{q(+)}(x, \xi, t) = H^{q}(x, \xi, t) - H^{q}(-x, \xi, t)$. Because of the model-dependency of this procedure we refrain from doing it here. We point out that a straightforward, however still model-dependent, access to the nucleon tomography is a feature of the CFF Ans\"atze proposed in our previous analysis \cite{Moutarde:2018kwr}.

\subsection{Subtraction constant}
\label{sec:results:sc}

The subtraction constant $C_{H}$ is evaluated with our results on the CFF $\mathcal{H}$ and the dispersion relation introduced in \refsec{sec:theory_framework}. More precisely, the imaginary part of the CFF $\mathcal{H}$ is integrated according to Eq. \eqref{eq:theory_framework:dr} in the range of $\epsilon < \xi' < 1$ and then subtracted from the corresponding real part. By studying $C_{H}$ as a function of $\epsilon$ the latter has been chosen to be $10^{-6}$, which introduces a little bias on $C_{H}$ comparing to $\epsilon = 0$. This bias is estimated to be smaller than $1\%$.

\begin{figure*}[!ht]
\begin{center}
\includegraphics[width=\figWidthSmall]{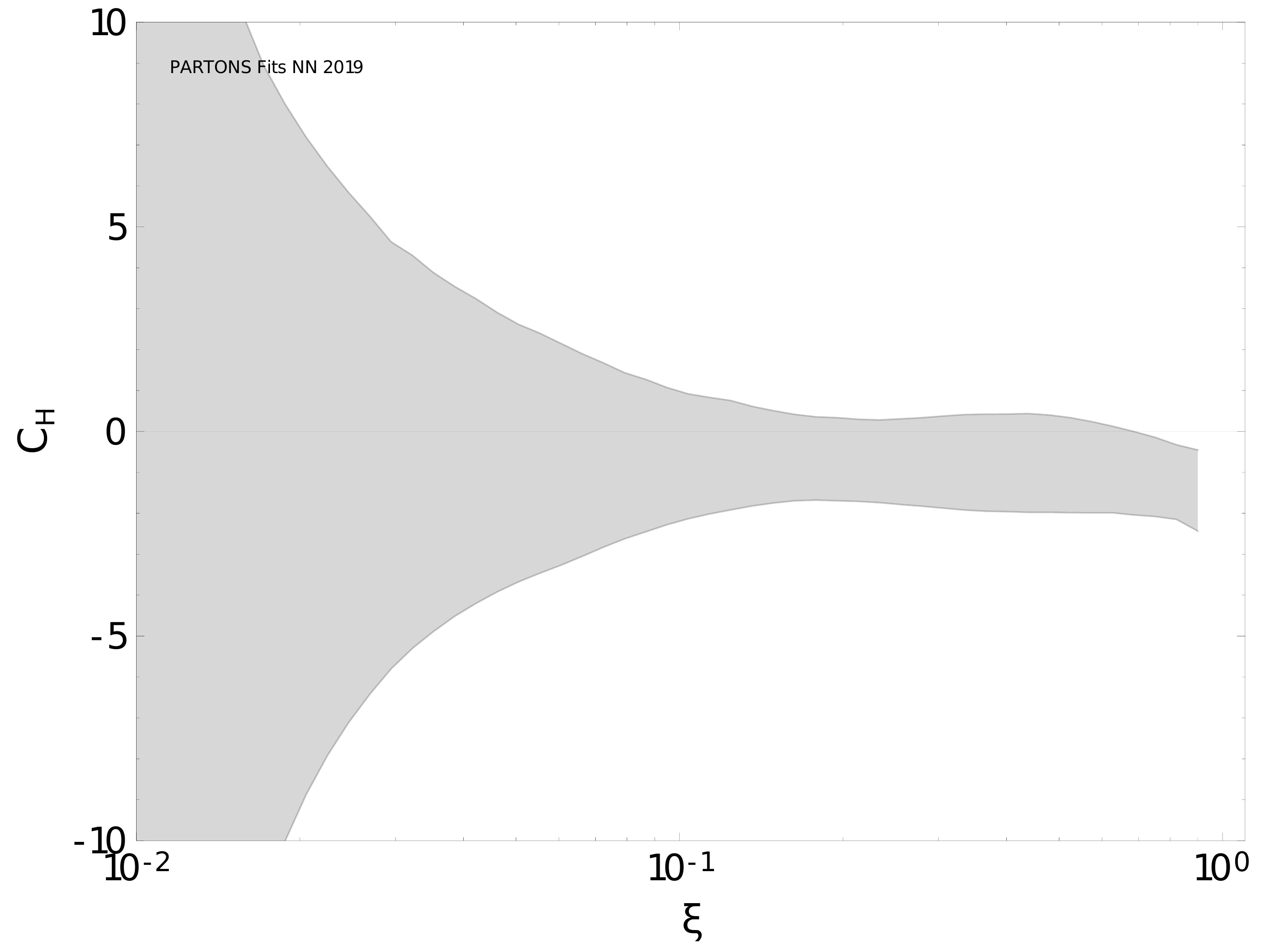}
\\
\includegraphics[width=\figWidthSmall]{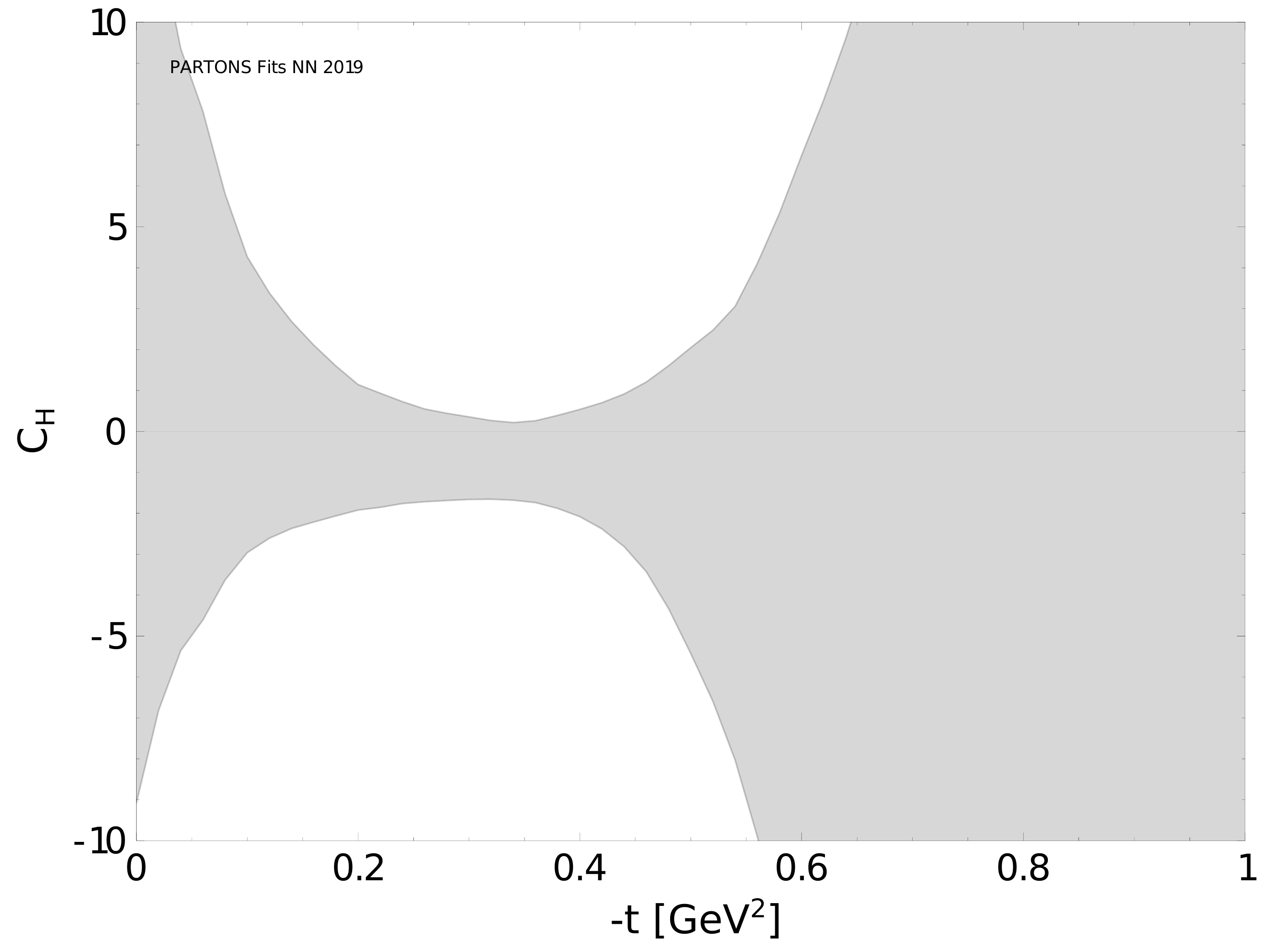}
\hspace{\figSpace}
\includegraphics[width=\figWidthSmall]{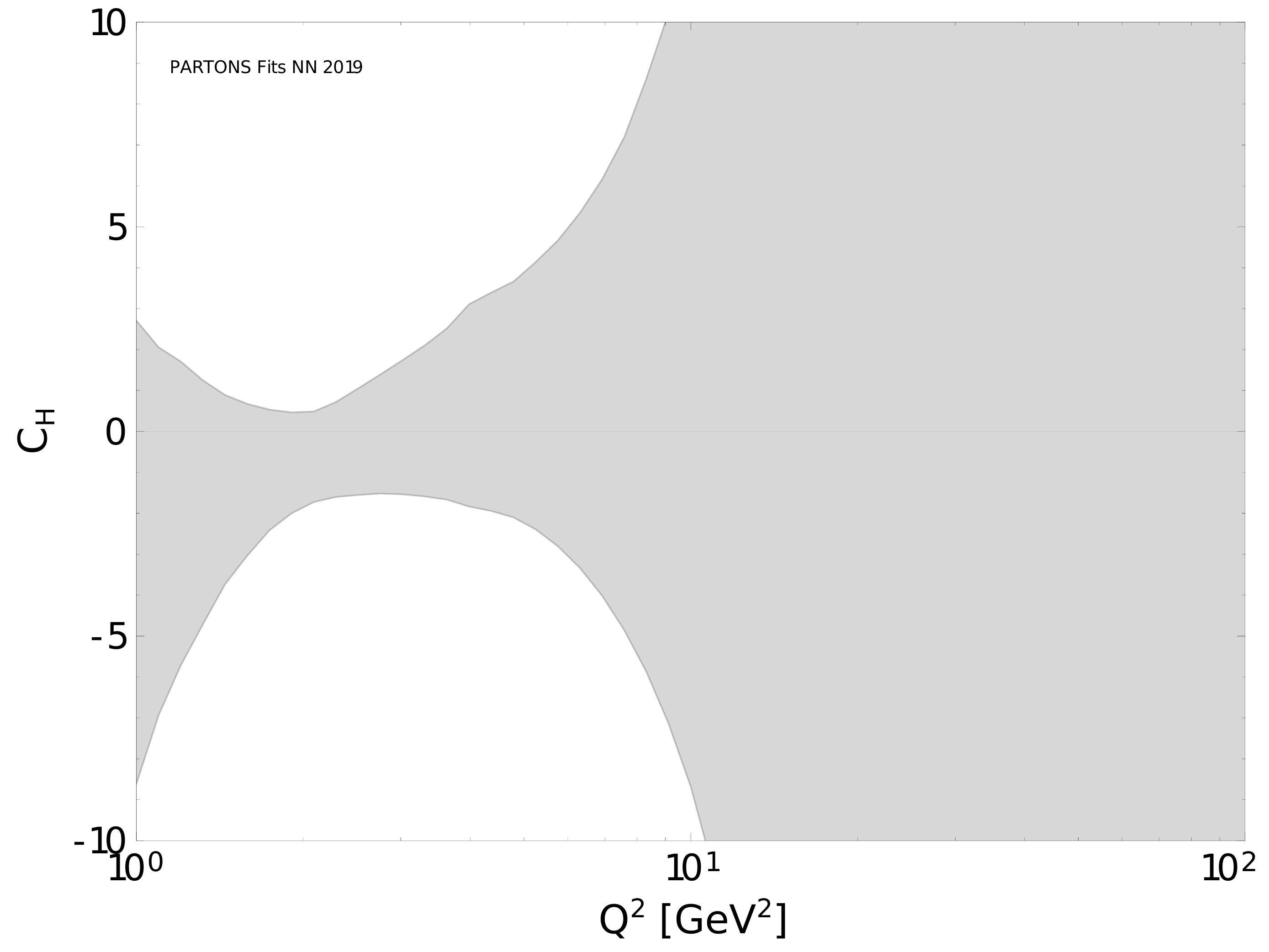}
\caption{Subtraction constant evaluated from the CFF $\mathcal{H}$ as a function of $\xi$ for $t = -0.3~\mathrm{GeV}^{2}$ and $Q^{2} = 2~\mathrm{GeV}^{2}$ (top), as a function of $-t$ for $\xi = 0.2$ and $Q^{2} = 2~\mathrm{GeV}^{2}$ (bottom left) and as a function of $Q^{2}$ for $\xi = 0.2$ and $t = -0.3~\mathrm{GeV}^{2}$ (bottom right).}
\label{fig:results:sc}
\end{center}
\end{figure*}

The subtraction constant for a given $Q^2$ and $t$ should be the same independently on the value of $\xi$ used in its extraction. This is demonstrated in Fig. \ref{fig:results:sc} (top), where $C_{H}$ is shown as a function of $\xi$ for example kinematics of $t = -0.3~\mathrm{GeV}^{2}$ and $Q^{2} = 2~\mathrm{GeV}^{2}$. We consider this test as a proof of consistency between the parameterizations of the real and imaginary parts of the CFF $\mathcal{H}$. 

The subtraction constant as a function of $t$ and $Q^{2}$ is shown in Fig. \ref{fig:results:sc} (bottom) for $\xi = 0.2$. As expected, the uncertainties are large in domains sparsely covered by data, in particular for $t \rightarrow 0$ and $Q^{2} \rightarrow \infty$, which are important for a direct interpretation of $C_{H}$ in terms of the energy-momentum tensor. However, the uncertainties obtained in domains covered by data are encouraging, so our result coming from a model-independent extraction can provide important constraints on models of $C_{H}$. Our findings are consistent with the recent observation of \refcite{Kumericki:2019ddg}.

\section{Summary}
\label{sec:summary}

In this paper we report the extraction of CFF parameterizations from proton DVCS data. The extracted quantities are the most basic observables as one can unambiguously access by exploring the DVCS process. We analyze the data in the region in which the interpretation in the language of GPDs is applicable. In this analysis, a dispersion relation is used to access the DVCS subtraction constant, which is related to the mechanical forces acting on partons in the nucleon. Also, the nucleon tomography that describes a spatial distribution of partons inside the nucleon is discussed.

The extraction of CFFs is done with the help of the artificial neural networks technique, allowing for an essential reduction of model dependency. The presented analysis includes such elements as the training of neural networks with the genetic algorithm, the careful regularization to avoid over-fitting and the propagation of experimental uncertainties with the replica method. The work is done within the PARTONS framework \cite{Berthou:2015oaw}.

The results of this analysis include in particular unbiased CFF parameterizations extracted in the three-dimensional phase-space of $(\xBj, t, Q^{2})$, with a reliable estimation of uncertainties. In addition, a direct extraction of the subtraction constant from the experimental data is presented.

The analysis is complementary to our previous one \cite{Moutarde:2018kwr}, where CFF parameterizations were constructed with the basic GPD properties acting as Ansatz building blocks. Although a physically motivated Ansatz gives more insight into GPD physics, its inherent model-dependency introduces an extra theoretical uncertainty, which usually can be only roughly estimated. The situation is opposite for the analysis presented in this paper -- the extraction of CFFs is unbiased, but the link with GPD physics is not as straightforward as in the case of previous analysis. Therefore, both analyses, which are performed on a similar data set, provide a complete picture of the DVCS process.

This work provides a benchmark for a powerful tool to be used in future GPD analyses. This tool allows in particular for a nearly model-independent estimation of the impact of future experiments, in particular those to be performed in the foreseen electron-ion facilities. CFF parameterizations that are presented in this paper can be used for a fast generation of DVCS cross sections, which may be useful \eg for Monte Carlo generation. Studies of Timelike Compton Scattering and higher twist contributions are also possible within the proposed framework. 

\onecolumn

\newpage

\vfill

\vfill

\vfill

\newpage

\vfill

\vfill

\vfill

\newpage


\vfill

\vfill

\vfill

\newpage

\vfill

\vfill

\vfill

\newpage


\vfill

\vfill


\vfill


\vfill

\vfill


\vfill

\twocolumn

\clearpage

\begin{acknowledgements}
The authors would like to thank C. Weiss and K. Kumerički for fruitful discussions. 
This work was supported by the Grant No. 2017/26/M/ST2/01074 of the National Science Centre, Poland 
and was done in part under CRADA agreement No. 2017S002 between Jefferson Lab,
USA and National Centre for Nuclear Research, Poland. The project is co-financed
by the Polish National Agency for Academic Exchange and by the COPIN-IN2P3
Agreement. The computing resources of {\'S}wierk Computing Centre, Poland are greatly acknowledged.
\end{acknowledgements}

\bibliographystyle{spphys}
\bibliography{bibliography,bibliography_data}

\begin{thebibliography}{10}
\providecommand{\url}[1]{{#1}}
\providecommand{\urlprefix}{URL }
\expandafter\ifx\csname urlstyle\endcsname\relax
  \providecommand{\doi}[1]{DOI \discretionary{}{}{}#1}\else
  \providecommand{\doi}{DOI \discretionary{}{}{}\begingroup
  \urlstyle{rm}\Url}\fi

\bibitem{Diehl:2015uka}
M.~Diehl, Eur. Phys. J. \textbf{A52}(6), 149 (2016).
\newblock \doi{10.1140/epja/i2016-16149-3}

\bibitem{Bacchetta:2016ccz}
A.~Bacchetta, Eur. Phys. J. \textbf{A52}(6), 163 (2016).
\newblock \doi{10.1140/epja/i2016-16163-5}

\bibitem{Ji:1996ek}
X.D. Ji, Phys. Rev. Lett. \textbf{78}, 610 (1997).
\newblock \doi{10.1103/PhysRevLett.78.610}

\bibitem{Ji:1996nm}
X.D. Ji, Phys. Rev. \textbf{D55}, 7114 (1997).
\newblock \doi{10.1103/PhysRevD.55.7114}

\bibitem{Polyakov:2002wz}
M.V. Polyakov, A.G. Shuvaev, {On 'dual' parametrizations of generalized parton
  distributions} (2002).
\newblock RUB-TP2-12-02, arXiv:hep-ph/0207153

\bibitem{Polyakov:2002yz}
M.V. Polyakov, Phys. Lett. \textbf{B555}, 57 (2003).
\newblock \doi{10.1016/S0370-2693(03)00036-4}

\bibitem{Polyakov:2018zvc}
M.V. Polyakov, P.~Schweitzer, Int. J. Mod. Phys. \textbf{A33}(26), 1830025
  (2018).
\newblock \doi{10.1142/S0217751X18300259}

\bibitem{Lorce:2018egm}
C.~Lorcé, H.~Moutarde, A.P. Trawiński, Eur. Phys. J. \textbf{C79}(1), 89
  (2019).
\newblock \doi{10.1140/epjc/s10052-019-6572-3}

\bibitem{Berthou:2015oaw}
B.~Berthou, et~al., Eur. Phys. J. \textbf{C78}(6), 478 (2018).
\newblock \doi{10.1140/epjc/s10052-018-5948-0}

\bibitem{Kumericki:2016ehc}
K.~Kumericki, S.~Liuti, H.~Moutarde, Eur. Phys. J. \textbf{A52}(6), 157 (2016).
\newblock \doi{10.1140/epja/i2016-16157-3}

\bibitem{dHose:2016mda}
N.~d'Hose, S.~Niccolai, A.~Rostomyan, Eur. Phys. J. \textbf{A52}(6), 151
  (2016).
\newblock \doi{10.1140/epja/i2016-16151-9}

\bibitem{Hassoun:1995:FAN:526717}
M.H. Hassoun, \emph{Fundamentals of Artificial Neural Networks} (MIT Press,
  Cambridge, MA, USA, 1995)

\bibitem{Kumericki:2011rz}
K.~Kumericki, D.~Mueller, A.~Schafer, JHEP \textbf{07}, 073 (2011).
\newblock \doi{10.1007/JHEP07(2011)073}

\bibitem{Moutarde:2018kwr}
H.~Moutarde, P.~Sznajder, J.~Wagner, Eur. Phys. J. \textbf{C78}(11), 890
  (2018).
\newblock \doi{10.1140/epjc/s10052-018-6359-y}

\bibitem{Burkert:2018bqq}
V.D. Burkert, L.~Elouadrhiri, F.X. Girod, Nature \textbf{557}(7705), 396
  (2018).
\newblock \doi{10.1038/s41586-018-0060-z}

\bibitem{Kumericki:2019ddg}
K.~Kumerički, Nature \textbf{570}(7759), E1 (2019).
\newblock \doi{10.1038/s41586-019-1211-6}

\bibitem{Kroll:1995pv}
P.~Kroll, M.~Schurmann, P.A.M. Guichon, Nucl. Phys. \textbf{A598}, 435 (1996).
\newblock \doi{10.1016/0375-9474(96)00002-4}

\bibitem{Belitsky:2012ch}
A.V. Belitsky, D.~Müller, Y.~Ji, Nucl. Phys. \textbf{B878}, 214 (2014).
\newblock \doi{10.1016/j.nuclphysb.2013.11.014}

\bibitem{Diehl:2007jb}
M.~Diehl, D.{\relax Yu}. Ivanov, Eur. Phys. J. \textbf{C52}, 919 (2007).
\newblock \doi{10.1140/epjc/s10052-007-0401-9}

\bibitem{Burkardt:2000za}
M.~Burkardt, Phys. Rev. \textbf{D62}, 071503 (2000).
\newblock \doi{10.1103/PhysRevD.62.071503, 10.1103/PhysRevD.66.119903}.
\newblock [Erratum: Phys. Rev. \textbf{D66}, 119903 (2002)]

\bibitem{Burkardt:2002hr}
M.~Burkardt, Int. J. Mod. Phys. \textbf{A18}, 173 (2003).
\newblock \doi{10.1142/S0217751X03012370}

\bibitem{Burkardt:2004bv}
M.~Burkardt, Phys. Lett. \textbf{B595}, 245 (2004).
\newblock \doi{10.1016/j.physletb.2004.05.070}

\bibitem{goodfellow2016deep}
I.~Goodfellow, Y.~Bengio, A.~Courville, \emph{Deep Learning}.
\newblock Adaptive Computation and Machine Learning series (MIT Press,
  Cambridge, MA, USA, 2016)

\bibitem{Mitchell:1998:IGA:522098}
M.~Mitchell, \emph{An Introduction to Genetic Algorithms} (MIT Press,
  Cambridge, MA, USA, 1998)

\bibitem{Prechelt2012}
L.~Prechelt, \emph{Early Stopping --- But When?} (Springer Berlin Heidelberg,
  Berlin, Heidelberg, 2012), pp. 53--67.
\newblock \doi{10.1007/978-3-642-35289-8_5}

\bibitem{Srivastava:2014:DSW:2627435.2670313}
N.~Srivastava, G.~Hinton, A.~Krizhevsky, I.~Sutskever, R.~Salakhutdinov, J.
  Mach. Learn. Res. \textbf{15}(1), 1929 (2014)

\bibitem{Goloskokov:2005sd}
S.V. Goloskokov, P.~Kroll, Eur. Phys. J. \textbf{C42}, 281 (2005).
\newblock \doi{10.1140/epjc/s2005-02298-5}

\bibitem{Goloskokov:2007nt}
S.V. Goloskokov, P.~Kroll, Eur. Phys. J. \textbf{C53}, 367 (2008).
\newblock \doi{10.1140/epjc/s10052-007-0466-5}

\bibitem{Goloskokov:2009ia}
S.V. Goloskokov, P.~Kroll, Eur. Phys. J. \textbf{C65}, 137 (2010).
\newblock \doi{10.1140/epjc/s10052-009-1178-9}

\bibitem{nissen03}
S.~Nissen, {\emph{Implementation of a Fast Artificial Neural Network Library
  (fann)}}.
\newblock Tech. rep., Department of Computer Science University of Copenhagen
  (DIKU) (2003).
\newblock \urlprefix\url{http://fann.sf.net}

\bibitem{fahlman:faster}
S.E. Fahlman, in \emph{{P}roceedings of the 1988 Connectionist Models Summer
  School} (San Francisco, CA: Morgan Kaufmann, 1989), p.~38

\bibitem{Dupre:2016mai}
R.~Dupre, M.~Guidal, M.~Vanderhaeghen, Phys. Rev. \textbf{D95}(1), 011501
  (2017).
\newblock \doi{10.1103/PhysRevD.95.011501}

\bibitem{Dupre:2017hfs}
R.~Dupré, M.~Guidal, S.~Niccolai, M.~Vanderhaeghen, Eur. Phys. J.
  \textbf{A53}(8), 171 (2017).
\newblock \doi{10.1140/epja/i2017-12356-8}

\bibitem{Defurne:2015kxq}
M.~Defurne, et~al., Phys. Rev. \textbf{C92}(5), 055202 (2015).
\newblock \doi{10.1103/PhysRevC.92.055202}

\bibitem{Defurne:2017paw}
M.~Defurne, et~al., Nature Commun. \textbf{8}(1), 1408 (2017).
\newblock \doi{10.1038/s41467-017-01819-3}

\bibitem{Akhunzyanov:2018nut}
R.~Akhunzyanov, et~al., Phys. Lett. \textbf{B793}, 188 (2019).
\newblock \doi{10.1016/j.physletb.2019.04.038}

\bibitem{Chekanov:2008vy}
S.~Chekanov, et~al., JHEP \textbf{05}, 108 (2009).
\newblock \doi{10.1088/1126-6708/2009/05/108}

\bibitem{Aktas:2005ty}
A.~Aktas, et~al., Eur. Phys. J. \textbf{C44}, 1 (2005).
\newblock \doi{10.1140/epjc/s2005-02345-3}

\bibitem{Aaron:2009ac}
F.D. Aaron, et~al., Phys. Lett. \textbf{B681}, 391 (2009).
\newblock \doi{10.1016/j.physletb.2009.10.035}

\bibitem{Airapetian:2001yk}
A.~Airapetian, et~al., Phys. Rev. Lett. \textbf{87}, 182001 (2001).
\newblock \doi{10.1103/PhysRevLett.87.182001}

\bibitem{Airapetian:2006zr}
A.~Airapetian, et~al., Phys. Rev. \textbf{D75}, 011103 (2007).
\newblock \doi{10.1103/PhysRevD.75.011103}

\bibitem{Airapetian:2008aa}
A.~Airapetian, et~al., JHEP \textbf{06}, 066 (2008).
\newblock \doi{10.1088/1126-6708/2008/06/066}

\bibitem{Airapetian:2009aa}
A.~Airapetian, et~al., JHEP \textbf{11}, 083 (2009).
\newblock \doi{10.1088/1126-6708/2009/11/083}

\bibitem{Airapetian:2010ab}
A.~Airapetian, et~al., JHEP \textbf{06}, 019 (2010).
\newblock \doi{10.1007/JHEP06(2010)019}

\bibitem{Airapetian:2011uq}
A.~Airapetian, et~al., Phys. Lett. \textbf{B704}, 15 (2011).
\newblock \doi{10.1016/j.physletb.2011.08.067}

\bibitem{Airapetian:2012mq}
A.~Airapetian, et~al., JHEP \textbf{07}, 032 (2012).
\newblock \doi{10.1007/JHEP07(2012)032}

\bibitem{Stepanyan:2001sm}
S.~Stepanyan, et~al., Phys. Rev. Lett. \textbf{87}, 182002 (2001).
\newblock \doi{10.1103/PhysRevLett.87.182002}

\bibitem{Chen:2006na}
S.~Chen, et~al., Phys. Rev. Lett. \textbf{97}, 072002 (2006).
\newblock \doi{10.1103/PhysRevLett.97.072002}

\bibitem{Girod:2007aa}
F.X. Girod, et~al., Phys. Rev. Lett. \textbf{100}, 162002 (2008).
\newblock \doi{10.1103/PhysRevLett.100.162002}

\bibitem{Gavalian:2008aa}
G.~Gavalian, et~al., Phys. Rev. \textbf{C80}, 035206 (2009).
\newblock \doi{10.1103/PhysRevC.80.035206}

\bibitem{Pisano:2015iqa}
S.~Pisano, et~al., Phys. Rev. \textbf{D91}(5), 052014 (2015).
\newblock \doi{10.1103/PhysRevD.91.052014}

\bibitem{Jo:2015ema}
H.S. Jo, et~al., Phys. Rev. Lett. \textbf{115}(21), 212003 (2015).
\newblock \doi{10.1103/PhysRevLett.115.212003}

\bibitem{Vanderhaeghen:1998uc}
M.~Vanderhaeghen, P.A.M. Guichon, M.~Guidal, Phys. Rev. Lett. \textbf{80}, 5064
  (1998).
\newblock \doi{10.1103/PhysRevLett.80.5064}

\bibitem{Vanderhaeghen:1999xj}
M.~Vanderhaeghen, P.A.M. Guichon, M.~Guidal, Phys. Rev. \textbf{D60}, 094017
  (1999).
\newblock \doi{10.1103/PhysRevD.60.094017}

\bibitem{Goeke:2001tz}
K.~Goeke, M.V. Polyakov, M.~Vanderhaeghen, Prog. Part. Nucl. Phys. \textbf{47},
  401 (2001).
\newblock \doi{10.1016/S0146-6410(01)00158-2}

\bibitem{Guidal:2004nd}
M.~Guidal, M.V. Polyakov, A.V. Radyushkin, M.~Vanderhaeghen, Phys. Rev.
  \textbf{D72}, 054013 (2005).
\newblock \doi{10.1103/PhysRevD.72.054013}

\end{thebibliography}

\end{document}